\documentclass[a4paper, 12pt]{article}
 
\usepackage[top=25truemm,bottom=25truemm,left=20truemm,right=20truemm]{geometry} 
\usepackage[english]{babel} \usepackage{tabularx,bm}
\usepackage{amsmath}

\usepackage{amscd}
\usepackage{amsfonts}

\usepackage{yfonts}

\newcommand{\bv}{{\bf v}}
\newcommand {\bp} {{\bf p}}
\newcommand {\bx} {{\bf x}}
\newcommand {\by} {{\bf y}}
\newcommand {\bk} {{\bf k}}
\newcommand {\bP} {{\bf  P}}
\newcommand {\bq}{{\bf q}}
\newcommand {\uu}{\underline}
\newcommand {\ba}{{\bf a}}

\newcommand {\cA} {\mathcal{A}}

\newcommand {\bra} {\langle}
\newcommand {\ket }{\rangle}
\newcommand {\cH} {\mathcal{H}}
\newcommand {\qp}{{\it q-p}}

\fontsize{12}{1pt} 

\usepackage[explicit]{titlesec}

\title { Quantum mechanics and quantum field theory.\\
Algebraic and geometric approaches.}
\author {
 Igor Frolov\\ Department of Mathematics,\\
 MEPhI \\(Moscow Engineering Physics Institute)\\ 115409, Kashirskoe shosse 31, Moscow, Russia \\
frolovi55@mail.ru\\
Albert Schwarz\\ Department of Mathematics\\ 
University of 
California \\ Davis, CA 95616, USA,\\ schwarz @math.ucdavis.edu \\}

\date{}							
 
\begin{document}
\maketitle 
{\bf Abstract}

This is a  non-standard exposition of the main notions of quantum mechanics and quantum field theory including some recent results.
  It is based on  the algebraic approach
where the starting point is a star-algebra and on the geometric approach where the starting point is a  convex set of states.  

Standard formulas for quantum probabilities are derived from decoherence. This derivation allows us to go beyond quantum theory in the geometric approach. Particles are defined as elementary excitations of the ground state (and quasiparticles as elementary excitations of any translation invariant state).
 The conventional scattering matrix does not work for quasiparticles (and even for particles if the theory does not have particle interpretation). The analysis of scattering in these cases is based on the 
 notion of inclusive scattering matrix, closely related to inclusive cross-sections. It is proven that the conventional scattering matrix can be expressed in terms of  Green functions (LSZ formula) and the inclusive
scattering matrix can be expressed in terms of generalized Green functions that appear in the Keldysh formalism of non-equilibrium statistical physics.

 The derivation of the expression of the evolution operator and other physical quantities in terms of functional integrals is based on the notion of the symbol of an operator; these arguments can be applied also in the geometric approach. This result can be used, in particular, to give a simple derivation of the diagram technique for generalized Green functions.

The notion of inclusive scattering matrix makes sense in the geometric approach (but it seems that one cannot give a definition of the conventional scattering matrix in this situation). 

The geometric approach is used to show that quantum mechanics and its generalizations can be considered as classical theories where our devices are able to measure only a part of observables.

\newpage

\tableofcontents

\section{Lecture 1} 
 
\subsection{Introduction} 

In the usual exposition of quantum mechanics, we live in Hilbert space and consider operators in this space. Self-adjoint operators correspond to observables. This is the approach that physicists use almost always, but it has its drawbacks. I will talk about other approaches. This is first of all the algebraic approach, where the starting point is an algebra of observables, an associative algebra with involution, in which the self-adjoint elements are observables. This approach is almost as old as quantum mechanics itself. Besides, I will talk about the geometric approach, in which the starting point is a set of states [1-7]. This approach I proposed a couple of years ago, and it is much more general than the algebraic approach. 
 
 The main thing, which, in my opinion, is not emphasized enough in the usual presentation of quantum mechanics (a little bit more is said in quantum field theory) is that the notion of a particle is not a primary notion in quantum theory. It is a secondary notion. Particles are elementary excitations of the ground state. Quasiparticles (also an important notion) are elementary excitations of any translation-invariant state. The basic notion that we have in the physics of elementary particles is the notion of scattering; this will be my main topic. I'll talk not only about the conventional scattering matrix ( related to scattering cross-sections) but also about
 the notion of an inclusive scattering matrix, which is closely related to the notion of inclusive scattering cross-sections. The scattering matrices can be expressed in terms of Green's functions by the well-known formula belonging to Lehmann, Simanzyk, and Zimmermann \cite{LSZ}, and the inclusive scattering matrices can be expressed in terms of generalized Green's functions, which first appeared in nonequilibrium statistical physics in  Keldysh formalism (see, for example, \cite {KELDYSH}). 

\subsubsection {Convex sets}

Before turning to physics, I want to say a few words about convex sets, which I will use many times. 

A convex set $\cal C$ is a subset of vector space that, together with every two points, contains a segment connecting these points.
The important thing is that in a convex set one can consider a mixture of points of the set. If we take some points of the set and ascribe a non-negative number to each point such that the sum of numbers equals one, then the sums of points $e_i\in \cal C$ with coefficients $p_i\geq 0, \sum p_i=1$ will also belong to the convex set. The sum $\sum p_ie_i\in \cal C$ is called the mixture of points $e_i$ with probabilities $p_i$. One can consider the numbers $p_i$ as weights, and then this sum will represent the center of gravity. Another important notion is that of the extreme point of a convex set. An extreme point is a point that does not lie inside any segment with ends belonging to the set. The extreme points of a polyhedron are vertices. For a ball, the extreme points lie on the boundary sphere. 

I will always assume that the vector space I am considering has some topology, there is a notion of limit and thus a notion of a closed set. I will assume that all convex sets I consider are closed;  then one can consider a mixture not only of a finite number of points but also a mixture of a countable number of points. In the latter case, the sum must be considered infinite. 

It is also possible to consider a mixture of points of any subset of a convex set if the subset is equipped by a probability distribution. If, say, on a sphere bounding a ball we have a probability distribution with probability density $\rho(\lambda)$, then we can consider the mixture of states on the sphere: just take the integral instead of the sum. If $\lambda$ are parameters describing points on the sphere, $\omega(\lambda)$ are points of the sphere, then we can consider such a mixture by taking the integral $\int\omega(\lambda) \rho(\lambda)d\lambda$ instead of the sum. If the convex set is compact, then each point is a mixture of its extreme points.

 \subsubsection {Quantum theory. Geometric approach}
 
Let me repeat what I need from quantum mechanics. For me, a state in quantum mechanics is a density matrix. A density matrix is a self-adjoint operator $K$, which is positive definite and whose trace is equal to one: $\mathrm{\mathrm{Tr}} K=1.$ The set of density matrices is convex, and its extreme points are called pure states. These pure states correspond to vectors in Hilbert space. Each normalized vector $\Psi$ corresponds to a density matrix, which is defined as an orthogonal projection on this vector: 
$$K_{\Psi}(x)= \bra x,\Psi\ket \Psi.$$

Note that if two vectors are proportional ($\Psi'=\lambda \Psi$), then the corresponding density matrices coincide: $K_{\Psi'}= K_{\Psi}$. 

The density matrix $K$ has a basis of eigenvectors $e_i$ with non-negative eigenvalues $p_i$ whose sum equals $1$. We can say that each of these vectors corresponds to a pure state, and the density matrix is a mixture of these pure states.  To check this we use representation in which the matrix $K$ is diagonal, then the diagonal elements are equal to  $p_i$. Since the trace is equal to $1$, their sum equals one: $\sum p_i=1$. Since the matrix is positive definite, the diagonal elements are non-negative: $p_i\geq 0$. Consequently, we can say that the density matrix is a mixture of pure states with probabilities $p_i$.

In textbook quantum mechanics, all this is told in reverse order, starting with pure states. Density matrices are defined as mixed states. 

I have discussed one way of representing the density matrix as a mixture of pure states. In fact, it can be done in an infinite number of different ways.

In the geometric approach, the starting point is the set of states. We assume that the set of states is a bounded closed convex subset of a topological linear space. It turns out that these assumptions are sufficient to develop a meaningful theory.

\subsubsection {Algebraic approach to quantum theory}

 Whereas in the geometric approach, the starting point is the space of states, in the algebraic approach the starting point is the algebra of observables $\cal A$. Recall that algebra is a vector space in which one can multiply elements with a distributive law; I will also require that the algebra is associative and is equipped with an involution (such an algebra is called $*$-algebra). A typical example of a $*$- algebra (for me the basic one) is an algebra of bounded operators in a Hilbert space. In this algebra, there is an involution $A\to A^*$, which corresponds to a transition to the adjoint operator. It has the property that if we pass twice to the adjoint operator, we return to the original one: $A^{**}=A$. If we take the adjoint operator to the product, it will be the product again, but in reverse order: $(AB)^*=B^*A^*$. Besides, involution is antilinear. In operator algebra these are simple properties, but for arbitrary associative algebras with involution ($*$-algebras) these are axioms. 
 
I always assume that there is some topology in algebra in which all operations are continuous. I usually won't talk about these topologies, first, because it takes time, and second, because different topologies can be equally sensible.  Sometimes it is necessary to have a norm in which the algebra is a Banach space, then one requires the inequality $||AB||\leq ||A||\cdot ||B||$ (this is the definition of a Banach algebra). Sometimes something else is required, such as that the algebra is $C^*$ -algebra.(This means that the norm of the product $A^*A$ is equal to the square of the norm of the operator $A$, that is, $||A^*A||=||A||^2$). I usually will not specify which topology is chosen. I want to emphasize that when considering a homomorphism or an automorphism of a $*$- algebra, I will always assume that it is continuous and agrees with involution. 

If there is an algebra with involution, the self-adjoint elements ($A=A^*$) correspond to physical quantities. Self-adjoint elements themselves do not form a subalgebra. The product of self-adjoint elements is not necessarily a self-adjoint element. This disadvantage of the algebraic approach was noticed as early as the 1930s; this led to the notion of Jordan algebra. Jordan noticed that although the product of self-adjoint elements is not self-adjoint, the anticommutator $A\circ B= AB+BA$, where $A$ and $B$ are self-adjoint, is again self-adjoint. He axiomatized this operation.
The theory of Jordan algebras was constructed in the 1930s, mainly in the famous work of Jordan, Wigner, and von Neumann \cite {JNW}. But although  Jordan algebra is a very beautiful and really useful object in many parts of mathematics, it has not yet had much use in physics. Now it has naturally appeared in the geometric approach and may come back to physics again. 

If we start from a $*$-algebra (an associative algebra with involution), we can define the notion of state: a state is a linear functional $\omega$ on  algebra $A$ that satisfies the non-negativity condition on the elements of the form $A^*A$:
$$\omega(A^*A) \geq 0.$$
We say that linear functionals corresponding to states are positive functionals.

States differing by a numerical factor are identified. It is often convenient to consider only states satisfying the condition $\omega(1)=1$ (normalized states).

 Now we can define the notion of the expectation value of a physical quantity in a given state. For the function $f(A)$ of $A\in \cal A$ the mathematical expectation  (the average)
is given by the formula:
$$\bra f(A) \ket_{\omega}=\omega (f(A)).$$
For a $C^*$-algebra, we can define $f(A)$ for any continuous function $f$ of a self-adjoint element $A$; then, knowing the expectation values for all continuous functions, we can define the notion of the probability distribution of a physical observable in a normalized state (a  state for which $\omega(1)=1$).

The notion of state alone is not enough: you also need the notion of evolution, because the goal of physics, like any science, is to make predictions. A physicist, first of all, considers the problem: if an initial state is known, what is the way to predict what will happen afterward?

In the algebraic approach, in order to define the notion of  evolution one should first consider the group $Aut (\cal A)$ of automorphisms of the $*$-algebra $\cal A$. Recall that automorphisms of $\cal A$ must always commute with involution, hence the group of automorphisms, naturally acting on linear functionals, transforms positive functionals into positive ones (states into states). In any approach to quantum theory, states must depend on time. There must be an evolution operator $U(t)$ that transforms a state at the initial moment into a state $\omega(t)$ at some other time $t$; in other words, $\omega(t)=U(t) \omega(0) $.

In the algebraic approach, we can assume that the operators $U(t)$ come from automorphisms of the algebra $\cal A$ denoted by the same symbol. As in textbook quantum mechanics, there is the Schr\"odinger picture, where the state evolves, and there is the Heisenberg picture, where the operator evolves. These two pictures are equivalent: 
$$\omega(t)(A)=\omega(A(t)).$$
(Observing the dynamics of a state  $\omega$  when an algebra element $A$ is fixed is the same as observing the dynamics of an algebra element when the state does not change.)

 In physics, the evolution operator is usually calculated from the equation of motion describing the same evolution operator, but over infinitesimal time. If there is invariance with respect to the time shift, then it can be argued that the operator describing a change over an infinitesimal time interval is itself independent of time. It has already been said that a change over finite time must be an automorphism of algebra, so changes over infinitesimal time intervals are infinitesimal automorphisms. Knowing an infinitesimal automorphism $H$, we solve the equation of motion $dU/dt=HU.$ The solution  obeying $U(0)=1$ can be written in the form  $U(t)=e^{Ht}\in \cal A$. As a result, we obtain a one-parameter group of automorphisms consisting of transformations of the form $e^{Ht}$ (evolution operators).  (In quantum mechanics textbooks  you will see the imaginary unit in the exponent - I do not write it, but, of course, this is irrelevant.)   The state  $\omega(t)=U(t) \omega(0) $ obeys the equation of motion 
 $$\frac {d\omega}{dt}=H\omega(t).$$
 
 The operator $H$ is an analog of Hamiltonian in quantum mechanics; I  say that  $H$ is a "Hamiltonian".
 
I have not given a formal definition of infinitesimal automorphism. One possible formal definition: an infinitesimal automorphism is a tangent vector to a curve in an automorphism group in a unit element of this group. We require a little more: that this curve be a one-parameter subgroup. 

It is important to note that an infinitesimal automorphism is a derivation. This means that it must satisfy the Leibniz rule: applying it to the product $xy$ one must first apply it to the first factor, leaving the second one unchanged, then to the second factor, leaving the first one unchanged: $A(xy)=(Ax)y+x(Ay)$. This follows instantly from the very definition of automorphism and the very definition of infinitesimal automorphism. If  $A$ is an infinitesimal automorphism, then $1+tA$ for small $t$ is already an automorphism. (More precisely, $1+tA$ plus something of higher order on $t$ is an automorphism.) If you apply the definition of automorphism, you just get the Leibniz rule. 

Conversely, if  $A$ is a derivation, that is, if the Leibniz rule is satisfied, and in addition, it is consistent with involution, that is, the condition $(Ax)^*=A(x^*)$ is satisfied, then we can hope that $A$ is an infinitesimal automorphism. In order to check that this is true, we need to write the equation 
$$dU/dt=AU,$$ where $U(t)$ is an element of the algebra $\cal A.$
If this equation has a solution with the initial condition $U(0)=1$, then $A$ is an infinitesimal automorphism. It can play the role of a  ``Hamiltonian'' (as in textbook quantum mechanics where any self-adjoint operator can play the role of a Hamiltonian). If the algebra is finite-dimensional, we can apply the existence theorem for solutions of differential equations. In this case, the notions of derivation and infinitesimal automorphism are equivalent. Since algebra in physics is infinite-dimensional, in our situation not every derivation defines an infinitesimal automorphism. It is necessary for the equation $dU/dt=AU$ to have a solution. 

 It is easy to check that derivations form a Lie algebra. The same is true for derivations that agree with involution. One can say that derivations consistent with involution form the Lie algebra of the group of automorphisms $Aut(\cal A)$. For the case of infinite-dimensional groups, the notion of Lie algebra is not very well defined, but nevertheless, it is an important notion that works in many cases.
 
 I considered the case when the equation of motion does not depend on time but this is not necessary. The ``Hamiltonian'' may depend on time, and then the equation of motion for the evolution operators has the form:
$$\frac {dU }{dt}=H(t)U(t).$$
 If the operator $H(t)$ does not depend on $t$, then the evolution operators form a one-parameter group:
 $$U(t+\tau)=U(t)U(\tau).$$ 

In textbook quantum mechanics a  density matrix $K$ corresponds to a linear functional $\omega (A)=\mathrm{Tr} KA$ on the algebra of bounded operators; this functional satisfies the condition $\omega(A^*A)\geq 0$. (It is easy to see that this condition follows from the positive definiteness of operator $K$). The evolution of the density matrix is described by an equation in which the right-hand side is a commutator with a self-adjoint operator ( up to a constant factor). This equation has the form 
 $dK/dt=H(K)$, where
 $H(K)=[\hat H, K]/i\hbar$,
 $\hat H$ is a  Hamiltonian of textbook quantum mechanics, and $H$ is a ``Hamiltonian".

Here we introduced the following notations: operators in Hilbert space are operators with a hat, and operators acting on density matrices are operators without a hat. According to Stone's theorem,  (not necessarily bounded) self-adjoint operators in Hilbert space correspond to one-parameter subgroups of the group of unitary operators.  In Stone's theorem, the subgroups are continuous in the strong sense.
(I will not explain what this is - I will not need it).  If a self-adjoint operator is bounded, then the corresponding one-parameter subgroup is differentiable in the sense of norm convergence.  In what follows I will not pay attention to these subtleties.

I want to say a few more words about the relation between the algebraic approach and the standard approach based on Hilbert spaces and explain why the algebraic approach is better. Suppose we have an involution-preserving representation of the algebra $\cal A$ by operators in the Hilbert space $\cal H$. In other words, consider an involution-preserving homomorphism of the algebra $\cal A$ into an algebra of operators. Let us denote the operator corresponding to the element $A$ of the algebra $\cal A$ by $\hat A$, then each normalized vector $\Phi \in \cal H$ specifies a normalized state $\omega$ of the algebra $\cal A$ by the formula  
$$\omega (A)=\langle \hat A\Phi,\Phi\rangle.$$
Moreover, each density matrix $K$ specifies a state according to the formula $\omega (A)=\mathrm{Tr} (K\hat A)$. 
In other words, it is possible to obtain states from vectors in Hilbert space. A natural question arises: can all states be obtained this way? The answer to this question is positive. Every state can be represented by a vector in Hilbert space, and this is the reason why physicists are able to work all the time in Hilbert space. 

Why is this inconvenient in many cases? This is because for the same algebra of observables it is necessary to consider different Hilbert spaces. For example, in statistical physics we consider equilibrium states. Each equilibrium state lies in its own Hilbert space. This is not always convenient. 

One Hilbert space, as a rule, is sufficient
 in quantum field theory, because there we usually consider a Hilbert space, in which the ground state lies. Its elements correspond to excitations of the ground state. In quantum field theory usually we consider only excitations of the ground state. However, it is impossible to use only one Hilbert space in quantum electrodynamics.

 Now I am going to prove that every state of an algebra with involution is represented by a vector from a Hilbert space. 
  I will construct a pre-Hilbert space $\cal E$ for each algebra $\cal A$ and a state $\omega$. ( Here it is convenient  to work with pre-Hilbert spaces.) I will construct a representation  $A\to \hat A$ of the algebra by operators in pre-Hilbert space $\cal E$  in such a way that some cyclic vector, which I denote by $\theta\in \cal E$, will correspond to the state $\omega$. ( This means that $\omega (A)=\langle \hat A\theta,\theta\rangle$.)  The fact that a vector is cyclic means that any other vector can be obtained from it using operators from algebra (all vectors have the form $\hat A\theta$, where $A\in \cal A$). 
 
 The construction I am going to explain is unambiguous  (up to equivalence), as will be seen from the proof. Let me assume that I already have such a representation. I can  define the scalar product in  $\cal A$ 
 by the formula 
$$\langle A,B\rangle=\omega (B^*A).$$
Knowing this scalar product in algebra, I can calculate the scalar product of vectors $\hat A\theta$ and $\hat B\theta$:
$$\bra \hat A\theta,\hat B\theta\ket=\bra \hat B^*\hat A\theta,\theta\ket= \bra\widehat{(B^*A)}\theta,\theta\ket= \omega (B^*A).$$
Since the vector $\theta$ is cyclic, each vector of the space has the form $\hat A\theta$.  Now I can say that there is a mapping $\nu: \cal A\to \cal E$, which transforms $A$ into $\hat A\theta$, and this mapping is surjective. It follows that $\cal E$ is obtained from $\cal A$ by factorization. We need to factorize over all vectors which give 0 in scalar product with any other vectors (zero vectors). 

Now we can answer the question: how to construct $\cal E$ from the algebra $\cal A$ and $\omega$?  One should take the algebra $\cal A$, introduce the scalar product in it as $\omega (B^*A)$, and factorize with respect to zero vectors. We obtain a pre-Hilbert space. ( The scalar product in $\cal A$ descents to a scalar product in quotient space).  I did two things: first, I built a pre-Hilbert space, and second, I proved that my construction is essentially unique, nothing else can be done. I derived it from cyclicity. This reasoning (the Gelfand-Naimark-Segal or GNS construction) is the most important element of what I am going to say in this lecture. I will use this construction many times. Instead of pre-Hilbert space, I can consider its completion 
  -Hilbert space  $\bar {\cal E}$ (then vector $\theta$ will be cyclic in a weaker sense: vectors of type $\hat A\theta$ will be dense in $\bar{\cal E}$).

To illustrate, let us take some stationary state (a state that does not change during evolution) and apply the GNS construction to it. Then we get some Hilbert space and a cyclic vector in it, which also will be stationary ( will not depend on time). 

Assertion: if we start with a stationary state, then the evolution operators $U(t)$ descend to the Hilbert space.

This is very easy to understand. In GNS construction we used $\omega (B^*A)$ as a scalar product. But this scalar product is invariant with respect to operators $U(t)$ because $\omega$ is invariant (that is, it is not changed by evolution operators). Since the scalar product is invariant, the operators $U(t)$ descend into unitary operators $\hat U(t)$. The operators $\hat U(t)$ form a one-parameter group. It has a generator (infinitesimal automorphism) $\hat H$, and this is what in physics is called the Hamiltonian. (Actually, this is not exactly true, because in physics the Hamiltonian is assumed to be a selfadjoint operator, therefore we need an imaginary unit in the definition: $\hat U(t)=e^{-i\hat H t}$.) 

I say that  $\omega$  is a  ground state if the spectrum of the operator $\hat H$ is non-negative.  Note that the ground state will have zero energy under this definition. This is consistent with the standard definition. If we apply the GNS construction to the algebra of bounded operators and the state corresponding to the eigenvector of the Hamiltonian with eigenvalue $E$, then the generator of the group $\hat U(t)$ constructed with the GNS construction is $\hat H-E$. In quantum field theory, we always say that we will count the energy from the ground state and ignore the infinite contribution to this energy. The algebraic GNS construction does this automatically.

I kept saying that I consider the algebraic approach to quantum mechanics. This approach works perfectly well in classical mechanics too. To show it, we will work in the Hamiltonian formalism and then we can repeat the same reasoning. The pure state is described by generalized momenta $p= (p_1,...,p_n)$ and generalized coordinates $q=(q^1,...,q^n)$ representing points in $2n$-dimensional space, which is called phase space. This is a pure state, but, just as in quantum mechanics, mixed states can also be considered. 

A mixed state is a probability distribution on a phase space or a positive measure on a phase space (the measure of the whole space is assumed to be equal to $1$). All these probability distributions form a convex set. Pure states are the extreme points of this set. A pure state is a probability distribution that is supported at exactly one point. It is described by a probability density, which is a delta function. Any function is a superposition of delta functions (in other words, any function can be represented as an integral of delta functions). This means that any probability distribution on a phase space corresponds to a probability distribution on pure states, and pure states can be identified with extreme points of the space of all states. 

In classical mechanics, every state can be represented in a single way as a mixture of pure states. This distinguishes classical mechanics from quantum mechanics, where a state can be represented as a mixture of pure states in many different ways. Later I will explain how quantum mechanics can be derived from classical mechanics by restricting the set of observables. It is natural from a physical point of view to assume that our devices can measure only a part of observables. I show that in such a situation,  classical mechanics can lead to quantum mechanics ( Lecture 10).

Next, I will recall the well-known Hamilton equation of motion 
$$\frac{dq}{dt}=\frac{\partial H}{\partial p},
\frac{dp}{dt}=-\frac{\partial H}{\partial q}$$
and the Liouville equation for the probability density function, which is written in terms of Poisson brackets as:
$$\frac{d}{dt}\rho(p, q, t)=\{H,\rho (p,q,t)\}.$$
The Poisson brackets in this case are defined by the formula
$$ \{f,H\}=-\frac{\partial f}{\partial p}\frac{\partial H}{\partial q}+\frac{\partial f}{\partial q}\frac{\partial H}{\partial p}.$$

In order to determine the evolution operator $U(t)$ it will be necessary to solve the equation 
$\frac{d}{dt}U(t)=LU(t),$ where $L\rho=\{H,\rho\}.$
This equation is equivalent to the Liouville equation.  To verify this, one must 
check that on pure states it reduces to the Hamiltonian equations. In classical mechanics, just as in quantum mechanics, you can alternatively follow how observables (instead of states) evolve. The observables 
 are real functions $f(p,q)$ on the phase space. One can easily deduce from Hamilton equations that the evolution of observables is governed by  the equation 

$$\frac{d}{dt}f(p(t), q(t))
=-\frac{\partial f}{\partial p}\frac{\partial H}{\partial q}+\frac{\partial f}{\partial q}\frac{\partial H}{\partial p}$$
or
$$\frac{d}{dt}f(p(t), q(t))=\{f,H\}.$$

 Now we see that we are in the same situation as quantum mechanics or, conversely, quantum mechanics is in the same situation as classical mechanics. We have observables that can be multiplied. They form an algebra $\cal A$. There is a notion of involution. (Involution is simply complex conjugation). Each state $\omega$ corresponds to a linear functional on the algebra of observables $\cal A$: you have to take the integral of the function with respect to the probability distribution
$$\omega(f)=\int f\omega.$$ 
This functional satisfies the positivity condition: $\omega (f)\geq 0$ if $f\geq 0$. Functions of the form $A^*A$ are certainly positive (just the square of the modulus), hence the functional $\omega (f)$ is a state in the sense of the algebraic approach.  We see that
 classical mechanics enters as a small piece into the algebraic approach to quantum mechanics, but there is a difference: in classical mechanics the algebra of observables is commutative. 
\newpage

\section{Lecture 2}

\subsection {Quantum mechanics as a deformation of classical mechanics. Weyl algebra}

Now I will try to explain how a mathematician could derive quantum mechanics from classical mechanics. He knows that quantum mechanics in the limit of small Planck constant reduces to classical mechanics. 
This means that quantum mechanics is obtained as a deformation, a small modification of classical mechanics. We should have a family of algebras ${\cal A}_{\hbar}$ that depends on the Planck constant $\hbar$. If Planck constant $\hbar$ is equal to zero we should obtain classical mechanics, i.e. we should get a commutative algebra with product $A\cdot B$. Let us assume that all these algebras are defined on the same vector space, i.e., addition and multiplication by a number are independent of the Planck constant, and multiplication   $A\cdot_{_\hbar}B$ of elements of the algebra ${\cal A}_{\hbar}$ depends on the Planck constant. Now consider a commutator in this algebra as a function of  the Planck constant: 
$$[A,B]_{\hbar}=A\cdot_{_\hbar}B -B\cdot_{_\hbar}A$$ 
Our main requirement is that this commutator tends to zero when the Planck constant tends to zero: $\hbar\to 0$. Let us assume that the dependence on the Planck constant is smooth. This means that the commutator can be represented as an expression linear with respect to $\hbar$ plus something of a higher order:
$$[A,B]_{\hbar}=i\{A,B\}\hbar+O(\hbar^2).$$ 
 It is easy to prove that the linear part
  has
 the same properties as the Poisson bracket. This means that the  operation $\{A,B\}$  is a derivation with respect to both arguments (satisfies Leibniz rule):
$$\{A\cdot B,C\}=\{A,C\}\cdot B+A\cdot\{B,C\}$$
and, in addition, it satisfies the axioms of the Lie algebra.
To prove this fact we use the following
properties of the  commutator in associative algebra: 
$$[A, B]=-[B,A],$$
$$[A,[B,C]]+[B,[C,A]]+[C,[A,B]]=0,$$
$$[AB,C]=[A,C]B+A[B,C].$$
These equations
must be satisfied  for each Planck constant $\hbar$. Let us decompose all these equations with respect to Planck constant. In the second equality, it is necessary to decompose to the second order, and in the rest - to the first order. Equating the leading terms with respect to $\hbar$, we obtain the desired properties.

 We have proven that in the limit  $\hbar \to 0$ we obtain classical mechanics from quantum mechanics. The commutator is related to the Poisson bracket in this limit. 

Let us see if it is possible to go the other way around. We have seen how quantum mechanics turns into classical mechanics. And now we wonder: can we get quantum mechanics from classical mechanics? 
To do this, I first describe all possible Poisson brackets in the case when the algebra $\cal A$ is an algebra of polynomial functions on some vector space with coordinates $(u^1,...,u^n)$. In order to calculate how the Poisson bracket works, you only need to know the Poisson bracket of  coordinates. This is because we deal with polynomials, and I have a property that allows you to calculate the Poisson bracket of products. A polynomial is a linear combination of the products of the coordinates, hence the bracket of two polynomials can be calculated. The result of the calculation is as follows:
\begin{equation}\nonumber
 \{A,B\}=\frac {1}{2}\sigma^{kl}(u)\frac{\partial A}{\partial u^k}\frac {\partial B}{\partial u^l,}
 \end{equation}
where $\sigma^{kl}(u)$ denotes the Poisson bracket of coordinates $u^k,u^l.$
 
If $\sigma$ is antisymmetric and independent of $u$, one can check that this expression satisfies the conditions imposed on the Poisson bracket. This is exactly the situation that arises when we deal with the standard Poisson bracket in the phase space.

 Now I can ask the question: how to deform the Poisson bracket to get a family of associative algebras? This problem is not easy. It was solved relatively recently by Kontsevich \cite {KONTSEVICH}. But in the case when the Poisson bracket of two coordinates does not depend on $u$, it is very easy to answer this question.  I will define the algebra ${\cal A}_{\hbar}$ as an associative algebra with generators 
 ${\hat u}^k$, obeying the relation: 

\begin {equation}\nonumber
{\hat u}^k\hat u^l-\hat u^l\hat u^l=i\hbar\sigma ^{kl}.
\end{equation}

Of course, this could also be done when $\sigma$ depends on $u$, but then we do not know whether we would get an associative algebra. If $\sigma$ does not depend on $u$, then we get an associative algebra, which is called a Weyl algebra. If we start with polynomials, this is the only 
 way to deform the Poisson bracket. I introduce involution in the Weyl algebra by assuming the generators ${\hat u}^k$ to be self-adjoint.

 We obtained commutation relations, which in a slightly different form are well-known from quantum mechanics textbooks. 
To show this, I require that the matrix $\sigma$ be non-degenerate. This is an antisymmetric matrix and, literally, it cannot be diagonalized, but it can be written in a suitable basis as 
a block-diagonal matrix consisting of two-dimensional blocks.
If we take advantage of this, we can reduce the commutation relations in Weyl algebra to the commutation relations 
$$\hat p_k\hat p_l=\hat p_l\hat p_k,\;\;\; \hat q^k\hat q^l=\hat q^l\hat q^k,\;\; \hat p_k\hat q^l-\hat q^l\hat p_k= \frac{\hbar}{i}\delta_k^l.$$
These are commutation relations for coordinates and momenta in the standard exposition of quantum mechanics. They are called canonical commutation relations (CCR).

Instead of self-adjoint generators, one can take other generators that are not self-adjoint but are adjoint to each other and satisfy the relations 
$\hat a_k \hat a_l=\hat a_l\hat a_k, \;\;\;\hat a^*_k\hat a^*_l=\hat a^*_l\hat a^*_k,\;\;\;\;\hat a_k\hat a^*_l-\hat a^*_l\hat a_k=\hbar \delta_{kl}.$
For example, one can take
$\hat a_k=\frac{1}{\sqrt{2}}(\hat q^k+i\hat p_k), \hat a^*_k=\frac{1}{\sqrt{2}}(\hat q^k-i\hat p_k).$

This is how the creation and annihilation operators are denoted, but, so far, it is a formal mathematical object. I introduced them formally and wrote commutation relations for them. These commutation relations are also called canonical commutation relations. 

Can I now say that the resulting algebra ${\cal A}_{\hbar}$ is a deformation of the commutative algebra? Formally, I can't, because when I defined the notion of deformation, I required that all these algebras be defined on the same space, otherwise it would be difficult to consider all of them simultaneously. My commutative algebra consisted of polynomials, and the new algebra consists of who knows what. But it can also be made to consist of polynomials. This is done very simply. 

The algebra   ${\cal A}_{\hbar}$ is generated by the elements $\hat q^k$ and $\hat p^k$, that is, its elements are sums of monomials composed of generators $\hat q^k$ and $\hat p^k$. Because of the commutation relations, I can shift all the $\hat q^k$ generators to the left and $\hat p^k$ to the right (or vice versa) and then remove the hats from them. Then we get a regular polynomial and I can say that the element from my algebra is represented by a polynomial, which is called a \qp-symbol. Now the algebra is defined on the space of polynomials. This is not a very good representation because it is not consistent with involution. Nevertheless, it is very useful in many cases. 

If we start with generators $\hat a_k, \hat a_k^*$, we can use the same idea: shift $\hat a_k^*$ to the left, $\hat a_k$ to the right, and you get what is called the normal form of a Weyl algebra element. Now you can remove the hats and get a polynomial, which is called the Wick symbol. Physicists do not use the term ``Wick symbol,'' but they use the words``normal form'' all the time.  The Wick symbols agree with involution. (Involution in algebra corresponds to a complex conjugation of polynomials.)

We can consider a Weyl algebra with an infinite number of generators.
So far we have considered the parameter $k$ to be discrete (although the number of $u_k$ or $a_k$ with the index $k$ could be infinite), but we can consider this parameter to be continuous. For example, consider an algebra with generators
$\hat a(k), \hat a^*(l)$
and relations
$$ \hat a(k) \hat a(l)=\hat a(l)\hat a(k),\;\;\; \hat a^*(k)\hat a^*(l)=\hat a^*(l)\hat a^*(k),$$
$$\hat a(k)\hat a^*(l)-\hat a^*(l)\hat a(k)=\hbar \delta (k,l).$$
In this case, instead of the Kronecker symbol we use its continuous counterpart: the $\delta-$function. Since the function $\delta (k,l)$ is a generalized function, the generators $\hat a(k), \hat a^*(l)$ must also be treated as generalized functions. 
A generalized function is a function that only makes sense under the integral sign. Only the elements $\hat a(f)=\int f(k)\hat a(k)dk$ and $\hat a^*(g)=\int g(l)\hat a^*(l)dl$ that represent formal integrals are meaningful. These elements depend linearly on $f$ and $g$, respectively, and satisfy the commutation relations:
$\hat a(f)\hat a(g)=\hat a(g)\hat a(f),\;\;\; \hat a^*(f)\hat a^*(g)=\hat a^*(g)\hat a^*(f),$ 
$\hat a(f)\hat a^*(g)-\hat a^*(g)\hat a(f)=\hbar \langle f, \bar g\rangle.$
For these relations to make sense, the scalar product $\langle f, \bar g\rangle$ must be defined. Since the scalar product depends on $g$ antilinearly, we take $\bar g$  in the last formula. For simplicity, I will usually assume that the index $k$ is discrete.

Transition to symbols is an operation that is closely related to the operation of quantization. What is quantization? Starting with classical Hamiltonian we want to obtain its quantum counterpart. If the Hamiltonian depends on $u_k$, then simply replacing $u$ by $\hat u$ creates a problem, in what order to put these generators? In the classics, the order is not important: $u_1u_2$ and $u_2u_1$ are the same, but if we go to quantum mechanics, to Weyl algebra, the result depends on the order. This is what is called "ordering ambiguity" and means that there is no unambiguous quantization procedure. It can be made unambiguous by choosing the notion of a symbol, but there are many symbols. 

There are such cases when the quantum Hamiltonian has a natural definition. It is, for example, a standard situation in classical mechanics when the Hamiltonian is represented as a sum of kinetic and potential energies. The kinetic energy depends only on momenta, and the potential energy depends only on coordinates, and then one can put hats and the construction will be absolutely unambiguous because quantum momenta commute with each other and quantum coordinates commute with each other.

In order to write the equation of motion in the Heisenberg picture there is a standard way: in the classical equations of motion in place of Poisson brackets one must write commutators: 
 $$\frac {\partial \hat u^k}{dt}= i[\hat H,\hat u^k].$$
In this formula, I take $\hbar=1.$
In what follows, Planck constant $\hbar$ will always be taken as equal to one, unless otherwise stated.

The equation of motion is meaningful if the Hamiltonian $\hat H$ is an element of a Weyl algebra. I have already explained that the equation of motion must contain an operation that satisfies the Leibniz rule. Such an operation is called derivation. The commutator of the form $D_h(a)=[h,a]$  satisfies the Leibniz rule $[h,ab]=[h,a]b+a[h,b]$ for any algebra and so as long as $\hat H$ is an element of a Weyl algebra, everything is fine. The only problem is that $\hat H$  is very often not an element of a Weyl algebra in the case of an infinite number of degrees of freedom. A typical example is the Hamiltonian of the form 
  
  $$\hat H=\sum \epsilon_k\hat a_k^*\hat a_k.$$
When the number of indices is infinite, it is an infinite sum.  This Hamiltonian does not belong to the Weyl algebra.
Nevertheless, one can formally take the commutator of the Hamiltonian $\hat H$ and $\hat a_k$. We obtain the equation of motion, which we will encounter more than once:

  $$\frac {d\hat a_k}{dt}=-i\epsilon_k \hat a_k, \;\;\; \frac{d \hat a_k^*}{dt}=i\epsilon_k\hat a^*_k.$$
Thus, in the case of an infinite number of degrees of freedom, the Hamiltonian is  simply a formal expression of the form

$$ \hat H=\sum \Gamma_{m,n}(k_1,...,k_m, l_1,...,l_n)\hat a_{k_1}^*...\hat a_{k_m}^*\hat a_{l_1} ...\hat a_{l_n},$$
which by itself does not have a meaning of an operator or of an element of the Weyl algebra, but, nevertheless, it makes sense under the commutator sign in the equations of motion.  This does not always happen, but there are very simple conditions when it makes sense. When a commutator of $\hat a_k$ or $\hat a_k^*$ with a product is taken, one should commute with each factor of that product. Due to the Kronecker symbol in CCR the commutator will get contribution only from coefficients where one of the indices coincides with $k$.  If for any index there exist only a finite number of nonzero coefficients in the Hamiltonian containing this index, the equation of motion makes sense.

\subsection {Quadratic Hamiltonians}
Let us  consider quadratic Hamiltonians of the form

$$H(u)=\frac{1}{2}H_{kl}u^ku^l.$$
Here the ordering plays no role because by changing the order we obtain an irrelevant constant. ( The Hamiltonian is used only under the commutator sign, where the constant disappears.) Classical equations of motion and quantum equations of motion are exactly the same. Moreover, if we know how to solve the classical equations of motion, we immediately know how to solve the quantum equations of motion, because all equations of motion are linear, and the difference between classical and quantum mechanics arises only when the operators are multiplied. 

The same problem can be made even simpler, namely, it is possible to simplify the Hamiltonian. If we assume that the Hamiltonian is positive definite and the matrix $H_{kl}$ is nondegenerate, we can represent the Hamiltonian as a sum of squares.

Further, we can simplify the matrix $\sigma$, preserving the representation of the Hamiltonian as a sum of squares. In order to preserve this property, only orthogonal transformations should be taken. Note that $\sigma$ is an antisymmetric matrix. If an antisymmetric matrix is multiplied by the imaginary unit, we obtain a matrix that corresponds to a self-adjoint operator; it can be diagonalized. This diagonalization occurs in the complex domain, but we can say that along with each eigenvector, we have a complex conjugate eigenvector. We can consider two complex conjugate eigenvectors, take the real and imaginary parts, and obtain a representation of the matrix $\sigma^{kl}$ in block-diagonal form with two-dimensional blocks. These two-dimensional blocks will be antisymmetric matrices, so the Hamiltonian will take the form of a sum of Hamiltonians of the form
$$\hat H=\frac {1}{2}(\hat p^2+\epsilon^2\hat q^2),$$ 
where $[\hat p,\hat q]=\frac {1}{i}$.
This is an extremely important simplification, which can always be done for a positive quadratic Hamiltonian. I  discussed this in the case of a finite number of degrees of freedom. It is important to note that in the case of an infinite number of degrees of freedom, there exists a similar simplification. The only difference is that when a self-adjoint operator is diagonalized, a continuous spectrum may appear in addition to a discrete spectrum. I will come back to this.

Let us consider the Hamiltonian 
$\hat H=\sum \frac {1}{2}(\hat p_k^2+\epsilon_k^2(\hat q^k)^2)$  assuming that $\epsilon_k> 0$
and solve the corresponding equations of motion
 $$\frac {d\hat p_k}{dt}=-\epsilon_k^2 \hat q^k,\;\;\;\; \frac {d\hat q^k}{dt}=\hat p_k.$$ 
These are the standard equations of motion of harmonic oscillator exactly as in classical mechanics. They can be solved in dozens of ways, but the simplest way is to introduce new variables
$$\hat a_k= \frac{1}{\sqrt 2}(\sqrt {\epsilon_k}\hat q^k+\frac{ i\hat p_k}{\sqrt {\epsilon_k}}),\;\;\;\; \hat a_k^*=\frac{1}{\sqrt 2}(\sqrt {\epsilon_k}\hat q^k-\frac{ i\hat p_k}{\sqrt {\epsilon_k}}). $$
In these variables the equations of motion 
$$\frac {d\hat a_k}{dt}=-i\epsilon_k \hat a_k,\;\;\;\; \frac{d\hat a_k^*}{dt}=i\epsilon_k\hat a^*_k,$$ correspond to the Hamiltonian
\begin{equation}\label{eqHamOsc}
\hat H=\sum \epsilon_k \hat a_k^*\hat a_k.
\end{equation} 
They have a very simple solution:
 $$\hat a_k(t)=e^{-it\epsilon_k}\hat a_k(0),\;\;\;\; \hat a_k^*(t)= e^{it\epsilon_k}\hat a_k^*(0).$$

In the case of an infinite number of degrees of freedom, by virtue of the spectral theorem, one can also assume that everything is diagonal, but instead of the sum we get an integral and  the Hamiltonian will have the form:
$$\hat H= \int d\lambda \epsilon (\lambda) \hat a^* (\lambda)\hat a(\lambda).$$
In physics, $\lambda$  usually consists of continuous and discrete indices, and the integral involves both integration and summation over a discrete index.
In the special case when the theory is translation-invariant, we assume that the operators $\hat a^* (\bx)$ and $\hat a(\bx)$ depend on coordinates $\bx$, which can be shifted without changing the Hamiltonian.   This means that the Hamiltonian has the form
$$\hat H=\int d{\bf x}d{\bf y}\epsilon ({\bf x -\bf y}) \hat a^*({\bf x})\hat a({\bf y}),$$ 
where the coefficient depends only on the difference $\bx-\by$. 
(There may also be discrete indices in the expression in question - we should sum over these indices.)

It is possible to pass to the momentum representation (take the Fourier transform). Then the Hamiltonian will take a form:
$$\hat H=\int d{\bf k} \epsilon ({\bf k})\hat a^*({\bf k})\hat a({\bf k}).$$ 
We will consider translation-invariant Hamiltonians all the time, and this formula will be essentially used. 

 \subsection {Stationary states}

Now let us briefly discuss stationary (time-independent) states. If the evolution operators are denoted by $U(t)$, then a stationary state $\omega$ obeys $U(t)\omega=\omega$. 

If we work in the formalism of density matrices $K$, then, using the fact that the equations of motion for the density matrix are written as a commutator with the Hamiltonian $\hat H$, we conclude that the density matrix represents a stationary state if it commutes with the Hamiltonian. In particular, if the density matrix is a function of the operator $\hat H$, then the state is stationary. 

If we speak about pure state represented by a vector $\Psi$ in Hilbert space, the stationary state satisfies the condition $\hat H\Psi=E\Psi$, i.e. the stationary state is an eigenvector of the Hamiltonian. The Hamiltonian can be interpreted as the energy operator, and $E$ is the energy level. 
The vector $\Psi$  changes with time but the state does not change:
$$\Psi(t)=\hat U(t)\Psi=e^{-itE}\Psi\sim \Psi.$$

In the algebraic approach, the Hamiltonian can be a formal expression, but it is possible to apply the GNS construction to stationary state $\omega$ and obtain a Hilbert space in which there are unitary operators describing the evolution (time shift).  The generator of the time translation group has the meaning of the Hamiltonian - energy operator. Its eigenvalues can be interpreted as energy levels of excitations of the state $\omega$. More precisely, such interpretation will be perfectly correct when $\omega$ itself is stationary and translation-invariant (invariant with respect to both spatial and time translations). 

This can be illustrated as follows: a translation-invariant state can be represented as a horizontal line. Excitation must be perceived as a bump concentrated in a finite region on this horizontal line. The energy of the translation-invariant state is infinite, but the difference between the energy of the bump and the energy of the translation-invariant state can be finite.

An important and simple remark, which is explained in t quantum mechanics textbooks in a less general situation, is as follows. Consider a classical Hamiltonian $H(u)$ which has a minimum at a non-degenerate critical point. This means that the quadratic part in the Taylor expansion is positive definite;  there are no zero modes. The quantum Hamiltonian is quadratic in the first approximation; in appropriate coordinates, it will have the form: 
$$\hat H=\int d\lambda \epsilon (\lambda) \hat a^* (\lambda)\hat a(\lambda)+... ,$$
where the terms denoted as $...$ start with cubic terms in $\hat a^*, \hat a$ (there can be no linear terms because we are at the critical point). The higher order terms with respect to $\hat a^*, \hat a$ are also higher order terms with respect to the Planck constant. At least in semiclassical approximation, we can neglect these terms. If we are not working in the semiclassical approximation, we can use the perturbation theory with respect to omitted terms.

 \subsection {Fock space}

Let us consider representations of the Weyl algebra $\cal W$ (or, what is the same, representations of canonical commutative relations). Among these representations, there is one remarkable, the simplest one, which is called the Fock representation, and the space in which it lives is called the Fock space. The Fock representation is defined simply: in it there exists a cyclic vector $ |0\ket$ which is annihilated by all operators $\hat a_k$:
$$\hat a_k |0\ket=0.$$ (We denote by $\hat A$ the operator corresponding to  the element $A\in \cal W$)
 This condition unambiguously defines the representation (up to equivalence).
 
  Let us describe Fock representation more explicitly. 
  The vector $ |0\ket$ is cyclic. The definition of cyclicity depends on whether we live in pre-Hilbert or Hilbert space. If we live in pre-Hilbert space, we should get all vectors by applying algebra elements to a cyclic vector.  In other words, the space is the smallest set containing the cyclic vector and invariant with respect to all operators $\hat A$. 
  In Hilbert space, we should get all vectors by applying algebra elements to cyclic vectors and taking limits. In other words, the Hilbert space is the smallest closed set invariant with respect to operators $\hat A$ and containing the cyclic vector.
  
   Let us  act  by  creation operators $\hat a^*_k$ on $|0\ket$: 
\begin{equation}\label{eqfock}
(\hat a_{k_1}^*)^{n_1}...(\hat a_{k_s}^*)^{n_s} |0\ket.
\end{equation} 
Here  creation operators are applied many times. (A finite number of times, since only finite combinations of these operators exist in Weyl algebra).  Take all linear combinations of vectors (\ref {eqfock}).

Can we get something new if we apply the operator $\hat a_k$ to these expressions? The answer is no, we will not, because it is possible to move the operator $\hat a_k$ to the right using commutation relations so that it acts on $|0\ket$. Then the operator $\hat a_k$ will disappear due to the condition $\hat a_k |0\ket=0.$ Therefore only expressions of the form (\ref{eqfock}) and their linear combinations belong to Fock space, if we use the definition of cyclicity appropriate for pre-Hilbert spaces.

Further, it should be noted that  states (\ref{eqfock}) are eigenvectors of any Hamiltonian of the form 
$\hat H=\sum \epsilon _k \hat a_k^*\hat a_k$ with eigenvalues equal to
 $\sum n_k \epsilon_k$. 

How to check this?  It is easy to calculate the commutator of the Hamiltonian  $\hat H=\sum \epsilon _k \hat a_k^*\hat a_k$  with the operator $\hat a_l^*$:

$$\hat H\hat a_l^*=\hat a_l^*\hat H+\epsilon_l \hat a_l^*.$$

To calculate the action of the Hamiltonian  
$\hat H$  on the vector (\ref {eqfock}) is enough, using this relation, to move the Hamiltonian to the right.  Somewhat simpler formal reasoning is as follows: it is known how the operators $\hat a_k^*$ change with time - they are multiplied by a numerical factor. It follows that the vector (\ref {eqfock}) is also multiplied by a numerical factor; hence it represents a stationary state.

We obtained an orthogonal (but not orthonormal basis of Fock space, which consists of eigenvectors (\ref{eqfock}). (To check orthogonality we use the fact that the representation is compatible with involution.)

All these formulas can be applied to the case of a multidimensional harmonic oscillator. There operators 
$\hat a_k, \hat a_k^*$ are called operators of creation and annihilation of quanta. In a crystal atoms somehow interact with each other, but the crystal is in a stationary state that is  close to the ground state. At least in the first approximation the crystal is described by a quadratic Hamiltonian. For quanta in this situation, there is another name: phonons - quanta of sound. In the general case, we are dealing with a system of non-interacting bosons. The operators $\hat a_k, \hat a_k^*$ are called particle creation and annihilation operators and the numbers $ n_k $ in the formula for energy levels
$\sum n_k \epsilon_k$ are called  occupation numbers.

If we want the Fock space to be a Hilbert space, we have to take a completion.

There are advantages in both approaches. In pre-Hilbert space, the operators $\hat a_k, \hat a_k^*$ are defined everywhere. In the Hilbert space, these are unbounded operators, defined on a dense subset. However, some important states do not belong to the pre-Hilbert space. 

A pre-Hilbert Fock space can be represented as a space of polynomials. Indeed, there is the following formula for the basis:
$$(\hat a_{k_1}^*)^{n_1}...(\hat a_{k_s}^*)^{n_s } |0\ket.$$
To get a polynomial, we delete$ |0\ket$ and remove the hats. We get a monomial with respect to variables $a^*_k, a_k.$

A linear combination of such monomials is a polynomial. That is, each element of the Fock space (which we consider to be pre-Hilbert space) can be represented by a polynomial. It is easy to calculate that the scalar product in such a representation in the form of polynomials will be given by the formula:
\begin{equation}\label{eqscalar}
\bra F,G\ket =\int da^*da F(a^*)G(a^*)^*e^{-a^*a},
\end{equation}
where $F(a^*)$ and $G(a^*)$ are polynomials corresponding to some vectors. Note that $G(a^*)$ stands with a star. The star applied to $a$ twice is $a$ again, hence $G(a^*)^*$ is
a polynomial of $a$. 

Let us check that the scalar product is written in the form (\ref{eqscalar}). This can be calculated, but one can give simpler proof. The Fock space is uniquely defined by the existence of a cyclic vector which is annihilated by all annihilation operators provided that the creation and annihilation operators satisfy the required commutative relations.  Let us consider the space of polynomials with respect to  $ a_k^*$.  I define the operator $\hat a_k^*$ in this space as multiplication by $ a_k^*$ and the operator $\hat a_k$ as differentiation with respect to  $a_k^*$. It is easy to see that multiplication and differentiation satisfy the necessary commutation relations. (If we multiply by $a_k^*$, then differentiate and apply the Leibniz rule, we get just what we need. )
Thus we have commutation relations, there is also a cyclic vector that equals $1$.    It remains to check that $\hat a_k$ and $\hat a_k^*$ are adjoint to each other. (This is necessary for the involution to work correctly.) Indeed, in the formula 
(\ref{eqscalar}) one can apply integration over parts to make sure that  for this scalar product multiplication and differentiation are adjoint to each other. As a result, all properties of the Fock representation are satisfied, and so it is not necessary to compare the original scalar product with the new one: they necessarily coincide (up to numerical factor)

So far we have dealt with polynomials. But I still want to be able to work in Hilbert space.  We should take a completion; it consists of holomorphic functions with respect to $a_k^*$, but only holomorphic functions having a finite norm in our scalar product belong to this space. 

It is important to note that all this reasoning also applies to an infinite number of degrees of freedom. Although the integral is infinite-dimensional there, it still turns out to be well-defined. 

There is another way to describe the Fock space. Polynomials are related to symmetric functions depending on discrete arguments. We can always assume that the coefficients of quadratic form are symmetric with respect to indices. For a cubic polynomial,   the coefficients depend on three indices, and, again, we can impose the symmetry condition. It has to be imposed if you want to have an unambiguous representation. For polynomials of higher degree, the situation is similar. Therefore we can assume that there is a unique representation for every element of a pre-Hilbert Fock space in the form: 

$$\sum_n \sum_{k_1,...,k_n} f_n(k_1,...,k_n)\hat a_{k_1}^*...\hat a_{k_n}^* |0\ket,$$
where $n=0,1,2,...$, and the coefficients  $f_n$ are symmetric with respect to indices.
This means that the Fock space is represented as a sequence of symmetric functions depending on  an increasing number of discrete variables: 
$$f_0, f_1(k), f_2(k_1,k_2), f_3 (k_1, k_2, k_3),...\;.$$
While we are working with polynomials, there should be only a finite number of symmetric functions of $k_1, k_2, k_3,...$.  We can take a completion, and then we have to consider a sequence of functions $f_0, f_1(k),...$, which must satisfy the condition: the norm of this infinite sequence is finite. Usually, this sequence is written as a column (a Fock column). This construction also makes sense when $k$ is a continuous parameter. In quantum mechanics textbooks, the Fock space is usually defined as the space consisting of columns of symmetric functions. 

\subsection {Hamiltonians preserving the number of particles}

In quantum theory, the operator
$$ \hat N=\sum\hat a^*_k\hat a_k=\int d\lambda \hat a^*(\lambda)\hat a(\lambda),$$
which is a sum over all $k$ (or integral if there is a continuous index)  plays an important role. This operator has a physical meaning of the number of particles or the number of quanta when we deal with oscillators. If there is an eigenvector $X$ of $\hat N$ with some eigenvalue $N$, the operators $a_k^*$, acting on $X$, increase the eigenvalue by $1$, and $a_k$, on the contrary, decrease it by $1$( create a particle or destroy a particle). This means that the operator preserving the number of particles must contain the same number of creation and annihilation operators. 

Let us consider quadratic Hamiltonians possessing this property. 

The quadratic Hamiltonians preserving the number of particles are very important because near the minimum energy state (ground state) they play the main role. Let us take a quadratic Hamiltonian which conserves the number of particles: 
$$\hat H=\int dxdyA(x,y)\hat a^*(x)\hat a(y)=\bra a^*, A a\ket.$$ 
It contains  products of type $a^*a $, but products of type $a^* a^*$, $a a$, do not appear.  If operator $A$ has only a discrete spectrum, we can write the Hamiltonian in the form: 
$$\hat H=\sum \epsilon_k\hat a_k^*\hat a_k,$$ 
where
 $\hat a_k=\int dx \phi_k(x)\hat a(x)$ 
corresponds to the eigenfunction $\phi_k(x)$ of the operator $A$. 

Let us now take as $x$ and $y$ vectors in Euclidean space and assume that $A$ is written in the form:
$A=-\frac {1}{2m}\Delta+\hat U$, 
which appear in the Schr\"odinger equation. This is an  operator  obtained from the quantization of the classical Hamiltonian of  the form 
$\frac {\bp^2}{2m}+U(\bx)$.
In this case, the Hamiltonian $\hat H$ describes a system of non-interacting nonrelativistic identical bosons. If we add nonquadratic terms conserving the number of particles, we obtain the Hamiltonian of the system of interacting nonrelativistic identical bosons.

I want to point out that the content of this lecture is, in a sense, an explanation of how a mathematician could have guessed quantum mechanics, but didn't. The physicists guessed, of course. Look at the logic involved. The mathematician knows that the observables in classical theory are simply functions on phase space. He knows that quantum mechanics is a somewhat deformed classical mechanics: everything obeys classical laws, but in some situations, there must be corrections. On this basis, he says: let me deform the commutative algebra of functions: I will leave associativity, but introduce noncommutativity. The simplest deformation that we have is a Weyl algebra. ( We  know that the deformation is governed by a Poisson bracket and the Weyl algebra corresponds to the Poisson bracket with constant coefficients.)
 The mathematician guessed that one should use Weyl algebra, and after that, he realizes that in the Weyl algebra one should consider the simplest Hamiltonian, which describes the life near the ground state. This is the quadratic Hamiltonian. The energy levels of this Hamiltonian are given by the formula:
$$\sum \epsilon_k\hat n_k,$$ 
where $\hat n_k = 0, 1, 2, ...$ 
are occupation numbers. This formula describes the energy levels of a system of non-interacting identical particles. There is no mystery in the appearance of identical particles. From the mathematician's point of view, it is they that must appear. The non-identical particles do not necessarily exist, but the identical ones always exist
 because the simplest quadratic Hamiltonian already describes identical particles. 

In the nonrelativistic case, the one-particle energy levels $\epsilon_k$ must be obtained by quantization 
of the classical one-particle Hamiltonian $\frac {{\bf p}^2}{2m}+U({\bf x})$. Adding the non-quadratic terms, we get the Hamiltonian of the system of nonrelativistic identical bosons.

\subsection {Representations of Weyl algebra}

 We studied Fock representation of Weyl algebra. Are there any other representations of this algebra?  First of all, let us note that one can take a direct sum of two Fock representations, and this will be a new representation. Since the Fock representation, as is easy to see, is irreducible (there is no other representation inside it), the main question is what are the irreducible representations?

This question is poorly formulated. In a Hilbert space operators $\hat a_k^*,\hat a_k$ are defined on a dense domain. It is not the whole Hilbert space.
One  can change the domain, but leave the operator the same and ask yourself: is this the same representation or a different one? From a formal mathematical point of view, it is different. In fact, of course, it is the same. When dealing with unbounded operators, this problem always arises. It can be solved, but it is better to deal with bounded operators acting in Hilbert spaces.

In a representation of Weyl algebra one  can consider operators: 
$$V_{\alpha}=e^{i\alpha_k \hat u^k},$$
where the  exponent is a linear combination of self-adjoint operators $\hat u^k$ satisfying the commutation relations:
 $$\hat u^k\hat u^l-\hat u^l\hat u^k=i\sigma^{k,l}.$$ 
The coefficients in the exponent are chosen to be real, then the exponent is a  selfadjoint operator multiplied by $i$. We obtain a unitary operator $V_{\alpha}$, and a unitary operator is bounded; it can be extended to the whole Hilbert space. It is convenient to use these operators instead of unbounded operators  $\hat u^k$. There is a relation:
\begin{equation}\label{EX}
 V_{\alpha}V_{\beta}=e^{-i\frac{1}{2}\alpha\sigma\beta} V_{\alpha+\beta}.
 \end{equation}
 It is easy to check it using  the formula: 
$$e^Xe^Y=e^{X+Y}e^{\frac {1}{2}C},$$
which is true if the commutator of operators $X$ and $Y$ is a number, which is denoted here by the letter $C$:
$$[X,Y]=C.$$

If we do not want to deal with unbounded operators, we can work with these unitary operators and with the exponential form of commutation relations (\ref{EX}). This is the exponential form of the Weyl algebra. From the point of view of a physicist, it is the same algebra we dealt with, from the point of view of a mathematician it is not quite so.

There is a single irreducible representation of a Weyl algebra in the case of a finite number of degrees of freedom (finite number of generators).
I will give proof of this fact without using the exponential form of Weyl algebra (but working in Hilbert spaces).  It is not rigorous, but, from the point of view of a physicist, it is acceptable. 

The reasoning is as follows: consider the already mentioned particle number operator $\hat N=\sum \hat a_k^*\hat a_k$. In the case of a finite number of generators, it is a good operator. Let us take the eigenvector of this operator and start applying annihilation operators to it. (A mathematician can ask: how do you know that there is such an eigenvector, but a physicist probably won't do it). If all annihilation operators give zero acting on the eigenvector, I will say that this is exactly the Fock vacuum we need. If there is such an operator which does not give zero, then I apply annihilation operators
 again and again and again and do it until I have such a vector $ |0\ket$ that all annihilation operators give zero acting on it: $a_k |0\ket=0$. 

Such a vector necessarily exists because the particle number operator is positive definite and the annihilation operators decrease the number of particles. After that, I  take the subrepresentation containing the vector $ |0\ket$. I will apply to the vector $ |0\ket$ all operators $\hat a_k^*$ many times and, when I take linear combinations of the resulting expressions, I will have a space that is invariant with respect to all creation and annihilation operators. Taking a closure of this space I obtain a subrepresentation of my representation in a Hilbert space and it will be a Fock representation because it contains a cyclic vector that is annihilated by all $\hat a_k$.  My representation contains Fock representation, but it is irreducible, hence it coincides with Fock representation.

In the case of an infinite number of generators, this reasoning does not apply. I will now explain how to construct an example of representation that is not equivalent to Fock representation. I will construct new operators, which I will denote by the letter $\hat A_k$, in the Fock space. These are the old annihilation operators minus  a number:
 $$\hat A_k=\hat a_k-f_k,\;\;\; \hat A_k^*=\hat a_k^*-\bar f_k$$
 (for each $\hat a_k$ I subtract a different number). The commutation relations that  I need in  Weyl algebra are  satisfied. Thus, I again have a representation of canonical commutation relations, a representation of  Weyl algebra. 
 
Now I will try to solve the equation $\hat A_k\Theta=0$.
Its solutions are eigenvectors of the operators $\hat a_k$. We will meet these vectors many times. They are sometimes called Poisson vectors. It is easy to understand that the answer will be represented in the form
$$\Theta=e^{f\hat a^*} |0\ket,$$ where $f\hat a^*=\sum f_k\hat a_k^*$
or as  a function $e^{fa^*}$ if we represent elements of Fock space as functions of  $a^*.$

One can calculate the norm of a Poisson vector and the scalar product of two Poisson vectors. 
The scalar product is given by an integral; this integral is Gaussian, it it is easy to calculate it. The norm of $\Theta$ is finite if the sum $\sum |f_k|^2$ is finite. If the norm is finite, then the vector $\Theta$ belongs to the Fock space and 
 the representation is equivalent to the Fock representation, but if the norm is infinite, there is no vector in the Hilbert Fock space which is annihilated by the operators $\hat a_k$. Hence this representation is not equivalent to the Fock representation. Moreover, it cannot contain a subrepresentation equivalent to the Fock representation. In the case of a finite number of degrees of freedom, $\sum |f_k|^2$ is always finite, and this reasoning gives nothing there.

Now I want to generalize this construction. I will consider operators $\hat A_k$ defined slightly differently. Before, I just added $f_k$, but here I will also take linear combinations: 
$$\hat A_k=\Phi_k^l \hat a_l+\Psi_k^l\hat a_l^*+f_k$$
 $$\hat A_k^*=\bar \Psi_k^l \hat a_l+\bar\Phi_k^l\hat a_l^*+\bar f_k$$
in such a way that  new operators $\hat A_k,\hat A_k^*$ also satisfy  canonical commutation relations. This imposes some conditions on the coefficients.  The transition to $\hat A_k,\hat A_k^*$ is called a linear canonical transformation. The operators $\hat A_k,\hat A_k^*$ define a new representation of the Weyl algebra and, again, if the vector $\Theta$ which is the solution of the equation $\hat A_k\Theta=0$ belongs to Fock space, then the new representation of CCR is equivalent to the Fock representation. If this condition is not met, then the new representation is not equivalent to the Fock representation (see \cite {BEREZIN} for details).

 Linear canonical transformations are often useful. Sometimes they and their analogs in the fermionic case are called Bogoliubov transformations. 
\newpage
\section{Lecture 3} 
 
\subsection{Clifford algebra and Grassmann algebra}

Weyl algebra has close relatives that are defined by the same formulas as Weyl algebra, only instead of commutators we consider anticommutators:
 $$u^ku^l+u^lu^k=2h^{kl},$$
where $h^{kl}$ is an invertible matrix. A unital associative algebra with generators obeying these relations is called Clifford algebra.

A unital  associative algebra with generators obeying
$$u^ku^l+u^lu^k=0$$ is called Grassmann algebra.

 The Grassman algebra is not a special case of the Clifford algebra - here we have a zero on the right-hand side instead of an invertible matrix. The elements of the Grassman algebra can be considered polynomials of anticommuting variables.


Let us start with the Grassmann algebra with anticommuting generators $\epsilon^1,...,\epsilon ^n$:
\begin {equation}\nonumber
\epsilon ^i\epsilon ^j=-\epsilon ^j\epsilon ^i.
\end {equation}
It is denoted by $\Lambda _n$.
Every element of Grassmann algebra can be represented as a  sum of monomials  with respect to $\epsilon ^i$:

\begin{equation}\label{GR}
\omega=\alpha +\sum _i\alpha _i \epsilon ^i+\sum \alpha _{ij} \epsilon
^i\epsilon ^j + ... +\sum\alpha _{i_1,...,i_k}\epsilon 
^{i_1}
... \epsilon ^{i_k} + ...\end{equation}

This representation is not unique, however, we can get a unique representation requiring antisymmetry of coefficients 
$\alpha _{i_1,...,i_k}.$ Another standard representation of $\omega$ is based on the  remark that
each element of a Grassman algebra can be uniquely written as a sum of monomials   in  such a way  that in each monomial the indices increase:
$$\omega=\alpha +\sum _i\alpha _i \epsilon ^i+\sum _{i<j}\alpha _{ij} \epsilon
^i\epsilon ^j + ... +\sum _{i_1<...<i_k}\alpha _{i_1,...,i_k}\epsilon 
^{i_1}
... \epsilon ^{i_k} + ...+\alpha_{1,...,n}\epsilon^1...\epsilon^n.$$
 This is obvious: we can use anticommutation relations to move the smaller indices to the left, and there cannot be two matching indices. (If we take $i=j$  in these relations we obtain that the square of a generator is zero.)

The Grassman algebra, just like the usual algebra of polynomials, is ${\bf Z}$-graded:  $\Lambda_n=\sum _k\Lambda_n^k$ 
This means that there is a notion of degree: the degree of each monomial is the number of generators in that monomial; clearly the degrees add up when monomials are multiplied,
 as for ordinary polynomials. 

I already said that there is a notion of degree (${\bf Z}$-grading). From this ${\bf Z}$-grading one can get ${\bf Z_2}$-grading by saying that there are even and odd elements: $\Lambda_n^{even}=\sum_{k\geq 0} \Lambda_n ^{2k}$ and  $\Lambda_n^{odd}=\sum _{k\geq 0}  \Lambda_n^{2k+1}$. The ${\bf Z_2}$-grading is more important because it governs multiplication.
An even element commutes with anything. Two odd elements anticommute. 

We can say that an element of a Grassman algebra is a function of anticommuting variables; it is automatically a polynomial because there are only a finite number of monomials (in the case when we have a finite number of anticommuting variables). The analogy with functions is an important idea because it suggests that there must be an analysis in Grassmann algebra, and indeed there is. One can define differentiation $\partial_i=\partial /\partial \epsilon^i$ with respect to a variable $\epsilon^i$. To differentiate you have to delete that variable. If there is no variable $\epsilon^i$ in the monomial, then the derivative is zero.  For the case of anticommuting variables, there are notions of left derivative and right derivative. For certainty, we will consider the left derivative; this means that before deleting the corresponding variable should be moved to the left :
\begin {equation}\nonumber
\partial _i(\epsilon ^i \epsilon ^{i_1} ...\epsilon ^{i_n})=\epsilon
^{i_1}... \epsilon ^{i_n} ,
\end {equation}
\begin {equation}\nonumber
\partial _i(\epsilon ^{i_1} ... \epsilon ^{i_n})=0
\end {equation}
if $i\neq i_k.$

The notion of the derivative is related to the Leibniz rule. Here, too, there is a modification of Leibniz rule, 
\begin {equation}\nonumber
\partial _i(\omega \rho)=(\partial _i\omega)\cdot \rho +(-1)^{\bar 
{\omega}} \cdot\omega\cdot \partial _i\rho,
\end {equation}
where $\omega, \rho\in \Lambda_n$, and $\omega$ has  parity $\bar {\omega}$, that is, $\omega$ is either even ($\bar {\omega}=0$) or odd ($\bar {\omega}=1$). 
This is called a graded Leibniz rule. If this rule is satisfied, we speak of odd derivation; if the regular Leibniz rule is satisfied, we speak of even derivation

There is also a notion of integration $\int : \Lambda_n \rightarrow{\bf C}$. 
The integral of any monomial of non-maximum degree gives zero, and the integral of a monomial of maximum degree gives plus or minus one: 
$$\int \epsilon ^{i_1}...\epsilon ^{i_k}d^n\epsilon =0\ \ \ {\rm if }\ k<n,$$ 
$$\int \epsilon ^1...\epsilon ^nd^n\epsilon =1.$$ 
In my notations plus one is obtained when the generators are ordered in ascending order.  It is easy to understand that the integral of a derivative is always zero: 
\begin {equation}\nonumber
\int (\partial _i\omega)d^n\epsilon =0.
\end {equation}
This is because a derivative cannot contain a term of maximum degree. From this and Leibniz rule we can derive the rule of integration by parts:
$$\int(\partial _i\omega)\cdot \rho d^n\epsilon =-(-1)^{\bar {\omega}} \int
\omega\cdot \partial_i\rho d^n\epsilon. $$

I would like to emphasize that when we defined a Grassman algebra $\Lambda _n $, we fixed a system of generators $\epsilon ^1,..., \epsilon ^n $. Naturally, it is possible to take another system of generators (this is analogous to a change
 of variables) and then everything will change - the notion of differentiation will change, the notion of integration will change. For the change of variables, there is an analog of chain rule and an analog of the Jacobian.
 
I want to consider only the special case where the change of variables is linear: 
\begin {equation}\nonumber
\tilde {\epsilon}^i=\sum A_j^i\epsilon ^j.
\end {equation}

It is easy to check that one can obtain the integral with respect to the new variables from  the integral with respect to the old variables by multiplying the latter by $(\det A)^{-1}$   where $A$ stands for the matrix $A_j^i.$ (In conventional calculus we multiply by $\det A.$)

The next observation is that in a smooth function $f$ that depends on a real variable $x\in {\bf R}$, one can substitute
$x$ by an even element $\omega$ of the Grassman algebra. To define $f(\omega)$ we represent  $\omega$ in the form $\omega=a+\nu$,
where $a$ is a number and $\nu$ is a nilpotent element
(i.e., ${\nu^k=0}$ for some $k$).
Take the Taylor series expansion of the function $f(a+\nu)$  with respect to the nilpotent part:
\begin {equation}\nonumber
f(\omega)=f(a)+{f^{\prime}(a)\over 1!}\nu +...+{f^l(a)\over l!}\nu ^l+... 
\end {equation}
and note that because of nilpotency this Taylor series has a finite number of terms.

In particular, we can consider $e^{\omega}$.  As an example, consider the exponent of a quadratic expression:

 $$e^{\lambda_1\epsilon ^1\epsilon ^2+
\lambda_2\epsilon ^3\epsilon ^4 + ...+ \lambda_k\epsilon ^{2k-1}\epsilon ^{2k}}=e^{
\lambda_1\epsilon ^1\epsilon ^2}...e^{ \lambda_k\epsilon ^{2k-1}\epsilon ^{2k}}=$$
$$(1+
\lambda_1\epsilon ^1\epsilon ^2)...(1+ \lambda_k\epsilon ^{2k-1}\epsilon ^{2k}),$$ 
it follows that
\begin {equation}\nonumber
\int e^{\lambda_1\epsilon ^1\epsilon ^2+...+ \lambda_k\epsilon ^{2k-1}\epsilon
^{2k}}d^{2k}\epsilon=\lambda_1...\lambda_k.
\end {equation}

In the more general case where $\omega ={1\over 2}\sum a_{ij}\epsilon ^i\epsilon ^j$,
 it is possible to represent the antisymmetric nonsingular matrix $a$ in the block-diagonal form by changing the variables. This allows us to calculate the Gaussian integral: 

\begin {equation}\nonumber
\int e^{\omega}d^n \epsilon=(\det a)^{1/2}.
\end {equation}

The answer is almost the same as in the usual case where we had  $(\det a)^{-1/2}$ instead of $(\det a)^{1/2}$. 

We can consider functions that depend both on commuting and anticommuting variables. This simply means that in the expression (\ref{GR}) for a general element of Grassmann algebra coefficients can be considered as functions on commuting variables. Either polynomial functions or smooth functions can be considered.  By definition functions of $m$ commuting and $n$ anticommuting variables by definition are functions on the superspace $\mathbb {R}^{m,n}$.  

One can say that polynomial or smooth functions of commuting variables $x^ 1,...,x ^ m$ and anticommuting variables
$\epsilon^1,...,\epsilon^ n$   are elements of algebra $\mathbb {C}[x^1,...,x^m]\otimes\Lambda_n$ or algebra $C^{\infty}(\mathbb {R}^m)\otimes \Lambda_n$.

 We can define supervarieties assuming that commuting and anticommuting variables obey some equations. It is important that all summands in these equations are of the same parity. For example, we can impose the condition: $x^1+\epsilon^1\epsilon^2=0$. If the equations are polynomial we get algebraic supervarieties.   This is a generalization of ordinary algebraic varieties specified by polynomial equations in ordinary linear space. 
 
I would like to discuss a notion of a point in a supervariety.
 
Let us start with algebraic varieties determined by polynomial equations with integer coefficients. In this case, we can think of the variables $x^i$ as elements of any commutative ring.  This means that we
 can consider the same variety over different fields or over different rings. We define a notion of a point of an algebraic variety over  some ring as a solution for defining equations with variables belonging to the ring. For example, if we consider, say, a circle of imaginary radius $x^2+y^2+1=0$, such an object has no real points, but it has points over complex numbers.
 
  When we consider supervarieties we have a similar situation.  A $\Lambda$-point of supervariety as a solution for defining equations with variables belonging to Grassmann algebra $\Lambda$.   (It is necessary to substitute an even element of  Grassmann algebra  for an even variable and an odd element for an odd variable. In other words, it is necessary to preserve the parity when introducing $\Lambda$-points).  

The notion of $\Lambda$-point is very convenient. For example, it allows us to give a very simple definition of the notion of a Lie superalgebra. A Lie superalgebra is a ${\bf Z_2}$-graded algebra with some generalization of the Jacobi rule. It is not necessary to memorize this rule. We just need to say that $\Lambda$-points of a Lie superalgebra should constitute an ordinary Lie algebra. 

I  won't talk about this in detail. What I told you is a small piece of what is called supermathematics.

\subsection{Representations of Clifford algebra}
 
 The following reasoning, as I promised, will repeat exactly the reasoning of the previous lecture, only instead of ``Weyl'' I will say ``Clifford'' and instead of ``commutator'' I will say ``anti-commutator''. 
 
According to the definition of Clifford algebra, the anticommutator of generators is a symmetric nonsingular matrix $h^{kl}=h ^{lk}$: 
$${\hat u}^k\hat u^l+\hat u^l\hat u^k=2 h ^{kl}.$$
The same relations appear in  Dirac equation:  gamma matrices are generators of a Clifford algebra.

Let us consider a Clifford algebra with involution, assuming that the generators $\hat a_k, \hat a_k^*$ are obtained from each other by involution and satisfy equations: 
$$\hat a_k \hat a_l+\hat a_l \hat a_k=0,\;\;\hat a^*_k \hat a^*_l+\hat a^*_l\hat a^*_k=0,\;\;\hat a_k\hat a^*_l+\hat a^*_l\hat a_k=\delta_{kl},$$
which are called canonical anticommutational relations (CAR). They are obtained from the canonical commutation relations by replacing commutators with anticommutators.  I note that the anticommutation relations presented here are exactly anticommutation relations that are satisfied by differentiation and multiplication with respect to generators of Grassman algebra.

 As in the case of Weyl algebras, we can consider Clifford algebras with an infinite number of generators or with generators $\hat a(k), \hat a(k)^*$ that depend on a continuous parameter
 and satisfy the relations:
 $$ \hat a(k) \hat a(l)=-\hat a(l)\hat a(k),\;\;\; \hat a^*(k)\hat a^*(l)=-\hat a^*(l)\hat a^*(k),$$
$$\hat a(k)\hat a^*(l)+\hat a^*(l)\hat a(k)=\delta (k,l).$$

In  Clifford algebra  with generators $\hat a_k, \hat a_k^*$ we have the notion of a normal form.  
 Just as in the case of Weyl algebra, the generators $\hat a_k^*$ should be moved to the left, $\hat a_k$ goes to the right, but there is a small peculiarity. Previously, starting with the normal form, I could define the Wick symbol just by removing the hats and treating $ a_k, a_k^*$ as complex variables. You can't do that here - you get zero, but I can define the Wick symbol by saying that by removing hats I get anticommuting variables. In other words, from an element of Clifford algebra, I can get a polynomial of anticommuting variables. 
 
Consider the formal Hamiltonian
$$ \hat H=\sum \Gamma_{m,n}(k_1,...,k_m, l_1,...,l_n)\hat a_{k_1}^*...\hat a_{k_m}^*\hat a_{l_1}...\hat a_{l_n}.$$ 
The Hamiltonian must always be considered an even element. 
When the number of variables is infinite, the Hamiltonian is usually not an element of a Clifford algebra, but commutators with generators make sense if the same conditions as for a Weyl algebra are imposed.

The definition of the Fock representation of the Clifford algebra exactly repeats the one given for the Weyl algebra: we require the existence of a cyclic vector $ |0\ket$ for which the condition $\hat a_k |0\ket=0$ is satisfied. Due to the cyclicity condition, one can obtain a basis by applying the operators $\hat a_{k}^*$ to the vector $ |0\ket$. All elements of pre-Hilbert Fock space will be linear combinations of monomials of the form $\Pi(\hat a_k^*)^{n_k} |0\ket$.  

The only difference is that the numbers  $n_k$  (occupation numbers)  in these monomials can only be zero or one ($n_k=0,1$) because $\hat a_{k}^{*2}=0$.  This is what is called the Pauli principle (hence we are dealing with fermions).
 
Again, the elements of this basis are eigenvectors of any Hamiltonian of the form
$$\sum \epsilon _k \hat a_k^*\hat a_k$$
and eigenvalues are given by exactly the same formula as in the bosonic case: 
$\sum n_k \epsilon_k$.

The operators $\hat a_{k}^*$ are creation operators and $\hat a_{k}$ are annihilation operators for the same reasons as in the bosonic case. One can say that the Hamiltonian $\sum \epsilon _k \hat a_k^*\hat a_k$ describes non-interacting fermions.
 
Now recall that in the bosonic case to obtain a representation of elements of the Fock space by polynomials, I removed the hats in the monomials and obtained a polynomial of complex variables. Now I want to do a similar thing: remove the hats and obtain a polynomial of anticommuting variables. The form of the scalar product is exactly the same, only the integration will be over anticommuting variables.  As in the case of Weyl algebra,  in this representation the operator $\hat a_{k}^*$ acts as multiplication, $\hat a_{k}$ acts as differentiation; this gives the correct anticommutation relations. 

 The only thing left to check is that in the scalar product 
$$\bra F,G\ket =\int da^*da F(a^*)G(a^*)^*e^{-a^*a}$$
multiplication and differentiation are adjoint operators. This can be done by applying integration by parts.

If we consider polynomials alone, we get a pre-Hilbert Fock space, but we can take a completion to obtain Hilbert space. In the case of a finite number of degrees of freedom, it is not necessary to take the completion - there are only polynomials, but in the case of an infinite number of degrees of freedom, if we want to work in a Hilbert space the completion is necessary. 

In Fock space, vector can be represented as a sum of monomials with antisymmetric coefficients:
$$\sum_n \sum_{k_1,...,k_n} f_n(k_1,...,k_n)\hat a_{k_1}^*...\hat a_{k_n}^* |0\ket$$
while for the Weyl algebra, the coefficients are symmetric. If we are working in Hilbert space these sums can be infinite. In other words, a point of fermionic Fock space can be considered as a sequence of antisymmetric functions whereas in Fock representation of Weyl algebra it was a sequence of symmetric functions. This is the standard representation from quantum mechanics textbooks.  This representation also works when $k$ is a continuous parameter. 

We can, again, consider  operator 
$$ \hat N=\sum\hat a^*_k\hat a_k=\int d\lambda \hat a^*(\lambda)\hat a(\lambda)$$
(number of particles).
Again $\hat a^*_k$ increases the number of particles by one, $\hat a_k$ decreases it. 

Let us now pass to the consideration of operators which conserve the number of particles, as in nonrelativistic quantum mechanics. The simplest Hamiltonian is the quadratic Hamiltonian. 
$$\hat H=\int dxdyA(x,y)\hat a^*(x)\hat a(y)=\bra a^*, A a\ket.$$
If the operator $A$ has a discrete spectrum, then by diagonalizing it, we get the operator 
$$\hat H=\sum \epsilon_k\hat a_k^*\hat a_k,$$ 
where $\epsilon_k$ are eigenvalues, $\phi_k(x)$ are eigenfunctions of the operator $A$, and 
$$\hat a_k=\int dx \phi_k(x)\hat a(x).$$

In the last lecture I said that by taking $A=-\frac {1}{2m}\Delta+\hat U$, we obtain a system of non-interacting nonrelativistic bosons. Having the same operator, but using the canonical anticommutation relations, we obtain a system of non-interacting nonrelativistic fermions.
The only difference is that we have to assume that this operator acts on multicomponent functions of $x\in \mathbb{R}^3$. It has exactly the same form, but we should add summation over discrete indices.  

Which canonical relations should be taken depends on how the group  of rotations of three-dimensional space acts on the wave functions. The action of this group on discrete indices determines the spin of the particle.  The case of half-integer spin corresponds to fermions, we  must quantize using Clifford algebra, and in the case of integer spin  we get bosons, hence we should use Weyl algebra. Put another way, if the representation is irreducible, then everything is determined by the number of indices. If the number is odd, we are dealing with  Weyl algebra, if it is even, with Clifford algebra. 

In order to describe interacting particles in nonrelativistic quantum mechanics, it is necessary to add terms of higher order with an equal number of creation and annihilation operators; then these terms preserve the number of particles. 

In the case of a finite number of degrees of freedom, there exists a single irreducible representation of the Clifford algebra and it is isomorphic to the Fock representation. The proof of irreducibility of the Fock representation and its uniqueness is the same as in the case of a Weyl algebra with the difference that in the case of Clifford algebra, the proof is rigorous. In the previous case, it was not rigorous because the operators $\hat a_k$ and $\hat a^*_k$  were unbounded operators, but here all these operators are bounded. This follows from  the relation
$\hat a_k\hat a^*_k+\hat a^*_k\hat a_k =1,$  
in which both summands are positive definite. Their sum is equal to one,  hence each of the operators is bounded. 

If we have an infinite number of degrees of freedom, the number of operators $\hat a_k$ and $\hat a^*_k$ is infinite.  We can consider the canonical transformation by introducing new generators that still satisfy the anticommutation relations. It is very easy to construct examples of  canonical transformations. The anticommutativity conditions are symmetric with respect to $\hat a_k$ and $\hat a^*_k$. If we swap the creation and annihilation operators, the anticommutator does not change.

In Fock space, we take operators defined by the formula $\hat A_k = \hat a_k$ for $k \in I$,$\hat A_k = \hat a_k^*$ for $k \notin I$.  The operators  $\hat A_k, ,\hat A_k^*$ obey canonical anticommutation relations, hence they define a representation of the Clifford algebra. For this representation to be a Fock representation, we must have a cyclic vector $\Theta$ which obeys $\hat A_k\Theta=0$. 

Finding a solution to the equation for $\Theta$ is very easy.  We define $\Theta$ acting on $ |0\ket$ with those operators $\hat a_k^*$ that were changed ($k \notin I$). Acting on $\Theta$ with any operator $\hat A_k=\hat a_k^*$ with $k \notin I$, we get zero because operators $\hat a_k^*$ will appear twice with the same index.  The operators $\hat A_k=\hat a_k$ where $k\in I$  also give zero, because they can be transferred to $ |0\ket$. Thus we have a monomial $\Theta$ which satisfies the relation $\hat A_k\Theta=0$ for all $k$. 

If we changed only a finite number of operators then $\Theta$ is a finite monomial, it belongs to the Fock space. If, however, we changed  an infinite number of operators, we will get something that does not belong to the Fock space at all - a monomial of infinite degree. The new representation will not be equivalent to the Fock representation, since it does not have the cyclic vector that we need. Thus we have examples of non-equivalent representations. This construction goes back to Dirac (the famous Dirac sea).

Let us consider linear canonical transformations of the form:
$$\hat A_k=\Phi_k^l \hat a_l+\Psi_k^l\hat a_l^*,$$
$$\hat A_k^*=\bar \Psi_k^l \hat a_l+\bar\Phi_k^l\hat a_l^*.$$
Unlike the Weyl case, one cannot add numerical terms here, but everything else is the same.  If we require that new operators satisfy canonical anticommutation relations, we obtain a new representation of the Clifford algebra (see \cite{BEREZIN}). 

That's more or less all I wanted to say about Clifford algebra. 

In conclusion, I also wanted to add that the group of automorphisms of Clifford algebra is isomorphic to the orthogonal group. From this remark,
 one gets instantly what is called a spinor representation of the orthogonal group. 

  \subsection {Statistical Physics}

Turning to a new topic, let me remind some notions of statistical physics. 

Both in classical and quantum statistical mechanics there is a notion of an equilibrium state and in both cases, it is defined as a state of maximum entropy under given conditions. The expression `` conditions'' may have different meanings in different situations. 
If a state is represented by a density matrix, the entropy of the state is given by the formula
 $$S=-\mathrm{Tr} K\log K.$$

 If the density matrix is diagonal, then the diagonal elements $p_i$ of the matrix are interpreted as probabilities, and this formula gives the usual  expression for the entropy of probability distribution:
$$S=-\sum p_i\log p_i.$$ 

If the Hamiltonian $\hat H$ is given, we can fix the average energy (the expectation value of energy). It is possible to fix an admissible energy interval - then we get microcanonical distribution, but we fix the average energy $E=\mathrm{Tr} \hat H K$ (then we get canonical distribution). Maximizing the entropy for a given average energy, we obtain the density matrix  
\begin{equation}\label{eqextrem}
 \frac{e^{-\beta \hat H}}{Z}. 
\end{equation} 
To calculate the constant $Z$ we notice that the density matrix, by definition, must have a trace equal to $1$ and therefore this constant must be equal to the trace of the operator $ {e^{-\beta \hat H}}$:
$$Z=\mathrm{Tr} e^{-\beta \hat H}.$$ 
This expression is called the statistical sum or partition function. 

If the spectrum $\hat H$ is discrete with eigenvalues $E_i$, then
$$Z=\sum e^{-\beta E_i}.$$
The physical meaning of $\beta$ is the inverse temperature: $\beta=\frac{1}{T}$.  The expression for the statistical sum shows that for  $\beta \to \infty$ or, what is the same, $T\to 0$, only the term with minimum energy (the term corresponding to the ground state) contributes to this expression ( we assume that the ground state is non-degenerate). We see that in the case of zero temperature the equilibrium state is a pure state (ground state).

The derivation of the formula (\ref{eqextrem}) is based on the method of  Lagrange multipliers.  We assume that the average energy $ \mathrm{Tr} \hat H K=E$ is fixed; in addition, we know that the trace of a density matrix is equal to one: $ \mathrm{Tr} K=1$.  Introducing Lagrange multipliers $\beta$ and $\zeta$ we see that we should calculate stationary points  of the expression
\[L=-\mathrm{Tr} K\log K -\beta \mathrm{Tr}(\hat HK-E)-\zeta (\mathrm{Tr} K-1).\]
These points obey the equation
\[-\log K- 1-\beta \hat H-\zeta=0.\]
To verify this we must use the formula :
\begin{equation}\label{eqdeltaTr}
 \nonumber
 \delta \mathrm{Tr} \phi (K)=\mathrm{Tr} \phi'(K) \delta K
\end{equation}
for the variation of the trace of a function of $K$.

It is sufficient to check this formula for the case $\phi (K)=K^n$.
In the variation of the function $K^n$ we have  $n$ terms due to noncommutativity, but when we take the trace, all these terms become identical and we get an answer agreeing with (\ref{eqdeltaTr}).
 
Although the statistical sum $Z$ itself has no direct physical meaning, many physical quantities can be expressed in terms of $Z.$ In particular, we can calculate the average energy 
\[E=\bar H=-\frac{1}{\beta} \frac{\partial \log Z}{\partial \beta}\]
and entropy
\[S=\beta E+\log Z.\]

Free energy is defined by the formula $F=E-TS$; it is expressed in terms of the statistical  sum as follows:
\[F=-T\log Z.\]
Free energy is convenient because instead of calculating the maximum of entropy one can search for the minimum of free energy.  This immediately follows from the method of Lagrange multipliers.

As a rule, the physical quantities can be obtained as follows: we take some Hamiltonian, add something to it and see what we get in the limit when the added term tends to zero. In particular, if  the Hamiltonian $\hat H$ changes a little:
$$\hat H(\lambda)=\hat H+ \lambda A+\ldots,$$
then the new value of the statistical sum is
\[Z(\lambda)=Z+(-\beta)\lambda \mathrm{Tr} A{e^{-\beta \hat H}}+\ldots.\]
The derivative of $\log  Z$ with respect to $\lambda$ at $\lambda=0$ (and hence the derivative of free energy at this point)  is  controlled by the average value  of the added term
\[\bar A=\frac{\mathrm{Tr} A{e^{-\beta \hat H}}}{Z}.\]
(I will also use the alternative notation $\bra A\ket$ for this expression). 

Namely,
\begin{equation}
 \label{eq:50.3}
 \bar A=-T\frac{\partial \log Z}{\partial \lambda}=\frac{\partial F}{\partial \lambda}
\end{equation}
(the derivatives are calculated at the point $\lambda =0).$

Now let's move on to our main goal, to correlation functions. A correlation function is simply an average of a  product of some physical quantities $\bra A_1 \cdots A_n \ket$.  We can also assume that these physical quantities are time-dependent - they are Heisenberg operators and satisfy Heisenberg equations. Then the expression $\bra A_1(t_1)\cdots A_n(t_n)\ket$ is also called a correlation function. You can consider correlation functions for any state - not necessarily an equilibrium state, but if it is an equilibrium state, I write the inverse temperature value as the index: 
$$\bra A_1(t_1) \dots A_n(t_n) \ket _{\beta}.$$

If the Hamiltonian depends linearly on a set of parameters $\lambda_1,\dots , \lambda_k$, we can calculate the correlation functions by differentiating the free energy $F$.
For example, if $\hat H=\hat H_0+\lambda_1A_1+\dots \lambda_k A_k$ we have
\begin{equation}
 \label{eq:50.4}
 \frac{\partial^2 F}{\partial \lambda_i\partial \lambda_j}= \bra A_iA_j\ket -\bra A_i\ket \bra A_j\ket
\end{equation}
(The derivatives are calculated at the point $\lambda_i=0.$)
The RHS of~(\ref{eq:50.4}) is called a truncated correlation function.
Higher truncated correlation functions can be defined as higher derivatives of $F.$ They will be important later.

 I want to note that all statements about the statistical sum and related things are applicable very often in the case of a finite number of degrees of freedom, but in the case of infinite number of degrees of freedom, they usually are not applicable. In nonrelativistic quantum mechanics, they are applicable when we are in a finite volume. In infinite volume they do not work - the statistical sum is not well defined, and maximum entropy is infinite.  One should consider the statistical sum  and 
 correlation functions first in finite volume and after that one should take the limit of correlation functions. One cannot work directly in the infinite volume.

I should also note that in passing to an infinite volume one usually takes a limit not with a fixed number of particles, but  with a fixed density  of particles. In other words, when passing to the limit, the number of particles changes in proportion to the volume,  then the density of particles remains constant.

Suppose now that the set of correlation functions in the infinite volume is obtained - what to do with it? There is no Hilbert space in which the operators $A$ entering the definition of correlation functions are defined  in  infinite volume, but there are correlation functions. Usually in such a case  one can construct a Hilbert space from these correlation functions, applying some analog of the GNS construction. Physicists usually do this implicitly. They simply say ``Now we have an equilibrium state - it can be represented by a vector in a Hilbert space or a density matrix in a Hilbert space and there these operators $A$ act". In fact, it is necessary to apply a construction which in axiomatic quantum field theory is called ``reconstruction theorem''. There the role of correlation functions is played by Wightman functions. 

Now let me turn to the question: how can one deal in the algebraic approach with equilibrium states in a situation when it is impossible to use the maximum entropy principle? Here we can apply what is called the Kubo-Martin-Schwinger condition (KMS):
\begin{equation} \nonumber
 \bra A(t)B\ket _{\beta}=\bra BA(t+i\beta)\ket _{\beta}.
\end{equation}
This is the condition on the correlation functions of the observables $A$ and $B$ in the equilibrium state. It is easy to derive in the case of finite-dimensional Hilbert space because in this case, everything is well-defined. In such a case, there is a Heisenberg operator which involves $e^{iHt}$ in the definition, there is a density matrix in which $e^{-\beta H}$ appears. We can assume that the time $t$ in the expression for the evolution operator is a  complex number, and when it is purely imaginary with $Im(t)>0$, then we obtain the equilibrium state density matrix (up to a constant factor). This is an important observation: in some sense, we obtain statistical physics from quantum dynamics in imaginary time. This is what is called ``Wick rotation.''  In finite-dimensional case we can consider any complex time, if the dimension is infinite this is not true. However, if  the correlation function 
$\bra BA(t)\ket_{\beta}$ can be continued analytically 
into the strip $0\leq Im(t)\leq \beta$,  the KMS condition  makes sense. 

The KMS condition does not use the notion of entropy - you only need to know correlation functions. It also works in an infinite volume. One can consider the KMS condition as a definition of an equilibrium state.
 The equilibrium state, as I defined it earlier, is almost always unique, and here there is room for the equilibrium state to be non-unique. This non-uniqueness of the equilibrium state is related to the presence of phase transitions. The KMS condition is a replacement for the maximum entropy condition in the framework of algebraic quantum theory.

Now let's look at some examples.

The simplest example is the quadratic Hamiltonian.  

A positive definite quadratic Hamiltonian can be reduced to the form:
$$\hat H=\sum \epsilon_k\hat a^*_k\hat a_k, $$ 
describing non-interacting bosons. Non-interacting bosons from the formal point of view are the same as a multi-dimensional harmonic oscillator, for which one can easily calculate the equilibrium density matrix, statistical sum, average energy, etc. The statistical sum is equal to a product of statistical sums for different values of the index $k$. For each $k$ one can sum the geometric progression and then take the product:
$$Z=\Pi \frac{1}{1-e^{\beta \epsilon_k}}.$$
The average energy  is calculated by the formula: 
$$E=\bar H=\sum {\epsilon_i}{\bar n_i},$$
where $\bar n_i=(e^{\beta\epsilon_i}-1)^{-1}$ are the average occupation numbers.

For the case of fermions, there is no significant difference: 
$$Z=\Pi ({1+e^{\beta \epsilon_i}}),$$

$$\bar H=\sum {\epsilon_i}{\bar n_i},$$
where $\bar n_i=(e^{\beta\epsilon_i}+1)^{-1} $.
The main difference is that the expression for the occupation numbers we have a plus rather than a minus, so the average occupation  numbers are always less than one.

\newpage
\section{Lecture 4}

\subsection {Adiabatic approximation. Decoherence}

I will begin this lecture by explaining what happens when the Hamiltonian 
$\hat H(t)$ depends on time, but changes slowly (adiabatically). I assume that all energy levels $E_n(t)$ of the Hamiltonian $\hat H(t)$ where $t$ is fixed are different and depend on $t$ continuously and even smoothly. I denote the corresponding eigenvectors by $\phi_n (t)$. I assume that the time-dependent vector $\phi_n (t)$ changes slowly and its derivative over $t$ can be neglected.  We will show that if we start with the eigenvector of the Hamiltonian $\hat H(t=0)$, then during the evolution controlled by the slowly changing Hamiltonian $\hat H(t)$ it remains an eigenvector, but it will be an eigenvector of another Hamiltonian (of the Hamiltonian $\hat H(t)$ where $t$ is fixed).  We verify that in the adiabatic approximation
\begin{equation}
\label {AS}
\hat U(t)\phi_n(0)= e^{-i\alpha_n(t)}\phi_n(t),
\end{equation}
where the phase factor $e^{-\alpha_n(t)}$
 is defined by the equation: 
\begin{equation}
\nonumber
\frac {d\alpha_n(t)}{dt}=E_n(t). 
\end{equation}
To check this we differentiate  (\ref {AS}), and neglect the derivative of $\phi_n (t)$. This reasoning is not completely accurate because I assumed that $\phi_n (t)$ changes slowly over time, which is not obvious because the eigenvector is defined only up to a constant factor. 

Let us carry out a more careful consideration.  Let us assume that the Hamiltonian $\hat H(g)$ depends on some parameter or many parameters, which I denoted by $g$.  Suppose that the eigenvectors $\phi_n (g)$ and the eigenvalues $E_n(g)$ depend smoothly on $g$.  Let us assume that the parameter $g$ depends on time in such a way that the derivative of $g$ with respect to  $t$ can be neglected. 
The standard choice is as follows: let us fix a function $g(t)$  and construct a family $g_a(t)=g(at).$ Corresponding Hamiltonians $\hat H_a(t)=\hat H(g(at))$ and their eigenvectors $\phi_n(g(at))$  vary slowly for small $a$. Obviously, the derivative of these eigenvectors with respect to $t$ vanishes  in the limit $a\to 0$; this remark allows us to justify the 
 above reasoning. 

In order to generalize this reasoning to density matrices, we should  note that the dependence of the density matrix $K(t)$ on time is determined by the equation
$${{dK}\over{dt}}=H(t)K(t)= \frac {1}{i\hbar}[\hat H(t),K(t)],$$
where $H(t)$ is a commutator with the Hamiltonian $\hat H(t)$ (up to a constant factor). 
Now we have to consider the eigenvectors $\psi_{mn}(t)$ of the operator $H(t)$. They  can be expressed in terms of  the eigenvectors of the Hamiltonian $\hat H(t)$, which I have denoted as $\phi_{n}(t)$ :
$$\psi_{mn}(t)x=\langle x,\phi_n(t)\rangle \phi_m(t).$$
In the representation where the operator $\hat H(t)$ is diagonal, these will be matrices having only one non-zero entry equal to $1$.  Exactly the same reasoning as above determines the evolution  of these eigenvectors: 
\begin{equation}
\nonumber
U(t)\psi_{mn}(0)= e^{-i\beta_{mn}(t)}\psi_{mn}(t),\;\;\;
\frac {d\beta_{mn}(t)}{dt}=E_m(t)-E_n(t).
\end{equation}
( We  differentiate with respect to $t$ and neglect the derivative of the vector $\psi_{mn}(t)$.)  A very important remark:  when $m=n$ we can assume that the phase is zero: $\beta_{mm}=0$.

Let us represent the same in a slightly different notation, namely, let us take the density matrix and write it as a sum over the eigenvectors with some coefficients $k_{mn}$:
$$K=\sum k_{mn}\psi_{mn}.$$
Instead of considering the evolution of the eigenvectors $\psi_{mn}$ we can
consider the evolution of the coefficients $k_{mn}(t)$. The formulas are the same - 
coefficients $k_{mn}(t)$ get phase factors:
$$k_{mn}(t)= e^{-i\beta_{mn}(t)}k_{mn},\;\;\;
 \beta_{mn}(t)=\int_0^t(E_m(\tau)-E_n(\tau))d\tau.$$
 If the adiabatic Hamiltonian $\hat H(t)$ is such that at time $T$ it returns to what it was at time zero: $\hat H(T)=\hat H(0)$, then the diagonal elements of matrix $K$ do not change, but the non-diagonal elements do - they are multiplied by a phase factor. 

Let us fix now the Hamiltonian $\hat H$ describing a molecule or an atom or something bigger - any quantum system. Let's assume that the interaction with the environment changes the Hamiltonian; the new Hamiltonian $\hat H(t)$ may depend on time, but we assume that it changes slowly.  We can imagine that a cosmic particle flies not very close to our molecule. The particle generates an electric field; this means that the Hamiltonian governing the molecule changes. If this particle flies far enough we can assume that the change is adiabatic. 

My favorite example: you are doing an experiment, and in the next room, someone turned on a microwave.  Then your experiment is affected by an adiabatic electric field.
Another example: we know that we live in a world where there is microwave cosmic radiation, which also generates some electromagnetic field. It's very small, but it's there nonetheless.

 We do not know these adiabatic perturbations, but we know that the diagonal elements of the density matrix are not affected by adiabatic perturbation, and the non-diagonal elements of the density matrix acquire phase factors, which we, of course, do not know, because we do not know the perturbation.

What I said can be interpreted in a different way. One can consider linear combinations of the form $\alpha_0\phi_0+\alpha_1\phi_1$ of two or more eigenvectors of the Hamiltonian $\hat H=\hat H(0)$.In the evolution of this state, phase factors will appear: $\alpha_k(t)= e^{-iE_k t}\alpha_k$. These phase factors always appear, they are predictable, but if an adiabatic perturbation is imposed, then phase factors become unpredictable ( absolute values of coefficients $\alpha_k(t)$ remain constant).  Before, the two eigenvectors were coherently changing over time, but now this coherence has disappeared. This is what is called decoherence. 

Now I want to explain how from these very simple considerations one can get a standard recipe for probabilities in quantum theory. Let us assume that the same molecule interacts with the environment and there is a random adiabatic perturbation $\hat H(t)$ of the Hamiltonian  $\hat H$. This means that there is a Hamiltonian that depends on some parameters $\lambda\in \Lambda$  and there is some probability distribution on $\Lambda$. Let us assume that the adiabatic perturbation acts in the time period from 0 to $T$. Then, as I said before, the entries  $k_{mn}{\lambda},T)$ of the density matrix $K_{\lambda}(T)$ in the $\hat H$ -representation get phase factors $ C_{mn}(\lambda,T)$. The phase factors $ C_{mn}(\lambda,T)$ are equal to $1$ for the diagonal entries and non-trivial for other entries. Since the Hamiltonian is random, the density matrix should be averaged over the perturbation, that is, the phase factors for the non-diagonal entries should be averaged. It is quite clear that the averaging these phase factors results in something that is less than $1$ by absolute value. By imposing some conditions, it is easy to check that the average of the non-diagonal matrix entries will be equal to zero.

 The formal proof is as follows. I have already said that we can include our Hamiltonian into some family of Hamiltonians $\hat H(g)$, where $g$ belongs to some parameter set denoted by $\Lambda$ ($g\in \Lambda$). 
Let us assume that all these perturbations are such that $g(0)=0, g(1)=0$ and the dependence of the Hamiltonian on time is defined by the formula $\hat H(g(t))$. I define the adiabatic Hamiltonian as follows:

$$\hat H_{\alpha}(t)= \hat H(g(\alpha t)),$$
where $\alpha\to 0$.
This will stretch time. If before it varied from zero to one, now it varies from zero to $T=\alpha^{-1}$. If we denote by $E_n(g)$ the eigenvalues of the Hamiltonian $\hat H(g)$, the values of phase factors $e^{-i\beta_{mn}(t)}$ at $t=T$ will be determined by the following formula: 
$$\beta_{mn}=\int _0^T d\tau(E_m(g(\alpha\tau))-E_n(g(\alpha\tau))).$$

By substituting $\alpha\tau=\tau'$ we obtain: 

$$\beta_{mn}= \frac {1}{\alpha}\int _0^1 d\tau'(E_m(g(\tau'))-E_n(g(\tau')).$$

Now let us use the Riemann-Lebegue lemma;
$$\int e^{ikx}\rho(x)dx\to 0$$  
at $k\to \infty$ if $\rho(x)$ is absolutely integrable.

 If the probability distribution on the set $\Lambda$ is not too bad (with a decent probability density distribution), then the coefficients $k_{mn}$ of the density matrix will vanish if $ m \ne n$.
As a result, the density matrix $K$ becomes diagonal in $\hat H$-representation  due to interaction with random adiabatic perturbation.  (If the initial density matrix corresponds to a pure state this effect is known as the collapse of the wave function.)

Let us now denote the diagonal matrix elements by the letters $p_n$. In the usual approach, $p_n$ are the probabilities of different states. In our case, the averaged density matrix is a mixture of pure states with probabilities $p_n$. This is  the usual formula for the probabilities of different pure stationary states (of eigenstates of the Hamiltonian $\hat H$) in a given mixed state. In particular, if we started with a pure state  we obtain the standard formulas of the theory of measurements for the probabilities  of different energy levels.

Notice that usually decoherence and the collapse of the wave function are derived from interaction with a macroscopic classical system.
Here the same statements were derived from the random adiabatic interaction.  Planck constant   and classical systems were not used in the proof. 

 \subsection {Geometric approach to quantum theory. Decoherence.}
So far we have considered ordinary quantum mechanics, but now we will consider the geometric approach to quantum theory. I have already said that there is an algebraic approach, where the starting point is an algebra of observables - associative algebra with involution. Self-adjoint elements of the algebra are physical observables.  States are defined as positive linear functionals on this algebra, that is, the notion of state is secondary. Let us assume that the notion of state is primary. We start with some set of states.  Let us ask ourselves what should we require of it.

The first thing I need is the notion of a mixture of states-not a mixed state, but a mixture of states so that any states can be mixed with some probabilities, with some weights. This is an absolutely necessary requirement and it is not specific to physics. You can, for example,  consider mixed strategies in economics or in game theory. In order to be able to mix, the set must be convex. Let us assume that a convex set is a subset of some vector space and then the notion of mixing will be defined in an obvious way: if $x_i$ are points of the convex set, $p_i$ are non-negative numbers whose sum is equal to $1$, then the mixture of these points is given by the formula $\bar x=\sum p_ix_i$
(the numbers $p_i$ are treated as weights or probabilities).

What else should be required? If we mix a finite number of states, then nothing else is needed. But if we want to mix an infinite number of states, then we need the notion of a limit.  In other words, we need a topology in the vector space $\cal L$ in which the convex set ${\cal C}_0$ lies. It must be a topological vector space. For simplicity, we assume that it is a Banach space (a  complete normed space).  We assume that  ${\cal C}_0$ is a closed convex subset of  $\cal L$; then we can mix any number of states.

We also require that the set of states be bounded.  This is essentially the only requirement necessary
 for the development of the theory. 

We also need the notion of an evolution operator $\sigma (t)$, which transforms a state at time zero to a state at some time $t$. (After all, what is the most important thing not just in physics - in science?  You should be able to predict the future.) The evolution operator should map the set of states into itself.  The time $t$ can be negative, hence the evolution operator should be an invertible transformation of the set of states. 
 
Let us assume that the evolution operator is a linear transformation. More precisely, we assume that it can be extended as a linear operator to a vector space $\cal L$ containing the set ${\cal C}_0$.  Let us introduce the notion of a group of automorphisms of the set of states ${\cal C}_0$. This is a group of invertible linear operators in the ambient space $\cal L$ which map the set ${\cal C}_0$ onto itself. It is natural to assume that the evolution operator will be an automorphism. Sometimes we require that evolution operators belong to some subgroup $\cal V$ of the 
group of automorphisms. This is necessary, for example, in classical mechanics.

Next, we should write the equation of motion.  Usually, we think that we know the change in the system over an infinitesimal time. This is called the equation of motion.  In the most general form, the equation of motion can be written as follows:
\begin{equation}\label{eqEM}
 \frac {d\sigma}{dt}=H(t)\sigma(t),
 \end{equation} 
 where $H(t)$ is a linear operator.

Formally we can say that $H(t)$ is determined from this equation, but in physics, we usually assume that the operator $H(t)$ is known and we need to find the evolution operator. Let us call the operator $H(t)$ ``Hamiltonian'' (in quotation marks). If ``Hamiltonian'' does not depend on time, the evolution operator is an exponent of ``Hamiltonian'' : $\sigma(t)=\exp(Ht)$. Everything is as in ordinary quantum mechanics, I just did not write the imaginary unit. (For a time-independent operator $H$ in Banach space the exponent can be defined as a solution of the equation (\ref{eqEM}). )

From equation (\ref{eqEM}) it follows that the ``Hamiltonian'' belongs to the Lie algebra of the group 
$\cal V$ (it is the tangent vector of the group $\cal V$ in the unit element of the group).  The group $\cal V$ is, generally speaking, infinite-dimensional, so the notion of a tangent vector depends on the choice of topology in the group, but I will not pay attention to these subtleties. It is reasonable to require that for a time-independent ``Hamiltonian'' the equation (\ref{eqEM}) has a solution.

Now I have the notion of states, I have the equations of motion, but I also need the notion of observable. Before, I was going from observables to states, and now I want to go from states to observables. What is observable? First of all, we must have an operator $A$ satisfying the same conditions as on the ``Hamiltonian''.
 In other words, the exponent $\sigma_A(t)=\exp (At)$, which can be treated as a one-parameter subgroup in $\cal V$, should be well defined.
 We need also a functional $a$ which is invariant with respect to the operators $\sigma_A(t)=\exp (At)$. (This condition is equivalent to the condition $a(Ax)=0.$) 
The function $a$ determines the expected value of the observable.
 
In particular, in the case of ordinary quantum mechanics $\cal V$ is a group of unitary operators. It acts on density matrices by the formula $U(K)=\hat UK\hat U^{-1}$, where $\hat U$ is a unitary operator. If $\hat A$ is a self-adjoint operator (not necessarily bounded), then the exponent $e^{i\hat A t}$ can be treated as a one-parameter subgroup of the group $\cal V$. This way we obtain all one-parameter subgroups continuous in strong topology (Stone's theorem). This means that we can identify self-adjoint operators with elements of the Lie algebra of the group of unitary operators.  (There are small difficulties here due to the fact that the commutator of unbounded self-adjoint operators is not always well-defined.  These difficulties arise always and have no relation to the geometric approach. To overcome these problems we assume that in topological Lie algebra, the commutator is defined only on a dense subset. )
  
  The ``Hamiltonian'' is expressed in terms of the self-adjoint operator $\hat H$ by the formula
 $$H(K)= \frac {1}{i}[\hat H,K],$$
where $K$ is the density matrix. (We assume that $\hbar =1$). 

Similarly, for any self-adjoint operator $\hat A$ we define an operator on  density matrices by the formula:
\begin {equation}\label{eqA}
  A(K)= \frac {1}{i}[\hat A,K].
 \end{equation}
An observable is  a pair 
$(\hat A,a)$,
where $\hat A$ is a self-adjoint operator acting on the density matrices by the formula (\ref{eqA}) and the functional on the density matrices $a$ is given by the formula $a(K)=\mathrm{Tr} (\hat AK)$. 
\vskip .1in

In the algebraic approach, the group $\cal V$ consists of automorphisms of the associative algebra $\cal A$ with involution  ($*$-algebra); in particular, a self-adjoint element $A$ of the algebra $\cal A$ defines an infinitesimal automorphism (a Lie algebra element of the group $\cal V$) by the formula $\alpha_A(X)=i[A,X]$.  Recall, that a  linear functional $\omega$ on algebra  $\cal A$ is called positive if $\omega (X ^ * X) \geq 0$ for all $X\in \cal A$; I denote the set of all positive functionals by $\cal C$.  

The set of states ${\cal C}_0$ consists of normalized positive functionals (positive functionals satisfying the condition $\omega(1)=1$). This is a bounded convex set; the group $\cal V$ naturally acts on it.  An observable is a pair $(A,a)$, where $A$ is a self-adjoint element of the algebra $\cal A$ treated as an infinitesimal automorphism, and $a$ is a linear functional on ${\cal C}_0$ mapping the state $\omega$ to the number $a(\omega)=\omega (A)$.

In the geometric approach, we also have decoherence. The proof of this is pretty much the same as before. I will start with a time-independent ``Hamiltonian'' denoted by the letter $H$. Then the evolution operator is an exponent $\sigma (t)=e^{tH}$. I will assume that the operator $H$ is diagonalizable, that is, there exists a basis $(\psi_j)$ consisting of eigenvectors of the operator $H$:
$$ H\psi_j=\epsilon_j \psi_j.$$

I have assumed that the evolution operator maps the set of states into itself and that the set of states is bounded. Under these conditions,  the operators $\sigma(t)$ are uniformly bounded  (the norms of these operators are bounded by the same number).

 The eigenvalues of the operator $H$  are purely imaginary because a function of the form $e^{\epsilon t}$ is bounded only if $\epsilon$ is purely imaginary. If there is a Jordan cell of size greater than one the exponent $\sigma (t)=e^{tH}$ will not be bounded, hence such Jordan cells cannot  appear in $H.$ 
 
 In finite-dimensional case, these statements imply that $H$ is diagonalizable.
In the infinite-dimensional case, they are not sufficient for diagonalizability, but nevertheless one should expect that the operator $H$ is diagonalizable. This is what I will assume.

Now I repeat my reasoning. I include my ``Hamiltonian'' into the family $H(g)$ of ``Hamiltonians'', with eigenvalues $(\epsilon_j(g))$ and eigenvectors $(\psi_j(g))$ that smoothly depend on $g$:
$$ H(g)\psi_j(g)=\epsilon_j (g)\psi_j(g),$$

I assume that these eigenvectors constitute a basis that coincides with the basis  $\psi_j$ of eigenvectors of $H=H(0)$ for $g=0.$

Further, I say that $\psi_j$ is a robust zero mode if $\epsilon_j(g)\equiv0$, that is, the eigenvalue is zero at any $g$. 

Now I want to do the same thing as before - I will assume that the interaction with the environment is determined by a random  ``Hamiltonian" $H(g(t))$, where $g$ depends on $t$, and will assume that these random ``Hamiltonians'' are adiabatic (they change slowly so that the derivative of $g$ with respect to $t$ can be neglected). 
Then the  vector  $\sigma (t)\psi_j $ is an  eigenvector of the ``Hamiltonian' $H(g(t))$ where $t$ is fixed:
 $$ \sigma (t)\psi_j= e^{\rho_j(t) }\psi_j(g(t)),$$
 where $\frac {d\rho_j}{dt}=\epsilon_j(g(t)).$
 To prove this, we differentiate the right-hand side of this expression by $t$ applying the equations of motion and neglecting the derivative $\dot g(t)$. I do not write imaginary unit, in these formulas  but $\epsilon_j(g)$ itself is purely imaginary.

The phase factor does not appear for the robust zero modes. If adiabatic perturbation acts for some finite time, then at the end the robust zero modes do not change.  All other modes acquire phase factors.

Everything is very similar to ordinary quantum mechanics. For a nonrobust zero mode taking an average of phase factors with respect to random perturbation, we get zero, and for the robust zero modes, nothing changes. 

In ordinary quantum mechanics, the robust zero modes are the diagonal elements of the density matrix in $\hat H$ - representation. (They are zero modes because all diagonal matrices commute with each other. It is easy to check that they are robust.) 

In general situation an arbitrary state $x\in {\cal C}_0$ can be represented as a linear combination of eigenvectors of the operator $H$; the interaction with the environment kills all modes except the robust zero modes. I  denote by $P'$ the operator killing all but the robust zero modes; we can say that all observable physics lies in its image (in the projection on the robust zero modes).
Next, we must represent the state $P'x\in {\cal C}_0$  as a mixture of pure robust zero modes
$P'x=\sum p_i u_i.$  (Notice that this representation is not necessarily unique.)
The coefficients $p_i$ in this expansion can be interpreted as probabilities. In ordinary quantum mechanics, the usual probabilities are obtained this way. In the general case 
coefficient $ p_i $ is the probability of a pure robust zero mode $u_i$ in the state $x.$

The Hamiltonian $H$ corresponds to the physical quantity $(H, h)$, where the function $h$ must be identified with the energy. The coefficient $ p_i $ should be interpreted as the probability of finding the energy $h(u_i)$ in the state $x.$ (It is assumed that all numbers $h(u_i)$ are different. If some of them are the same, then to get the probability that the energy is $h$, we have to sum all the coefficients $p_i$ for which $h(u_i)=h$.)

If all zero modes are robust, we can write a simple formula for the operator: $P'=P$, where $P$ is the operator that kills all non-zero modes:
$$P=\lim _{T\to \infty}\frac{1}{T}\int _0^T\sigma(t)dt.$$
If we decompose $\sigma(t)$ by eigenvectors in this integral, then non-trivial phrase factors appear for the nonzero modes; averaging them over time we get zero.

I have explained how the probabilities for ``Hamiltonians'' appear. They are interpreted as probabilities of different energy levels. One can repeat the same reasoning by considering other observables represented as  pairs $(A,a)$, where $A\in \cal V$, $a$ is a functional obeying $a(Ax)=0$. It is possible to define the notion of a robust zero mode for any observable. 

I will give a slightly different, more general, definition of the robust zero mode. When it is said that $x$ is a robust zero mode, first of all, it is necessary to require that it is a zero mode: $Ax=0$. If $A$ is slightly changed (replaced by a close element $A'$ of the group $\cal V$), then one must require that $A'$ has a zero mode $x'$ close to the original zero mode $x$. 

Everything that was said for the energy can be repeated for an arbitrary observable. It is necessary to consider the projection $P_A$ on the space of zero modes of $A$: 
  $$P_A=\lim_{T\to\infty}\frac {1}{T}\int_0^T dte^{At}.$$ 
(I assumed that all zero modes are robust). After that, it is necessary to represent the projection on the zero modes as a mixture of extreme points:
$$P_A(x)=\sum p_i u_i.$$
The coefficients $p_i$ will be interpreted as probabilities of values $a(u_i)$ in the state $x$.

\subsection {L-functionals}

I have now explained that there is a formalism in which the starting point is a set of states. There is a question that a physicist should ask: is this formalism convenient, is it convenient to calculate in this formalism? I am going to answer this question now. I will work in the framework of the algebraic approach, assuming that the algebra is  the  Weyl algebra with generators $\hat u^i$ and relations

$$\hat u^k\hat u^l-\hat u^l\hat u^k=i\hbar\sigma^{k,l}.$$

I want to introduce
the notion of an L-functional corresponding to density matrix $K$ in any representation of a Weyl algebra by the formula:

$$L_K(\alpha)=\mathrm{Tr} e^{i\alpha_k \hat u^k}K=\mathrm{Tr} V_{\alpha} K,$$
 where $V_{\alpha}=e^{i\alpha_k \hat u^k}=e^{i\alpha \hat u}$ are operators that we have already considered,
$\alpha_k$ are real numbers corresponding to the generators of Weyl algebra. Recall that we can assume that Weyl algebra is generated by operators $V_{\alpha}$. If $u_k$ is self-adjoint and $\alpha_k$ is real, then these generators are unitary and satisfy the relations
\begin{equation}\label {WW}V_{\alpha}V_{\beta}=e^{-i\frac{\hbar}{2}\alpha\sigma\beta} V_{\alpha+\beta},
\end{equation} 
where $\alpha\sigma\beta=\alpha_k\sigma^{k,l}\beta_l$. 

This is the so-called exponential form of Weyl algebra.

An important property of L-functionals is that they unite all representations of the Weyl algebra. The problem of non-equivalence of representations of the Weyl algebra in the L-functional formalism completely disappears. In the formula for an L-functional a unitary operator is multiplied by an operator from the trace class, and therefore the trace is well-defined. 

I can introduce the space $\cal L$ of all linear functionals on a Weyl algebra.  We can assume that an L-functional defines a linear functional on a Weyl algebra: $L_K\in \cal L$. ( This follows from the observation that every element expressed in terms of generators $V_{\alpha}$ by a finite number of addition and multiplication operations is a linear combination of generators $V_{\alpha}$).
 As it is easy to check, the functional $L_K$ is positive and, moreover, normalized (on the unit element of the algebra it is equal to one).  In what follows I will identify 
L-functionals with positive functionals on the Weyl algebra.

Let us define operations in the space $\cal L$ under consideration. These operations are defined for the space of linear functionals on any $*$-algebra.  

There are two operations on linear functionals for every element $A$ of the algebra. One can define an operation on a functional $\omega (x)$, where $x$ is an element of the algebra, by multiplying $x$  by $A$ from the right. Another operation is obtained by multiplying $x$ by $A^*$ from the left.
I denoted the first of these by the same letter $A$, and the other one  by $\tilde A$ :
$$(A\omega)(x)=\omega (xA),\;\; 
(\tilde A\omega)(x)=\omega (A^*x).$$

  Let us ask the question: does the functional  $\omega$ remain positive when acted on by these operators? The answer is no, but if we apply the combination $\tilde A A$, then the positive functionals are mapped into positive functionals (which are not necessarily normalized). Recall that positive functionals must be non-negative on elements of the form $x=B^*B$. It is easy to check that the operation $\tilde A A$ transforms $x \to A^*xA= (BA)^*x BA$. We see that positivity is preserved. 

This is an important statement. We denote by $\cal C$ the space of all positive functionals that are not necessarily normalized. The operator $\tilde A A$ acts in it. 

The next observation is that operators of the form $\tilde A$ always commute with operators of the form $A$.  This is because one multiplies from the left and the other multiplies from the right. 

Further, if the operator has the form 
$A=e^{it H}$, then $\tilde A=e^{-it\tilde H}$.
If the equations of motion are written as 
\begin{equation}\label{eqEMotion}
  id\sigma/dt=(\tilde H-H)\sigma,
\end{equation}
then the evolution operator will have the form
$e^{-it(\tilde H-H)}.$ 
According to the previous observation, this expression can be represented in the form $\tilde A A$. That is if the equations of motion are written in the form (\ref{eqEMotion}), then $\sigma$ acts on the cone of positive functionals (cone of not necessarily normalized states). I will use this fact.
 
Let us now turn to the question of the form of the equations of motion in the case of a Weyl algebra. It is easy to calculate the operators $V_{\beta}$ and $\tilde V_{\beta}$ acting in the space $\cal L$ :
 $$(V_{\beta} L)(\alpha)= e^{+i\frac{\hbar}{2}\alpha\sigma\beta}L(\alpha+\beta), 
 \;\; (\tilde V_{\beta} L)(\alpha)=
 e^{+i\frac{\hbar}{2}\alpha\sigma\beta}L(\alpha-\beta),$$
Let us assume that $\hat H$ is represented as an integral
$$\hat H=\int d\beta h(\beta) V_{\beta}.$$
This operator will be self-adjoint if $h(-\beta)=h(\beta)^*$.
Introducing the Planck constant we should replace $\hat H$ with 
$\frac{1}{\hbar}\hat H.$
The equation of motion in which $\hat H$ plays the role of the Hamiltonian
will take the form:
  $$i \hbar \frac {d\sigma}{dt}=(\tilde {\hat H}-\hat H)\sigma.$$
The equation in L-functionals can be written in the form:
$$i\hbar \frac {dL}{dt}=\int d\beta h(-\beta)e^{+i\frac{\hbar}{2}\alpha\sigma\beta}L(\alpha-\beta) -\int d\beta h(\beta)e^{+i\frac{\hbar}{2}\alpha\sigma\beta}L(\alpha+\beta) =$$
$$=-\int d\beta h(\beta)(e^{+i\frac{\hbar}{2}\alpha\sigma\beta}-e^{-i\frac{\hbar}{2}\alpha\sigma\beta})L(\alpha+\beta).$$

As a result, we arrive at the formula: 
$$\frac {dL}{dt}=-\int d\beta h(\beta)\frac {2\sin(\frac{\hbar}{2}\alpha\sigma\beta)}{\hbar} L(\alpha+\beta).$$

It is clear from this formula  that the equation of motion for L-functionals has a limit when $ \hbar \to 0.$

\newpage
\section{Lecture 5}

 \subsection {Functional integrals}
This lecture is devoted, first of all, to functional integrals widely used in quantum mechanics. I will use Berezin's idea \cite {BERCO},\cite {BS}, which allows us to apply functional integrals not only in the conventional approach to quantum mechanics, but also in the geometric approach, and in L-functional formalism.

In quantum mechanics, a physical quantity is represented as a functional integral - an infinite-dimensional integral, where the integrand includes an exponent of the action functional multiplied by something. The action depends on a curve (more precisely, on a function $q(\tau)$, the graph of which we consider as a curve): 

$$S[q(\tau)]=\int_0^t d\tau L(q(\tau),\dot q(\tau)).$$

In this expression $q(\tau)$ is in square brackets to emphasize that $S$ is not a function, but a functional that depends on a curve, not on a point of a curve.

Matrix element 
$\langle q_2|\hat U(t)|q_1\rangle$ 
of the evolution operator $\hat U(t)= e^{-\frac {it\hat H}{\hbar}}$, in the coordinate representation is expressed in terms of a functional integral with an integrand
$$e^{\frac {i}{\hbar}S[q(\tau)]} $$
 (we integrate over a set of  curves $q(\tau)$ with  given starting point and end point: $q(0)=q_1, q(t)=q_2$.)

What does the word ``functional integral'' mean? An integral of the type I have described can be approximated by finite-dimensional integrals: under the sign of the functional integral, replace the ordinary integral with, say, integral sums and take the limit. The problem is that, unlike ordinary calculus, where the approximation scheme is irrelevant, the value of functional integral depends on the choice of approximation. Moreover, the limit usually is either infinite or zero, and we should do something to get a finite answer.

What I am about to discuss can be made rigorous up to some point. Rigorous statements can be proven for Gaussian integrals (for integrals of quadratic exponents q, possibly multiplied by a polynomial). To calculate such integrals in the finite-dimensional case, one can use the formula: 
$$\int \exp(\frac {i}{2}\langle Ax,x\rangle+i\langle b,x\rangle)dx=(\det A)^{-\frac 1 2}\exp(-\frac{ i}{ 2} \langle A^{-1}b,b\rangle).$$
(This formula must contain a constant factor, which we do not write, having included it in the definition of the measure. The operator $A$ can be complex, but one should impose some conditions to guarantee that the integral makes sense).
By differentiating this expression with respect to $b$, we conclude that an integral of the form 
$$\int P(x)\exp (\frac {i}{2}\langle Ax,x\rangle)dx,$$
where $P(x)$ - some polynomial can be expressed in terms of determinants, and at least for the case of elliptic operators there is a quite good theory for computing such determinants. Again, a determinant of an elliptic operator must be approximated by something, and then infinite terms of the expansion for the logarithm of the determinant must be discarded. 

If t$W(x)$ has the form 
$W(x)=Q(x)+gV(x)$, where $Q(x)$ is a quadratic expression, and we consider the constant $g$ to be small, then the normalized functional integral of the form

$$\frac {\int e^{W(x)}dx}{\int e^{Q(x)}dx}$$
can be calculated in the framework of perturbation theory (as a series with respect to $g$). Every term of the series is represented as a  sum of Feynman diagrams. In the more general case where $W(x)$ is represented as a quadratic on $x-x_0$ expression plus terms that contain only monomials of higher degree with respect to $x-x_0$, perturbation theory is also well defined.

After these preliminary words, let's move on to the construction of functional integrals. I will construct a functional integral for the exponent of an operator acting in Banach (not necessarily Hilbert) space or, more generally, I will solve the equation of motion:
\begin {equation}
\nonumber
\frac {d\sigma}{dt}=H(t)\sigma (t)
\end {equation}
in terms of functional integrals. The considerations below generalize the approach suggested by F. Berezin \cite{BERCO},\cite {BS}.

The operator $H(t)$ in the equation of motion can be a function of $t$, but, to simplify the formulas, I will assume that $H$  does not depend on $t$.
In ordinary quantum mechanics, the equation for the evolution operator is exactly the same, but in this case, $H$ is a self-adjoint operator up to multiplication by an imaginary unit.

Let us denote the space in which the operator $H$ acts by the letter $\cal L$. My main tool is the notion of a symbol of an operator. We already considered some symbols but now I would like to give a very general definition.  

A symbol of an operator is a function defined on some measure space (or, more generally, a function defined somewhere where there is a notion of integration).   
The symbol of an operator $A$ is denoted $\underline A$.  I impose the following conditions on symbols.  The symbol of the operator $A=1$ is equal to $1$. The symbol $\underline A$ must depend linearly on the operator $A$. The product of the operators $C=AB$ must correspond to some operation on symbols, which I denote by $*$, that is, $\underline C=\underline A *\underline B$. 

The following simple arguments lead immediately to the functional integral. I will use the standard formula for the exponent:
$$ \sigma (t)= e^{tH}=\lim _{N\to \infty} (1+\frac {tH}{N})^N.$$
For large $N$
$$\underline{1+\frac {tH}{N}}=e^{\frac {t}{N}\underline H}+O(N^{-2}).$$
The error will be of the order $\frac {1}{N^2}$, hence for  $N\to \infty$, this correction term can be neglected. We obtain
$$\underline{\sigma (t)}= \lim_{N\to \infty}I_N(t),$$
$$I_N(t)=e^{\frac {t}{N}\underline H}*...*e^{\frac {t}{N}\underline H}$$
($N$  factors).

My claim is that I derived a representation of the evolution operator as a functional integral. I need only give examples of symbols and decipher the meaning of `$`*$''.

I want to emphasize one thing that, in my opinion, is important; I believe it is underestimated by physicists. So far I have given trivial arguments that can be made rigorous. It is not necessary to talk about functional integrals - it is sufficient to investigate the functions $I_N(t)$. One can, for example, apply the stationary phase method and obtain more or less the same results as in the language of functional integrals.

Later I will introduce a large class of operators for which the operation $*$ is written in a simple form. This class of operators includes \qp-symbols and Wick symbols.  I think of a symbol as a function of two variables $\underline A(\alpha,\beta)$. For this large class of symbols, the expression for the symbol of a product looks like this:
\begin {equation}\label{eq28} 
\underline C(\alpha,\beta)=
\int d\gamma d\gamma'\underline A(\alpha,\gamma)\underline B(\gamma',\beta)e^{c(\alpha,\gamma)+c(\gamma',\beta)-c(\alpha,\beta)-r(\gamma',\gamma)},
\end{equation}
where $c(\alpha,\beta)$ and $r(\alpha,\beta)$ are some functions. 

Later I will construct a large class of symbols obeying (\ref{eq28}), but for now I postulate that there is such a formula for the symbol of the product. Knowing the formula for the product of two operators, we can write the formula for the product  of operators $A_1,...,A_N$: 
$$\underline {C}(\alpha,\beta)= \int d\gamma_1d\gamma'_1...d\gamma_{N-1} d\gamma'_{N-1} \underline A_1(\alpha,\gamma_1) \underline A_2(\gamma'_1,\gamma_2)...\underline A_n(\gamma'_{N-1},\beta) e^{\rho_N},$$
Where
$$\rho_N=c(\alpha,\gamma_1)+c(\gamma'_1,\gamma_2)+...+ c(\gamma'_{N-1},\beta) -c(\alpha,\beta))-r(\gamma'_1,\gamma_1)-...-r(\gamma'_{N-1},\gamma_{N-1}).$$ 

The result is the product of symbols multiplied by the exponent of some expression, which is denoted by $\rho_N$. 
Returning to the expression $I_N(t)$ that approximates  the evolution operator, we can represent it as:
$$
I_N(t)=\int d\gamma_1d\gamma'_1...d\gamma_{N-1}d\gamma'_{N-1}e^{\rho_N}\exp(\frac {t}{N}(\underline H(\alpha,\gamma_1) +\underline H(\gamma'_1,\gamma_2)+...+\underline H(\gamma'_{N-1},\beta)).$$

What I have described is a very broad scheme. There is one very concrete example, which is the \qp-symbol. If we consider the kernel of the unit operator in the sense of mathematics (or the matrix of the unit operator, as physicists say), it is a $\delta$-function: in the coordinate representation $\langle \bq_2|1|\bq_1\rangle=\delta (\bq_1-\bq_2)$, and in the momentum representation $\langle \bp_2|1|\bp_1\rangle=\delta(\bp_1-\bp_2)$. We want the symbol of the unit operator to be equal to one, not to the $\delta$-function.  This is very easy to do. The Fourier transform of the $\delta$-function is a constant, so in order to get a symbol equal to $1$ for the unit operator, we must take the Fourier transform and multiply it by a constant factor. Since the matrix of unit operator in coordinate representation is a $\delta$-function of $\bq_1-\bq_2$, we take the Fourier transform of the matrix element
 $\langle \bq_2|A|\bq_1\rangle$ with respect to  $\bq_1-\bq_2$:
$$\underline A^{q-p} (\bq,\bp)= \int d{\bf y}\langle {\bf y}|A|\bq\rangle e^{i\bp(\bq-{\bf y})}.$$

Hence I can define the \qp-symbol as the Fourier transform of a matrix element with respect to the difference between the arguments. Obviously, one can take the inverse Fourier transform and express the matrix elements $\langle \bq_2|A|\bq_1\rangle$ in terms of the \qp-symbol. 
Since we know how to express the matrix element of the product in terms of the matrix elements of the factors, we can calculate the \qp-symbol of the product of two operators. The answer is given by the formula (\ref{eq28}), 
where the functions $c(\alpha,\beta)$ and $r(\alpha,\beta)$ are  scalar products (up to a constant factor):
\begin {equation}\label{eq29} 
c(\bq,\bp)=r(\bq,\bp)=-i\bp \bq.
\end{equation}

The definition of the \qp-symbol that has just been given differs from the definition given earlier. This definition applies to any operator, as long as the integral makes sense. If $A$ is a differential operator, then it is easy to understand that the definition of \qp-symbol which I have just given agrees with the previous one. Let me remind the old definition. If there is a differential operator with polynomial coefficients, then it can be written as a polynomial of  coordinate operators $q^j$ and momentum operators $\hat p_j=\frac {1}{i}\frac {\partial}{\partial q^j}$. (The coordinate operators are on the left, and the momentum operators are on the right.) Then the hats should be removed from the operators. You will get a polynomial function called the \qp-symbol of the operator. 

Note that I considered Planck constant to be equal to $1$. Sometimes it is convenient to keep it in the formulas.

Turning to ordinary quantum mechanics, we find that the formula for $I_N$ takes the form
\begin{equation}\label{eqIN}
I_N(\bq,\bp,t)=\int \prod_1^{N-1}d\bq_{\alpha}d\bp_{\alpha}\exp(i\sum_1^N\bp_{\alpha}(\bq_{\alpha}-\bq_{\alpha-1})-\frac{it}{N}
\sum_1^N \uu H(\bp_{\alpha},\bq_{\alpha-1})),
\end{equation}
where $\bp_N=\bp, \bq_0=\bq_N=\bq$.

Thus, we obtained the evolution operator as a limit of finite-dimensional integrals. 

It would be quite reasonable to stop there and  study this representation, but it is also possible to say the words ``functional integral.'' To do this, note that the expression that stands in the exponent is the integral sum for some integral.  This integral is well known to physicists - it is the action functional
 $$S[\bp(\tau),\bq(\tau)]
 =\int_0^t (\bp(\tau)\frac{d}{dt} \bq(\tau)-\uu H(\bp(\tau),\bq(\tau))d\tau.$$
  
Everything is already very close to what I want. Let's get even closer. What was written is the \qp-symbol for the evolution operator. 
We can say that the \qp-symbol is the functional integral of the exponent
$$e^{iS[\bp(\tau),\bq(\tau)]},$$
where  the functional $S$ depend on  a pair of functions $\bp(\tau),\bq(\tau)$ satisfying the boundary conditions
$$\bp(0)=\bp(t)=\bp,\bq(0)=\bq(t)=\bq,$$
when $\tau$ changes from $0$ to $t$. 

We already know that the \qp-symbol is a Fourier transform of a matrix element, hence we can go to matrix elements by doing the inverse Fourier transform. The result is the same integral, but with different boundary conditions:
$$\bq(0)=\bq_1, \bq(t)=\bq_2$$.

To arrive at the formula I mentioned above, consider the special case where the symbol $\uu H(\bp,\bq)$ is the sum of the quadratic function of $\bp$ (kinetic energy) and some function $V(\bq)$ (potential energy). Then it is easy to integrate over $\bp$ - it is a Gaussian integral. As a result, the matrix element of the evolution operator can be represented as a functional integral with an integrand of the  form 
$$ e^{iS[\bq(\tau)]}=e^{i(\int_0^t d\tau (T(\dot \bq(\tau))-V(\bq(\tau))}).$$

 Thus, I derived the functional integral with which I began and even in a more general form.
 
Notice, that in the expression (\ref{eqIN}) for $I_N$ there should be a constant factor tending to zero at $N \to \infty $ (this factor comes from the constant we threw out when we wrote the expression for the Gaussian integral). This zero factor is neglected here. This is done all the time in functional integrals because a good object is actually a quotient of two functional integrals.
\\

I want to construct a large number of examples. These examples generalize what Berezin called covariant symbols. 

I take two Banach spaces $\cal L$ and $\cal L'$. Let us fix a nondegenerate scalar product (pairing) between them.  ( I assume that the scalar product is linear with respect to the first argument and antilinear with respect to the second argument to alleviate the comparison with Hilbert spaces.)

So, I have two Banach spaces that are almost dual to each other in the sense that there is a nondegenerate scalar product between them. Then I take two systems of vectors $e_{\alpha}\in \cal L$, where $\alpha\in \cal M$, and $e'_{\beta}\in\cal L'$, where $\beta \in \cal M'$, in these spaces. They do not have to be linearly independent - they are not bases, but they should be overcomplete. This means that any vector can be expressed in terms of these vectors as a limit of linear combinations. (Poisson vectors are an example of such an overcomplete system.)

Now I would like to insert unity in the scalar product. This means that a scalar product $\langle l,l'\rangle$, should be expressed in terms of the scalar products $\langle l, e'_{\mu}\rangle$ and $\langle e_{\lambda},l'\rangle$  by some integration. This can always be done and even in different ways. I will assume that such a way is fixed:  
\begin{equation}\label{eqllprim}
\langle l,l'\rangle=\int \langle l, e'_{\mu}\rangle\langle e_{\lambda},l'\rangle e^{-r(\lambda,\mu)}d\lambda d\mu,
\end{equation}
where $r(\lambda, \mu)$ is a function on the space  $\cal M\times \cal M'$. ( We assume that    $\cal M\times \cal M'$ is a measure space with the measure $d\lambda d\mu$. This assumption can be weakened; it is sufficient to suppose that some functions can be integrated over    $\cal M\times \cal M'$.)

After that, I want to define the covariant symbol $\uu A(\alpha,\beta)$ of the operator $A$ acting in $\cal L$ (one may consider an operator in  $\cal L'$ with the same success). This symbol is defined by a formula:
$$\uu A(\alpha,\beta) =\frac {\langle Ae_{\alpha},e'_{\beta}\rangle}{\langle e_{\alpha}, e'_{\beta}\rangle}.$$

My main condition is satisfied - the symbol of the unit operator is equal to $1.$ It is easy to calculate the symbol $\uu C$ of the product of operators $C=AB$ using the relation:
$$\langle ABe_{\alpha},e'_{\beta}\rangle=\langle Be_{\alpha},A^*e'_{\beta}\rangle,$$
and the formula (\ref{eqllprim}) for $\langle l,l'\rangle $, where 
$l= Be_{\alpha}$, $l'=A^*e'_{\beta}$. Introducing the notation $\langle e_{\alpha},e'_{\beta}\rangle=e^{c(\alpha,\beta)}$, we obtain the following expression: 
$$\uu C(\alpha,\beta)=\int d\lambda d\mu \uu B(\alpha,\mu)\uu A(\lambda,\beta) 
\exp (-r(\lambda,\mu)-c(\alpha,\beta)+c(\alpha,\mu)+c(\lambda,\beta));$$
it agrees with the formula (\ref{eq28}) .

Note that the above construction is incredibly general. The vectors $e_{\alpha} $ and $e'_{\beta}$ could be chosen almost arbitrarily, the only requirement is that they constitute overcomplete systems.
 However, it is important to have simple expressions for the scalar product $\langle e_{\alpha},e'_{\beta}\rangle$ and for the function $r(\lambda,\mu)$.

If $\cal L=\cal L'$ is the Fock space, one can take as $e_{\alpha}$ the Poisson vectors $ e_{\alpha}=e^{\alpha \hat a^*} |0\ket$ which we already considered. In this case 
$c(\alpha,\beta)=r(\alpha,\beta)=\langle \alpha,\beta\rangle.$
This is easy to calculate because all integrals are Gaussian.

\subsection {L-functionals and functional integrals}

Let us now turn to the L-functionals. We defined them in the preceding lecture and identified them with positive 
functionals on Weyl algebra represented in exponential form (\ref {WW}).
In this definition, nice formulas can  be obtained by taking the Weyl algebra with  overcomplete system $e'_{\alpha}=V_{\alpha}$  as ${\cal L}'$ and the space
 of  linear functionals on the Weyl algebra with overcomplete system  $e_f$ of linear functionals obeying  $e_f(V_g)=\exp{\frac{i}{2}(f, g)}$ as $\cal L.$
 
In the present section, we will write  canonical commutative relations in  the form: 
\begin{equation}\label {CCC}
[a(k), a^+(k')]=\hbar\delta (k,k'),
[a(k), a(k')]= [a^+(k), a^+(k')]=0,
\end {equation}
where $k,k' \in M$. We assume that $M$ is a measure space. Recall that in these relations we are dealing with generalized functions; in other words, we should work with formal expressions  $fa=\int dk f(k)a(k)$, $ga^+=
\int dk g(k)a^+(k)$. In the discrete case, the integral is understood as a  sum and the  $\delta$- function as the Kronecker symbol.

 In this section we  define 
   the L-functional corresponding to a density matrix $K$ in the representation of CCR by the formula:
\begin{equation}
\label {LLL}
L_K(\alpha^*,\alpha)=\mathrm{Tr} e^{-\alpha a^+}e^{\alpha^*a}K,
\end{equation}
An easy formal calculation shows that
 \begin{equation} \label {EXP}
W_{\alpha}W_{\beta}= e^{-(\alpha^*,\beta)} W_{\alpha+\beta} ,
\end{equation} 
 where
 \begin {equation}\label {WWW}
 W_{\alpha}=e^{-\alpha a^+}e^{\alpha^*a}
 \end{equation}

In the preceding section, we wrote an exponent of a linear expression in the definition of  L-functional corresponding to a density matrix in a representation of CCR. This exponent can be considered as a unitary operator in representation space, hence the L-functional is well-defined. Here we are writing a product of exponents. This difference is not significant ( a numerical factor), but with the new definition, I can say that L-functional is just a generating functional for correlation functions. 

When $\alpha$ is a square-integrable function, the expression (\ref {LLL}) is well-defined because the numerical factor is finite. However, we do not assume that $\alpha$ is  square integrable (this is important for applications to string theory). We modify slightly the definition of $\cal W$  assuming that it is generated by
  $a(k), a^*(k),$ and $W_{\alpha}$ obeying  (\ref {CCC}), (\ref{EXP}). For appropriate topology in $\cal W$ the multiplication is defined on a dense subset and we can consider $\cal W$ as a version of Weyl algebra.

Let us denote by $\cal L$ the space of continuous linear functionals on $\cal W.$  Such a functional is completely determined by its values on generators $W_{\alpha}$: we denote these values by 
${\bf L}(\alpha^*,\alpha).$  In other words, a linear functional $L$  can be represented by a non-linear functional  ${\bf L}(\alpha^*,\alpha).$

The action of the Weyl algebra $\cal W$ on the space $\cal L$ is realized by the operators $b$ and $b^+$, whose action on functionals $L_K$ corresponds to the multiplication of the density matrix by the operators $a^+$ and $a$ from the right:
$$b(k)L_K = L_{Ka^+(k)}, \;\;\; b^+(k)L_K = L_{Ka(k)}.$$
It is easy to check that these operators satisfy the canonical commutative relations and can be represented in the following form
$$b^+(k)=-\hbar c_2^+(k)+c_1(k), \;\;\; b(k)=-c_2(k), $$
where $ c^+_i (k)$ are multiplication operators by $\alpha_k^*$ for $i=1$ and by $\alpha_k$ for $i=2$, and $c_i(k)$ are derivatives  with respect to $\alpha^*(k)$ and $\alpha (k)$. 

An alternative action of $\cal W$ on $\cal L$ is realized by operators, whose action on functionals $L_K$d corresponds to the multiplication of the density matrix by operators $a$ and $a^+$ from  the left:
$$\tilde b(k)L_K = L_{a(k)K},\;\;\;\tilde b^+(k)L_K = L_{a^+(k)K} $$
These operators also satisfy the canonical commutative relations. They can be represented in the form:
$$\tilde b^+(k)=\hbar c^+_1(k)-c_2(k),\;\;\;\tilde b(k)=c_1(k).$$ 

Thus, there are two commuting actions of Weyl algebra  on $\cal L$.  (In the terminology of physicists we have a doubling of fields.)

Consider now the formal Hamiltonian $\hat H$: 
\begin{equation}
\nonumber
\hat H=\sum _{m,n}\sum _{k_i,l_j} H_{m,n}(k_1, ...k_m|l_1,...l_n)a^+_{k_1}...a^+_{k_m}a_{l_1}...a _{l_n}
\end{equation}
expressed in terms of creation and annihilation operators and presented in the normal form (i.e., all the creation operators are moved to the left). 
Recall that in the algebraic approach, formal Hamiltonians may not make sense as operators, but the corresponding equations of motion may make sense.
The Hamiltonian $\hat H$ induces two formal operators acting on $\cal L$:
\begin{equation}
\nonumber
\hat H=\sum _{m,n}\sum _{k_i,l_j} H_{m,n}(k_1, ...k_m|l_1,...l_n)b^+_{k_1}...b^+_{k_m}b_{l_1}...b _{l_n}
\end{equation}
\begin{equation}
\nonumber
\tilde H=\sum _{m,n}\sum _{k_i,l_j} H_{m,n}(k_1, ...k_m|l_1,...,l_n)\tilde b^+_{k_1}...\tilde b^+_{k_m}\tilde b_{l_1}...\tilde b _{l_n}
\end{equation}
One of them is denoted by the same symbol, the other is denoted by the tilde symbol. And now we can write the equation of motion for the L-functional  $L(\alpha^*,\alpha)$:
\begin{equation}
\nonumber
i\hbar\frac{dL}{dt}= {H} L=\tilde HL - \hat HL,
\end{equation}
where we introduced  the notation $ H=\tilde H - \hat H.$

If we consider a translation- invariant Hamiltonian, then in the momentum representation the coefficients of $H_{m,n}$ contain $\delta$-functions $\delta (k_1+...+k_m-l_1-...-l_n)$( they express the momentum conservation law).

The equations for the L-functional can make sense even when the equation of motion in Fock space is ill-defined. The problem of non-equivalence of different representations of canonical commutation relations disappears for  L-functionals. 

The standard perturbation theory suffers from divergences when we live in an infinite volume. (Sometimes they are characterized as trivial volume divergences.) If there are 
 no UV divergences the formalism of L-functionals leads to well-defined perturbation theory even in infinite volume.

Returning to the expressions for $ H$, we find ourselves in a  familiar setting. 
Indeed, we have representations of two Weyl algebras (we may consider them as a representation of one larger Weyl algebra). In quantum field theory textbooks usually, everything is done in the framework of perturbation theory. We write the evolution operator in the interaction representation as a  $T$-exponent of the interaction Hamiltonian;  applying the Wick lemma we derive the diagram techniques. In the case of L-functionals, there are no essential changes. We have the same commutation relations, we can apply exactly the same techniques.  

The only thing that has changed is that the number of fields has doubled and that we are not working in Hilbert's space. However, the fact the notion of  Hilbert space was not used anywhere in the derivation of diagram techniques and therefore in the formalism of L-functionals all standard techniques from the course of quantum field theory work. Thus, the formalism of L-functionals from the computational point of view is not worse than the usual one. In fact, it is better. As I explained, it solves problems with trivial volume divergences. 
It is also significantly better if we consider the adiabatic approximation. This is true not only for L-functionals, but also in general for geometric approach to quantum theory.

Let us consider a family of ``Hamiltonians''  $H(g)$ in this framework. For example, we can consider a family 
$H(g)=H_0+gV$ that is used in perturbation theory.
I argue that if we consider the expression $\omega (g(t))$, which is the stationary state for the ``Hamiltonian'' $H(g(t))$ for fixed $t$ and $g(t)$ varies slowly (adiabatically), then $\omega (g(t))$ is a solution of the equation of motion for the non-stationary ``Hamiltonian'' $H(g(t)$.  This is a trivial fact:
in adiabatic approximation, we can neglect the derivative $\dot g(t)$ in the equations of motion.

Using this observation I will express the stationary states of the "Hamiltonian" $H(g)=H_0+gV$ in terms of stationary states of the  "Hamiltonian" $H_0.$

Let us consider  the Hamiltonian $H_0+ge^{-\alpha |t|}V$  and corresponding evolution operator
$\sigma_{\alpha}(t,-\infty).$ Applying adiabatic approximation in the limit $\alpha\to 0$ and assuming that  $\omega(0)$ is a stationary state of $H_0$ we can check that

\begin{equation}\label{eqomegaG}
\omega(g)=\lim_{\alpha\to 0}\sigma_{\alpha}(0,-\infty)\omega(0)
\end{equation}
is a stationary state of the "Hamiltonian"   $H(g)=H_0+gV.$

A similar formula  (with a phase factor) is true in ordinary quantum mechanics. In ordinary quantum mechanics, this formula usually applies to get the ground state for the coupling constant $g$  from the ground state of $H_0$. In our approach, it is possible to apply this formula in much more general situation. For example, if the Hamiltonian is translation-invariant, one can apply this formula to any translation- invariant stationary state of the free Hamiltonian $H_0$. 

 This approach is very natural in both equilibrium and nonequilibrium statistical physics. It is possible to take as $\omega(0)$, say, an equilibrium state for some temperature, then applying this process,we obtain an equilibrium state, though at a different temperature, but with the same entropy ( because the adiabatic process does not change the entropy). 
 
The Keldysh formalism of non-equilibrium statistical physics is closely related to the formalism of L-functionals. One can say that the formalism of L-functionals provides justification of Keldysh formalism.

Along with the  operator $\sigma_{\alpha}(0,-\infty)$ one can consider the evolution operator $\sigma_{\alpha}(+\infty,-\infty)$ (the adiabatic S-matrix). If the adiabatic parameter $\alpha$ tends to zero, the adiabatic S-matrix   in the formalism of L-functionals multiplied by some factors (which I do not want to describe here) tends to an operator that I am calling an inclusive scattering matrix. (The inclusive scattering matrix  can be used to calculate the inclusive cross section.)  We will discuss the inclusive scattering matrix and inclusive cross-sections later. I just want to say that the usual scattering matrix in quantum mechanics can also be obtained from the adiabatic evolution operator. 
 Taking limit $\alpha\to 0$  in adiabatic S-matrix for bare particles we obtain the scattering matrix of physical (dressed) particles. The masses and the wave functions are renormalized in this process, but the coupling constant is not renormalized.

Now I want to explain some things that I will discuss later from a different perspective.

It is well known (I will prove this later) that the scattering matrix is expressed in terms of Green's functions. This is what is called the Lehmann-Symanzik-Zimmermann (LSZ) formula \cite {LSZ}. But to calculate an inclusive scattering matrix, I have to consider what can be called  generalized Green's functions. 

The standard Green's function is defined as a chronological product:
$$M=T(B^*_1({\bx}_1,t_1)\dots B_n^*({\bx}_m,t_m))= T(B^*)$$
averaged over some state (in other words Green's function is an expectation value of chronological product in this state) . In a chronological product, the times are decreasing. One can consider the anti-chronological product  (times increase):
$$N= T^{opp}(B_1({\bx}'_1,t'_1)\dots B_n({\bx}'_n,t'_n)) = T^{opp}(B).$$ 

To calculate  generalized Green's functions  we take the chronological product of some operators, multiply it by the anti-chronological product of other operators, and  take the average (expectation value)  in some state $\omega$:
$$G_{mn}=\omega (MN).$$
This is the generalized Green's function in a given state. These Green's functions  appear in the Keldysh formalism and, as will now be explained, they also appear in the  formalism of L-functionals ( and in general in the algebraic approach). 

Recall that for any $*$-algebra $\cal A$  any element $B\in \cal A$ specifies two operators in the space $\cal L$ of continuous linear functionals on $\cal A.$ One of them transforms a linear functional $\omega(A)$ into the functional $\omega(AB)$, the second one transforms $\omega (A)$ into the functional $\omega (B^*A).$ The first of these operators is denoted by the symbol $B$, the second operator is denoted $\tilde B.$.

If $B=B_1B_2$  then
$(B\omega)(A)=\omega (AB_1B_2)$, hence 
$$ (B_1B_2)\omega= B_2(B_1\omega)$$
(the order is changed because $B$ acts from the right.)
 
It follows that
\begin {equation}\label{MN}
(T(\tilde B B)\omega)(x)=\omega (T(B^*)xT^{opp}(B))=\omega (MxN).\end{equation}

Taking $x=1$ we obtain that the generalized Green functions can be expressed in terms of an analog of usual Green functions in the formalism of L-functionals. This means that we
can apply the technique of calculating ordinary Green's functions to compute generalized Green's functions. This gives  Feynman diagrams for generalized Green's functions.
 
 One can  solve the equation of motion for L-functionals  in terms of functional integrals by applying  the methods of  Section  5.1 and assuming that $\cal L'$ \ $=\cal W.$ We define covariant symbols of operators acting in $\cal L$
using systems of vectors $e_f\in \cal L$  and vectors $e'_{f'}\in \cal L'$ that are defined in the following way. We assume that  $e_f\in \cal L$ corresponds to a non-linear functional $e_f(W_{\alpha})=\exp i((f,\alpha^*)+(\alpha,f^*))$ and that $e'_{f'}=W_{f'}$. It follows that
$\bra e_f, e'_{f'}\ket=\exp i((f,f^{*'})+(f',f^*)).$ 

It is easy to calculate the covariant symbol of the operator $H:$

$$
\begin{gathered}
 \uu H(f,f') =
i\sum_{m,n}\int\prod_{
\genfrac {}{}{0pt}{2}{1\leq i\leq m}{1\leq j\leq n}} (dk_i dl_j) 
 H_{m,n}(k_1,...k_m|l_1,...,l_n)\times
\\
 \left(\; f^*(k_1)...f^*(k_m)f'(l_1)...f'(l_n)-f'^*(k_1)...f'^*(k_m)f(l_1)...f(l_n)\;\right).
\end{gathered}
$$

This allows us to get a representation of the symbol of the evolution operator in terms of functional integrals.

\newpage
\section{Lecture 6}
\subsection{Solitons as classical analogs of quantum particles}

I am going to define quantum particles and quasiparticles. The basic statement is that the notion of a particle is secondary. I define a particle as an elementary excitation of the ground state. One can also consider an elementary excitation of any stationary translation-invariant state, then the elementary excitation is a quasiparticle.

First of all, I want to talk about the classical analogs of these notions. Namely, I will discuss the notions of soliton and of generalized soliton. Consider a translation-invariant Hamiltonian in an infinite-dimensional phase space which consists of vector-valued functions $f(\bx)$, where $\bx\in \mathbb {R}^d$ are spatial coordinates. I assume that spatial translations act as shifts of these coordinates, and the time translations are specified by a Hamiltonian which is invariant with respect to spatial translations. Suppose that the corresponding equation of motion has the form
\begin{equation}\label{NEQ}
\frac {\partial f}{\partial t}= Af+B(f),
\end{equation}
where $A$ is a linear operator and $B$ stands for  non-linear part. We assume that the nonlinear part is at least quadratic; then for small $f$ the linear part dominates. In particular, we can say that $f\equiv 0$ is a solution, and in its neighborhood, we can neglect the nonlinear part. 

Now we define a soliton as a solution that has the form $s({\bx}-{\bf v}t)$.  We can represent the solution $f\equiv 0$ as a horizontal straight line, then the soliton (solitary wave) is a bump moving uniformly without changing the shape. There is also the notion of a generalized soliton. This is a bump that moves, with a constant average speed, but at the same time it can pulsate, it can change its shape. I will not talk about this notion in detail.

If the theory is Lorentz-invariant, one can apply a Lorentz transformation to the soliton and again obtain a soliton. Solitons walk in families - solitons with different velocities.  The same is true for Galilean invariance and Galilean transformations. In these cases, we have a family of functions  $s_{\bf p}({\bf x}-{\bf a})$ that is invariant with respect to temporal and spatial translations (here $\bp$ denotes the momentum of soliton). This
 family can be considered as a symplectic manifold. A family of generalized solitons also can be considered a symplectic manifold that is invariant with respect to temporal and spatial translations; the coordinates on this manifold are the data characterizing a (generalized) soliton.

 I require that the soliton has finite energy.  For a translation-invariant solution, the notion of energy is meaningless - for it we can talk about energy density, but I will consider its energy equal to zero and count the energy of the soliton from the energy of the translation-invariant state. The fact that the energy is finite means, roughly speaking, that the soliton is more or less concentrated in some finite region.
  
In a paper with Fateev and Tyupkin \cite{FTS} written almost fifty years ago we conjectured that for very many systems and for almost all initial conditions with finite energy the behavior of solution for times tending to plus or minus infinity can be described in the following way. If there are no solitons in the theory then asymptotically the solution obeys a linear equation.  In general, we obtain a few solitons and something else which satisfies a linear equation, at least approximately. This is well known for integrable systems in the case $d=1$ (see, for example, \cite {FADD})\cite{FADK}); we have conjectured that this is true without the assumption of integrability in any dimension. I don't think any mathematician has read our paper, but this hypothesis has also been expressed in other papers. Soffer \cite{SOFFER},\cite {LSOF} calls it ``grand conjecture'', Tao \cite {TAO}calls it ```soliton resolution conjecture''. Nevertheless, the hypothesis remains a hypothesis. So far it remains outside the limits of existing mathematics, except for the case when solitons do not exist (see, for example,\cite{STRAUSS}). 

This conjecture can be justified as follows.  Let us take as an initial condition a field concentrated in some domain. In this case, we should expect what in old quantum mechanics textbooks was called the spreading of wave packets. That is, if the initial data were concentrated somewhere, then later the solution spreads to a larger domain. The energy is conserved, so this spreading causes the amplitude of the wave to decrease. If it really does decrease all the time, then, as I said, in the case of small amplitudes the nonlinear part can be neglected and the solution of the nonlinear equation approaches the solution of the linear one.

Of course, this does not happen if there is a soliton in the theory. The height of the bump remains the same, but we can expect that in the end, we have some solitons or generalized solitons plus a tail that approximately satisfies the linear equation. Of course, this reasoning is not proof, but it is convincing.

There is no doubt that we should impose some conditions to prove the above conjecture. In particular, one should require that the translation-invariant state $f\equiv  0$ is stable and the solitons are stable, otherwise, the solution can blow up.   However, it is natural to think that the conjecture is true in many cases.

 In this picture, one must think that there is a notion of soliton scattering. There are solvable models of dimension 1+1 (one space dimension and one time dimension); for such models, this statement is a theorem.  Two solitons collide, we see something that does not resemble any solitons ('' a mess"), and then the same solitons arise again.  This is specific for integrable models. The standard situation in non-integrable models is a bit different: after the collision, we get some solitons plus a ''tail".  The tail asymptotically behaves as a  solution of a linear equation. The solitons that we obtain after collision in general do not coincide with the original solitons.

Now I will try to give some formal definitions. Let us denote the space of possible initial data by the letter $\cal R$. We conjectured that for a dense set of initial data we can define a mapping $D^+(t):{\cal R}\to{ \cal R}_{as}$ of initial data at the moment $t$  to asymptotic data at $t\to +\infty$. (The asymptotic data characterize the solitons and the asymptotic behavior of the tail.)  I could also consider the asymptotic data at $t\to -\infty$ to get a mapping $D^-(t):{\cal R}\to{\cal R}_{as}$.

Now I assume that there is also an inverse mapping, i.e. one can find initial conditions from the asymptotics or at least prove that the given asymptotics is obtained from some initial conditions. That is, I want to consider inverse operators
$ S(t,+\infty) = (D^+(t))^{-1} $ and
$ S(t,-\infty) = (D^-(t))^{-1} $.

 To find these operators we should determine from the asymptotic data the solutions of the equation and thus the initial data. This does not seem to be difficult, but nobody has done it. I  think it's an interesting and not very difficult task: to construct a solution from asymptotic data. In the quantum case, the solution to this problem is well known - it is what is called the Haag-Ruelle scattering theory; I will explain a generalization of this theory ( see \cite {AH} for the version of Haag-Ruelle theory that is close to our approach). 

Now I can define what should be called a non-linear scattering matrix:
$$S=S(0,+\infty)^{-1} S(0,-\infty):{\cal R}_{as}\to {\cal R}_{as}.$$
Roughly speaking, we set the initial conditions at minus infinity and watch the asymptotics at plus infinity. One can hope that one can obtain the non-linear scattering matrix from the quantum scattering matrix in the limit  $\hbar \to 0.$ ( More precisely, one should  expect that the inclusive scattering matrix has a limit as $\hbar\to 0$; this limit should be related to the non-linear scattering matrix.)

Concluding the discussion, I want to say that the classical soliton should be considered a model of a quantum particle. In quantum field theory, the notion of a particle is an asymptotic notion: if two particles collide,  we get ''a mess", which then disintegrates into particles. Notice,
that the analogy with solitons makes it obvious that the existence of identical particles is not surprising.

The following reasoning further emphasizes the analogy of solitons with quantum particles.  Consider a phase space and a Hamiltonian, that is, there is a symplectic manifold  $\cal M$ (that can be identified with the space of initial data $\cal R$) and an evolution operator. Assume that spatial translations act on $\cal M$ and temporal translations are described by a translation-invariant Hamiltonian. Formally this means that on the symplectic manifold $\cal M$ we have an action of the commutative group $\cal T$ of spatial and temporal translations. Now let us take a stationary translation-invariant point $m\in \cal M$ of this symplectic manifold. 

In the previous picture, such a point was the solution $f\equiv 0$. This solution is translation-invariant and stationary. 

Let us define an excitation of a translation-invariant stationary state as a state with finite energy (recall that we assume that the energy of a translation-invariant state is equal to zero).

  I will define an elementary symplectic manifold  $\cal E$ as such a  symplectic manifold where in Darboux coordinates $\bp, \bx$ the spatial translations act simply as shifts ${\bx}\to \bx+\ba$, while $\bp$ remains unchanged. We consider a Hamiltonian that is invariant with respect to these translations. This means that it depends only on $\bp$. Let us denote it as $\epsilon (\bp)$. Then the time translations are transformations ${\bx}\to {\bx}+{\bf v}({\bp})t$, where  $\bf v$ is calculated by the formula ${\bf v}({\bp})=\nabla \epsilon (\bp)$.

Suppose now that $\cal M$ is realized as a space of vector-valued functions $f(\bx)$ where ${\bx}\in \mathbb {R}^d$ and the spatial translations act as shifts ${\bx}\to \bx+\ba.$ If we take a symplectic embedding of the elementary symplectic space $\cal E$ into the set of excitations in $\cal M$ and require that the embedding commutes with the space-time translations, then we get a family of solitons. 
This is very simple to explain. The symplectic embedding of elementary symplectic space  $\cal E$ maps  the point $(\bp, 0)$  into some function $s_{\bp}(\bx) $ depending on $\bp$. Since the embedding $\cal E\to \cal M$ commutes with  spatial translations, the point $(\bp,\ba)$ maps into a shifted function $s_{\bp}(\bx+\ba)$. The condition that the mapping $\cal E\to \cal M$ commutes with time shifts means that the function $s_{\bp}({\bx}-{\bf v}({\bp})t)$ satisfies the equation of motion.

 \subsection{Particles and quasiparticles}
 
Now let us introduce the notion of a particle and a more general notion of a quasiparticle. The difference is only that a particle is an excitation of the ground state, while a quasiparticle is an excitation of any translation-invariant stationary state. In order to define the notion of a particle, I need the notions of spatial and temporal translations.

In ordinary quantum mechanics, if we consider evolution, 
 we need to have the notion of time translations $T_{\tau}$. To define a notion of particle  I need also spatial translations $T_{\ba}$ that act on states and commute with temporal translations.   In the geometric approach, the space of states is the basic object, but here it is convenient to consider non-normalized states. Recall that in the algebraic approach, the states are positive functionals normalized by the condition $\omega(1)=1$. Discarding the normalization condition we obtain a cone, which I denote by $\cal C$. The state is now defined only up to a numerical factor. I will talk about this cone of non-normalized states all the time.  Space-time translations must act on this cone.

Let us now denote the commutative group of space-time translations as $\cal T$. In the algebraic approach, this group should act by automorphisms of the algebra $\cal A$.    The group of automorphisms of an algebra (and hence the group $\cal T$) acts on $\cal C$ (recall that we always assume that automorphisms agree with involution).

I use the standard notation, $A(\tau, {\bx}) =T_{\tau}T_{\bx}A$ 
for an element $A\in \cal A$ shifted in time and space.  The translation-invariant stationary state $\omega$ in the algebraic approach satisfies the condition $\omega (A(\tau, {\bx}))=\omega(A).$  Standard examples of such a state are ground states and equilibrium states. 

In particular, we can consider the Weyl algebra $\cal A$ with generators $\hat a^*({\bx}), \hat a({\bx})$ obeying CCR and assume that the spatial translations simply shift the argument, while the temporal translations are defined by a formal Hamiltonian, which is expressed in terms of $\hat a^*({\bx}), \hat a({\bx})$ with some coefficient functions depending only on the differences $\bx_i-\bx_j$. This ensures translational invariance. I will also require that the coefficient functions decrease rapidly. Then the equation of motion makes sense.  

I can do a Fourier transform and go to the momentum representation. Then the argument is denoted by $\bk$ and a spatial translation is realized as multiplication by $\exp( i\bk\ba)$. The time translations will be again determined by the Hamiltonian. The condition that the functions in the coordinate representation depend on the difference leads to $\delta$-functions corresponding to the momentum conservation, and the requirement that the coefficient functions decrease rapidly means that the functions in the momentum representation will be smooth (after the $\delta$-function is omitted).

In the geometric approach, when the group of space-time translations acts on the cone of states, I  define a translation-invariant stationary state as a state that does not change for spatial and temporal shifts. This will be the basic object for me.

Now I want to define the notion of excitation of a translation-invariant stationary state as an analog of the previously introduced notion of a state with finite energy.  When a soliton goes to infinity, we stop seeing it.  Formalizing this observation we say that a state  $\sigma$ is an excitation of translation-invariant state $\omega\in \cal C$ if
$(T_{\bf a}\sigma)(A)$  tends to ${\rm const} \cdot\omega(A)$  in the limit ${\bf a}\to \infty.$ 

The constant appears here because the state is defined only up to a numerical factor. 

The notion of excitation is a general notion that can be applied in both geometric and algebraic approaches.  

In the algebraic approach, a pre-Hilbert space $\cal H$ (I want to live in a pre-Hilbert space) can be constructed from a translation-invariant stationary state using the GNS construction.  In this space there is a cyclic vector $\theta$ corresponding to the state $\omega$. Recall, this means that $\omega$ is a positive functional  
$$\omega(A)=\langle \hat A\theta,\theta\rangle,$$
where $\hat A$ is the operator that corresponds to $A$ in representation space of $*$-algebra $\cal A$.    

Spatial and temporal translations $T_{\ba}$ and $T_{\tau}$ descend to the pre-Hilbert space $\cal H$ as unitary operators. 
Translations act in the algebra $\cal A$ as automorphisms of the algebra, and we constructed the pre-Hilbert space by factorizing the algebra in some way. This allows us to define these operators in $\cal H$ as unitary operators. 
Next, we define the energy and momentum operators as infinitesimal translation operators in time and space:
$$T_{\ba}=e^{i \hat\bP\ba}, \;\;\; T_{\tau}=e^{-i\hat H \tau}.$$  

In the algebraic approach, the elements of the pre-Hilbert space $\cal H$ can be identified with the excitations of the state $\omega$. The physical meaning of the GNS construction is that starting with some translation-invariant state we can construct the space $\cal H$ in which the excitations live. This, in fact, is the explanation of why this construction is so important in physics. 

What I have claimed is not always true. I  should demand the cluster property to justify my claim.

Let us imagine a ferromagnetic.  If spin has some direction at the origin of the coordinate system, then the same direction of spin will be everywhere, at least statistically. This is the case when there is no correlation decay. In a more standard situation at larger distances, the spin no longer remembers what the spin was at the origin. This is what is called cluster property. 

Mathematically this can be formulated as follows.  Let us take $\omega( A(\tau,{\bx})B)$, where $A, B$ are two algebra elements.  Then the cluster property implies that$$\lim_{\bx\to \infty} \omega( A(\tau,{\bx})B)=\omega (A)\omega(B)$$

This is the simplest form of cluster property. Later I will formulate it in a more general way. At this point, I need only the following generalization. Let us take three elements $B', A$ and $B$. If one of these elements is shifted to infinity then
\begin{equation}\label {BAB}\lim_{{\bx}\to \infty} \omega( B'A(\tau,{\bx})B)=\omega (A)\omega(B'B)
\end{equation}


Any element of $\cal H$ can be represented as $B\theta$ where $B\in \cal A$. Then for the state $\sigma (A)$  corresponding to the vector $B\theta$ we have
\begin{equation}\label{eqBAB}
\sigma (A)= \langle\hat A \hat B\theta,\hat B\theta \rangle=\omega (B^*AB).
\end{equation} 

 It follows from (\ref{BAB}) that
$$(T_{\bx}\sigma)(A)=\sigma (A(0,{{\bx}})=\omega (B^*A(0,{\bx})B)\to \omega (A)\omega (B^*B),$$
as $\bx\to\infty.$
This means that all elements of the pre-Hilbert space $\cal H$ correspond to excitations. In the algebraic approach I will only consider such excitations.  In fact, I could start here - I could define the notion of excitation this way: take $\omega$, apply the GNS construction, and take the elements of the pre-Hilbert space $\cal H$. This construction gives excitations in the algebraic approach.

 I will now define the notion of elementary excitation of a translation-invariant state. An elementary excitation of the ground state is what is called a particle in quantum field theory.  Elementary excitations of a translation-invariant state are called quasiparticles. Since I will consider both cases, I will speak about elementary excitations, but I may also use the terms ``particle'' or ``quasiparticle''. 

 In the algebraic approach, we live in a Hilbert space, which is obtained with the GNS construction.  I want to understand what should be called a particle in this situation.  First of all  I notice that it is necessary to be able to talk about a particle having momentum $\bp$. A particle could have other quantum numbers - they will just appear as discrete indices, which do not bother me in any way. 
I will denote the vector describing a particle with momentum $\bp$ by $\Phi(\bp)$.    This means
that

\begin{equation}\label{eqPhiP}
\hat{\bf P}\Phi(\bp)=\bp\Phi ({\bp}) 
\end{equation} 

The energy of this state is some function $\epsilon(\bp)$, which is called the dispersion law:

\begin{equation}\label{eqPhiE}
\hat H\Phi(\bp)=\epsilon(\bp)\Phi(\bp). \end{equation} 
Note that (\ref{eqPhiP}), (\ref{eqPhiE}) can be rewritten as
\begin {equation}\label{eqPhiPP}
 T_{\ba}\Phi(\bp)=e^{i\bp\ba}\Phi(\bp),
 \end{equation}
 \begin{equation}\label{eqPhiEE}
 T_{\tau}\Phi(\bp)=e^{-i\epsilon(\bp) \tau} \Phi (\bp)
\end{equation}
It is important to note that $\Phi(\bp)$ is not an element of Hilbert space (it has infinite norm), but a generalized vector function.  To obtain an element of Hilbert space we must consider an integral of $\Phi(\bp)$ with some test function $\phi(\bp):$

\begin{equation}\label{eqPhiphi}
 \Phi (\phi)=\int d\bp \phi(\bp) \Phi(\bp),
\end{equation} 

 This will be a well-defined vector.  It is convenient (but not necessary) to impose normalization condition

\begin{equation}\label{eqNormd}
\langle \Phi(\bp), \Phi (\bp')\rangle=\delta (\bp-\bp')
\end{equation} 
 (normalization on $\delta$-function).
For vectors $\Phi(\phi)$ the normalization condition implies that

\begin{equation}\label{eqNormphi}
\langle \Phi(\phi),\Phi (\phi')\rangle=\langle \phi,\phi'\rangle.
\end{equation}

Let us define an elementary space $\textgoth {h}$ as a subspace of the space of square-integrable functions $\phi_a ({\bx})$ taking values in the space $\mathbb {C}^r$.  The elements of this space can be considered as test functions; generalized functions are considered as linear functionals on $\textgoth{h}.$   In what follows we assume for definiteness that $\textgoth{h}$ consists of smooth functions decreasing faster than any power (in other words, $\textgoth{h}$ is the Schwartz space $\cal S$).

We will define the action of spatial and temporal translations on this space. 
The action of spatial translations on the test functions in the $x$-representation is a shift of the argument; in the momentum representation, this action is multiplication by the exponent $e^{i\bk\ba}$. We can deduce the formula for time shift from the requirement that the time translations commute with spatial translations. In momentum representation the time translation $T_{\tau}$ is represented as multiplication by the exponent $e^{-iE ({\bk})\tau}$ where $E(\bk)$ is a Hermitian matrix of dimension $r\times r$. We can diagonalize this matrix, then we get multiplication by scalar phase factors. This means that we can always restrict ourselves to the case $r=1$,

The elementary excitation of a translation-invariant stationary state $\omega$ can be defined as an  isometric mapping $\sigma$ of the elementary space $\textgoth {h}$ to the set of excitations.   This map should commute with spatial and temporal translations. 

It is important to note that for the scalar case $r=1$ this definition is equivalent to the definition above. Indeed, elementary excitation has been defined as a function $\Phi(\bp)$ which is an eigenvector for momentum and energy (\ref{eqPhiP}, \ref{eqPhiE}). There are also normalization conditions (\ref{eqNormd}). It follows from these conditions that  $\Phi (\phi)$ is a mapping of the space of test functions to the excitation space.  The fact this mapping is an isometry follows from the normalization condition. 
 The formulas (\ref{eqPhiP}, \ref{eqPhiE}) ensure that translations of the vector $\Phi(\bp)$ correspond to translations of the function $\phi(\bp)$ in both space and time. Thus, in the algebraic approach for $r=1$ we can consider the space of test functions  as elementary space and define 
$\sigma (\phi)$ as $\Phi (\phi).$

In the geometric approach, I have to consider cones as sets of states. If we started with  the theory in the algebraic approach, then the function $\phi$ from the elementary space is mapped to the state
$$(\sigma'(\phi))(A)=\langle A\Phi(\phi), \Phi(\phi)\rangle.$$
Here $\sigma'$ is a quadratic (or rather, hermitian) mapping of the elementary space to the cone of states that commutes with all translations.

This remark suggests that in the geometric approach one should define the elementary excitation as a  mapping of the elementary space to the cone of states that commutes with spatial and temporal translations.
 
If $\Phi(\phi)$ in the algebraic approach belongs to the pre-Hilbert space $\cal H$ and $\theta$ is a cyclic vector in this space, then $\Phi(\phi)$ is obtained by applying some element from algebra $\cal A$  to the cyclic vector: $\Phi(\phi)=B(\phi)\theta$. Then one can easily verify  the formula  
\begin{equation}\label{eqSI}
\sigma'(\phi)= L(\phi)\omega,
\end{equation}
 where $L(\phi)=\tilde {B}(\phi)B(\phi)$.
(Let me remind you that in the algebraic approach, an element $B\in \cal A$  specifies two operators  on the space of functionals: one corresponds to the multiplication of the argument by $B^*$ from the left (it is denoted $\tilde B$) , the other corresponds to multiplication of the argument from the right (it is denoted by the same letter $B$) .)

It is convenient to include the existence of the operator $L(\phi) $ satisfying the relation (\ref{eqSI}) in the definition of elementary excitation in the geometric approach.

I have already said that only translational invariance is important in the definition of scattering, but if, say, we are dealing with a Lorentz-invariant theory, it is natural to assume that in the algebraic approach, the vector $\theta$ is Lorentz-invariant. Then the entire Poincaré group acts in the space $\cal H$. The elementary space should carry a representation of Poincar\'e group and the map of the elementary space into $\cal H$ should agree with representations of this group. In local quantum field theory, Lorentz-invariant particles are defined as irreducible representations of the Poincaré group. This definition agrees with the above definition.
 
Now let us make the following observation. Let us consider a translation-invariant Hamiltonian of nonrelativistic quantum mechanics. In this case the Hamiltonian is invariant under Galilean transformations, and the energy of elementary excitation is given by the usual formula: 
$\epsilon(\bp)=\bp^2/2m+const$.

If we take the operator $\hat a^*(\bp)$ and apply it to the translation-invariant Fock vacuum $|0\ket$, we obtain an elementary excitation of the Fock vacuum:
$$\Phi(\bp)=\hat a^*(\bp) |0\ket.$$
It is a particle, but besides such a particle there are other particles that also satisfy the imposed conditions. They are called bound states. 

What is a bound state? The Hamiltonian acts on states with any number of particles and preserves this number. Take $n$  particles and separate the motion of the center of inertia. The Hamiltonian in this space can have normalizable eigenstates. They are called bound states. 

 Equivalently we can try to solve the equations (\ref {eqPhiP}),  (\ref {eqPhiE}) for $\bp=0.$ Then the solution will contain the $\delta$-function of the sum of momenta: 
$$\int d\bp_1...d\bp_n\Psi(\bp_1,...\bp_n)\delta(\bp_1 +...+\bp_n)\hat a^*(\bp_1)...\hat a^*(\bp_n) |0\ket.$$
If the function $ \Psi(\bp_1,...\bp_n)$ is square-integrable we obtain a bound state.
It is easy to understand that  the generalized function 
$$\Phi(\bp)=\int d\bp_1...d\bp_n \Psi(\bp_1,...\bp_n)\delta(\bp-\bp_1-...-\bp_n)\hat a^*(\bp_1)...\hat a^*(\bp_n) |0\ket,$$
can be regarded as an elementary excitation.  From my perspective, such bound states  (composite particles) are no worse than elementary excitations with $n=1$.  The general theory, which I will present, gives, in particular, a description of the scattering of composite particles. One can prove that non-relativistic quantum mechanics has interpretation in terms of particles in the sense of Section 7.2 (see, for example, \cite {HUNSIG}).

In the present section, we defined a notion of $stable$ elementary excitation. Notice, however, that particles can be unstable and quasiparticles are almost always unstable. This means that our requirements are satisfied only approximately. The theory of inclusive scattering matrix developed in the next lectures can be applied to unstable (quasi)particles if the lifetime of colliding (quasi)particles is much greater than the collision time. Notice that the conventional scattering matrix does not make sense for quasiparticles.

 \subsection {Asymptotic behavior of solutions of linear equations}
 
 Let  us consider solutions of  translation-invariant linear equation
 \begin{equation}\label{NNEQ}
\frac {\partial f}{\partial t}= Af,
\end{equation}
discarding the non-linear part in (\ref{NEQ}).

We assume that for fixed $t$ the function $f(\bx,t)$ is defined on $\mathbb{R}^d$, it takes  values in $\mathbb{C}^r.$

 It follows from translation invariance that after Fourier transform with respect to $\bx$ (i.e. in the momentum representation) the operator $A$ can be considered as an operator of multiplication of $f(\bp)$ considered as a column vector by an $r\times r$ matrix $A(\bp)$. If the operator $A$  is local  (represented as a polynomial of derivatives) then the matrix $A(\bp)$ is a polynomial. We do not assume locality, but we suppose that $A(\bp)$ is a smooth function of $\bp)$ ( then we can say that $A$ is quasi-local).
 
  The solution to (\ref {NNEQ}) in momentum representation has the form
 $e^{tA(\bp)}f(\bp).$ 
 
 Let us assume that the matrix $A(\bp)$  is diagonalizable and has purely imaginary eigenvalues.  Then the solution $f\equiv 0$ is stable. (This means that for an appropriate definition of the norm, the evolution operators $T_{\tau}$ are uniformly bounded:
 $|T_{\tau}||\leq C$ for all $\tau$. In particular, if $f$ is small at some moment it remains small as $\tau\to \pm \infty.$). 
 
 Let's consider  the solution to the  equation  (\ref {NNEQ}) in coordinate representation
 
 \begin{equation}\label {FX}
 f(\bx,t)=\int d\bp e^{i\bp\bx}e^{tA(\bp)}f(\bp).
 \end{equation}
 
 Notice that the same formula describes the behavior of a test function in the coordinate representation if we take
 $E(\bp)=-iA(\bp)$.
 
 We are interested in the behavior of the function (\ref {FX}) 
 as $t\to \pm \infty.$ Diagonalizing the matrix $A(\bp)$ we can reduce this problem to the case when $r=1$
 We assume that $A(\bp)=-i\epsilon (\bp)$ following the notations of the preceding section. To analyze the  behavior of the function
 \begin{equation}\label {FXR}
 (T_{\tau}f)({\bx})=f(\bx,\tau)=\int d\bp e^{i\bp\bx-i\tau\epsilon(\bp)}f(\bp)
 \end{equation}
 we notice that for large $|\tau|$ the phase is large and
 we can use the stationary phase method. It  leads to equations:
\begin{equation}\label{eqxt}
 \frac{\bx}{\tau}=\nabla \epsilon (\bp). 
\end{equation} 
 Clearly, we must consider only the situation when $\bp \in{ \rm supp }f$. (Recall that ${ \rm supp }f$-the support of the function $f$ in momentum space -is defined as the closure of the set of points where  $f(\bp)\neq 0$. We assume that the set ${ \rm supp }f$ is compact.)
 
Now 
I define the set $U$ as a neighborhood of the set of points $\bx$ where the condition (\ref{eqxt}) with $\bp\in  {\rm supp}f$ and $\tau=1$ is satisfied.
 Outside the set $\tau U$ the equation (\ref{eqxt}) has no solution, so the function $(T_{\tau}f)(\bx) $ is very small at $\bx\notin \tau U$.
  
   I say that $\tau U$ is the essential support of the function $f$ in coordinate representation for large $|\tau|$. One can prove that for $\bx\notin \tau U$ we have
  \begin{equation}\label{ESS} |(T_{\tau}f)({\bx})|< C_n (1+|{\bx}|^{2}+\tau^2)^{-n}\end{equation}
     for any integer $n.$ The proof can be based on a generalization of the Riemann -Lebesgue lemma. Recall that it follows from this lemma that the function $\int dkg(k)e^{ih(k)t}$ where $g(k)$ is a smooth function having compact support and $h(k)$ is a linear function tends to zero faster than any power of $t.$ ( To prove this statement we integrate by parts many times.) This statement can be easily generalized to the case when the function $h(k)$ is not linear but is smooth and does not have stationary points. Then locally this function can be made linear by means of a change of variables. Using a partition of unity  ( a representation of unity as a finite sum of functions  $g_a(k)$ that do not vanish only on small sets $U_a$ covering the support of $g(k)$) and linearizing $h(k)$ on the sets $U_a$ we obtain the generalization we need.  The Riemann-Lebesgue lemma and its generalization can be proven also in the multi-dimensional case; this allows us to verify the above estimate.

   When we apply the stationary phase method to the calculation of the integral (\ref{FXR}) we obtain a factor $(\det (\tau \mathtt{Hess}))^{-\frac 1 2}=\tau^{-\frac {d}{2}} (\det \mathtt{Hess})^{-\frac 1 2} $ where $\mathtt{Hess}$ denotes the matrix of second derivatives. This remark allows us to conjecture
   that  
   \begin {equation}\label {HESS}
    |f(\bx,\tau)|\leq \frac {C}{\tau^{\frac d 2}}.
    \end {equation}
    One can prove this conjecture by imposing some conditions on $\epsilon (\bp)$ (see, for example, \cite {SEGAL}). 
    
    The estimate (\ref{HESS}) can be used to analyze the problem of the existence of a solution of the non-linear equation  (\ref {NEQ}). To find a solution that behaves as $T_tg_-$ as $t\to-\infty$ we should 
  find a fixed point of  non-linear operator
  $${\cal B} (f)=T_tg_-+\int_{-\infty}^t T_{t-\tau}(Bf)(\tau)d\tau,$$
   where $f$ is considered as a function of $t$ taking values in appropriate space of functions of $d$
   variables.  One can use the contraction principle to prove the existence of the fixed point of $\cal B$
   under certain conditions \cite {STRAUSS}.

\newpage
\section{Lecture 7}

   \subsection{Multi-particle states}

In Section 6.2, we defined the elementary space $\textgoth {h}$ as the space of test functions $\phi_a ({\bx})$ where spatial translations act by shifting the argument. 
Test functions take values in $\mathbb{C}^r$.  For definiteness, we assume that the test functions belong to the space $\cal S$ of smooth fast decreasing functions.

In the momentum representation, spatial translations act as multiplication by $e^{i\bk\ba}$ and temporal translations as multiplication by $e^{-iE ({\bk})\tau}$.  (This follows from the assumption that time shifts are unitary operators commuting with spatial translations.) Here $E(\bk)$ denotes an hermitian $r\times r$ matrix. By diagonalizing the matrix $E(\bk)$, we can reduce the general case to the case $r=1$.

Let me now remind the notions of excitation and elementary excitation.  

The elementary excitation of a translation-invariant stationary state $\omega$ (quasiparticle) is given by a mapping $\sigma$ from elementary space $\textgoth {h}$ into the set of excitations. This mapping must commute with translations (both spatial and temporal).

 In the algebraic approach, the set of excitations
  is the pre-Hilbert space $\cal H$, which is obtained using the Gelfand- Naimark- Segal (GNS) construction applied to a stationary translation-invariant state $\omega$. The state $\omega$ is represented by a cyclic vector denoted by $\theta$. 

The map $\sigma$ transforms $\phi\in \textgoth{h}$ into a vector $\sigma (\phi)$, which was denoted by $\Phi (\phi)$ in Section 6.2. It is a mapping into the space $\cal H$, which means that there exists an element $B(\phi)$ from  the algebra  $\cal A$ which transforms the cyclic vector $\theta$ into  our vector 
\begin{equation}\label{eqPhiB}
\sigma (\phi)=B(\phi)\theta.
\end{equation} 

I also assume that the mapping $\sigma$ is isometric.

I want to emphasize that the operator $B(\phi)$ exists, but it is not unique, it must be chosen somehow. I will impose some conditions on it, which will allow me to develop the scattering theory.  In particular, I will require that it be linear in $\phi$. As I explained, each vector in the representation space of  the algebra $\cal A$ corresponds to a state (to a positive linear functional on $\cal A$).  The state  corresponding to $\sigma(\phi)$ can be represented  by the formula
$$(\sigma' (\phi))(A)= \langle A\sigma(\phi),\sigma(\phi)\rangle.$$

If a vector  $\sigma(\phi)$ is represented in the form (\ref{eqPhiB}), we have the following formula:
\begin{equation}\label{eqSigL}
\sigma'(\phi)= L(\phi)\omega,
\end{equation} 
where 
\begin{equation}\label{eqLBB}
L(\phi)=\tilde {B}(\phi)B(\phi).
\end{equation} 
So, in the algebraic approach I have some mapping $\sigma':\textgoth {h}\to\cal C$, acting according to the formula (\ref{eqSigL}).

In the geometric approach one must forget about algebra, but there remains a cone of all states $\cal C$.  By definition, the mapping $\sigma'$ of an elementary space $\textgoth {h}$ to the cone $\cal C$ defines an elementary excitation if it commutes with spatial and temporal translations.

 We postulate that, just as in the algebraic case, the mapping $\sigma'(\phi)$ is obtained according to (\ref{eqSigL}) by the action of some operator $L(\phi)$ on a translation-invariant stationary state $\omega$.  The mapping $\sigma(\phi)$ considered in the algebraic situation was linear, but the mapping $\sigma'(\phi)$ is not linear at all. Indeed, from the formula (\ref{eqLBB}) it follows that in the algebraic approach, $L(\phi)$ is a quadratic expression, or more precisely, a hermitian expression because it is linear in one variable and anti-linear in the other. (An expression $f$ is called hermitian if it can be represented in the form $f(x) =F(x,x^*)$, where $F(x,y)$ is linear in the first argument and antilinear in the second one.) It is natural to require that in the geometric approach, $L(\phi)$ satisfies the same conditions.
In what follows I will use the word ``quadratic'' instead of ``hermitian,'' but it should be understood that it is not really quadratic. 

If one prefers working with linear mappings, this can be done using the following general algebraic construction.
 For each linear complex space $E$ in the tensor product $E\otimes \bar E$ of this space by a complex conjugate one can construct a cone $C(E)$ as a minimal cone containing all elements of the form $e\otimes \bar e$. (The bar stands for complex conjugation).   
I call the cone $C(\textgoth {h})$ an elementary cone. It corresponds to the elementary space $\textgoth {h}$, and $\sigma'$ can be viewed as a linear mapping $\sigma':C(\textgoth {h}) \to \cal C$ of the elementary cone to the cone of states. 

To simplify the notations, I consider the case when the elementary space consists of scalar functions ( $r=1$).

Suppose that the support $\rm {supp}(\phi)$ of the function $\phi(\bp)$ in the momentum space 
 is a compact set. 
In that case, it is possible to find a bounded set $U_{\phi}$ for which
 all points of the form $\nabla \epsilon (\bp)$ where $\bp$ belongs to $\rm {supp}(\phi),$ are interior points (the function $\epsilon(\bp)$ is assumed smooth).
  
Then for large $ |\tau |$ and $\frac {{\bx}}{\tau}\notin U_{\phi}$  the function 
$$(T_{\tau}\phi)({\bx})= \int d\bk e^{i\bk\bx-i\epsilon (\bk)\tau}\phi(\bk)$$
obeys
$$ |(T_{\tau}\phi)({\bx})|< C_n (1+|{\bx}|^{2}+\tau^2)^{-n},$$
where $n$ is some integer (see (\ref{ESS}) in Section 6.3).

In other words, for large $|\tau|$  the function $(T_{\tau}\phi)({\bx})$ is small outside the set $\tau U_{\phi}$, which I call the essential support of the function $(T_{\tau}\phi)({\bx}).$ 

Let us now return to the general case where the elementary space $\textgoth {h}$
consists of vector-valued functions. We say that the set $\tau U_{\phi}$ is an essential support of the function
$$(T_{\tau}\phi)({\bx})= \int d\bk e^{i\bk\bx-iE (\bk)\tau}\phi(\bk)$$
if
$$ ||(T_{\tau}\phi)({\bx})||< C_n (1+|{\bx}|^{2}+\tau^2)^{-n},$$
 at large $ |\tau |$ and $ \frac {{\bx}}{\tau}\notin U_{\phi}.$ 

We say that functions $\phi$ and $\phi'$ do not overlap if the distance between the sets $U_{\phi}$ and $U_{\phi'}$ is positive; then the corresponding essential supports do not overlap, moreover, at large $\tau$ they are distant from each other. We say that $\phi_1,...,\phi_n$ is a non-overlapping family of functions if $\phi_i$ does not overlap with $\phi_j$ at $i\neq j$. {\it  We will always assume that there are many non-overlapping families of functions (more precisely
 linear combinations of non-overlapping families of functions
should be dense everywhere in the space of families of functions we are interested in)}. When $r=1$ this is satisfied, for example, when the function $\epsilon(\bp)$ is strictly convex.

What should be called a two-particle state in the algebraic approach? I want to note that when defining a one-particle space, I only needed spatial and temporal shifts, but now I need more. Before, I used the representation as (\ref{eqLBB}), 
to describe a one-particle state with wave function $\phi$. When there are two particles, however, $B$ must be applied twice:
$B(\phi)B(\phi')\theta$.
At least when $\phi$ and $\phi'$ have supports far apart in coordinate space, one can say that this vector describes a state of two distant particles. One must require that $B(\phi)$ and $B(\phi')$ almost commute with each other (then the particles will be bosonic) or almost anticommute (then the particles will be fermionic). This definition is given in terms of states described by vectors, but it is possible to give a definition in terms of states described by positive functionals on the algebra $\cal A$. For this purpose, we note that state, which corresponds to the vector $B(\phi)B(\phi')\theta$, can be written in the form
$L(\phi)L(\phi')\omega$,
where

$L(\phi)=\tilde {B}(\phi)B(\phi)$, $
L(\phi')=\tilde {B}(\phi')B(\phi').$

 In all cases, $L(\phi)$ almost commutes with $L(\phi').$

In the geometric approach, a two-particle state is written as $L(\phi)L(\phi')\omega$, where 
$L(\phi)$ almost commutes with $L(\phi').$

Thus, the distinction between bosons and fermions is smoothed out in the geometric approach. 

In what follows I will talk about bosons all the time, but the transition to fermions is trivial: one only has to replace commutators with anticommutators.

\subsection {Scattering; $in$- and $out$-states}

I would like to quote a wonderful statement by Bertrand Russell: 

{\it
The axiomatic method has many advantages over honest work.
}

 Everything in this section will be very simple, but, unfortunately, this simplicity is achieved at the expense of working exclusively in the axiomatic approach. I want to remind you that axioms of local quantum field theory were proposed by Wightman in the 1950s and until now there is no known example of non-trivial theory, about which it is proved that it satisfies all Wightman axioms in our three-dimensional space. A big step forward took place when many such theories were constructed in the formalism of conformal field theory in one-dimensional space, but to date, Wightman axioms in three-dimensional space have been verified only in the framework of perturbation theory. In my approach, the situation is slightly better, but nevertheless, the verification of necessary axioms remains a big problem. This will be discussed in the next lecture. Fortunately, at least in perturbation theory, everything is fine.

Let us consider the scattering of elementary excitations in algebraic and geometric approaches.  In the algebraic approach, I assume that the mapping  $\sigma$ of the elementary space $\textgoth{h}$  to the space $\cal H$ defining a particle or quasiparticle can be written in the form 
\begin{equation}\label{eqSigB}
\sigma (f)=B(f)\theta.
\end{equation} 
In the geometric approach, I assume the existence of a mapping 
$L:\textgoth {h}\to End (\cal L)$, where 
\begin{equation}\label{eqSigL2}
\sigma' (f)=L(f)\omega.
\end{equation} 
Both $\sigma$ and  $\sigma'$  should commute with translations.

In both approaches, I will define states describing the scattering process.

In the algebraic approach, we define  the  operator $B(f,\tau)$  by the formula:
$$ B(f,\tau)=T_{\tau}B(T_{-\tau}f))T_{-\tau}$$
(We must remember that the time-shift acts on an operator as a conjugation with the operator $T_{\tau}.$)

In the geometric approach, we   define the operator $L(f,\tau)$ by similar formula: 
$$L(f,\tau)=T_{\tau}L(T_{-\tau}f) T_{-\tau}.$$ 
It is easy to check that  $L(f,\tau)\omega$ does not depend on $\tau$.
To verify this we notice that
$$L(f,\tau)\omega=T_{\tau}L(T_{-\tau}f)\omega=T_{\tau}\sigma'(T_{-\tau}f)=\sigma'(f).$$
(We used the invariance  of $\omega$ with respect to time translations, the formula (\ref{eqSigL2}), and the fact that $\sigma'$  commutes with time translations.)

By exactly the same reasoning we can show that $B(f,\tau)\theta$ does not depend on $\tau$. 

We obtain that
$$\dot L(f,\tau)\omega=0,$$ 
$$\dot B(f,\tau)\theta=0,$$
where the dot at the top indicates the derivative with respect to $\tau$.  
This will be my main tool.

In the case of many particles, by analogy with  the definition of a single-particle state, I first apply $B(f_i,\tau)$ many times with different $f_i$ to $\theta$: 
$$\Psi (f_1, \cdots, f_n|\tau)=B(f_1,\tau)...B(f_n,\tau)\theta$$
to get a multi-particle state. Then I take the limit of the resulting expression at $\tau \to -\infty$: 

\begin{equation}\label{eqIntauB}
\Psi (f_1,\cdots,f_n|-\infty)= \lim _{\tau\to-\infty}\Psi (f_1, \cdots, f_n|\tau)
\end{equation} 
This limit (which lies in the Hilbert space $\bar{\cal H}$, the completion of the space $\cal H$) I will call $in$-state. A little later I will explain its physical meaning. 

In the geometric approach,  instead of $B(f,\tau)$ we take $L(f,\tau)$:
\begin{eqnarray}\label{eqInLam}
\Lambda (f_1,...,f_n|\tau)=L(f_1,\tau),...L(f_n,\tau)\omega \\
\Lambda (f_1,\cdots,f_n|-\infty)= \lim _{\tau\to-\infty}\Lambda (f_1, \cdots, f_n|\tau)
\end{eqnarray} 
We obtain an $in$-state lying in $\cal L$.  In the algebraic approach, it corresponds to the vector 
$$\Psi (f_1,\cdots,f_n|-\infty).$$

Applying the operator $T_{\tau}$ to $L(f,\tau')$ leads to a time shift in both argumrnts:
\begin{equation}\label{eqTtauL}
T_{\tau}(L(f,\tau'))=T_{\tau+\tau'}L(T_{-\tau'}f) T_{-\tau-\tau'}= L(T_{\tau}f,\tau+\tau').
\end{equation} 
This is a purely formal calculation. 

        The formula (\ref{eqTtauL}) implies that 
\begin{equation}\label{eqTtauLam}
T_{\tau} \Lambda (f_1,\cdots,f_n|-\infty)=\Lambda (T_{\tau}f_1,\cdots,T_{\tau}f_n|-\infty). 
\end{equation} 
If functions $f_1,...,f_n$ do not overlap, hence essential supports of functions $T_{\tau}f_i$ are far away in the limit $\tau\to-\infty$ , it follows from (\ref {eqTtauLam}) that in this limit $\tau\to -\infty$ the evolution of the $in$-state $\Lambda$ describes the process of scattering.

Usually one considers the scattering of particles with definite momenta. It is inconvenient to work with definite momenta in my approach because in such case the wave functions will be non-normalizable. We can consider the situation when the momentum lies in some narrow range, i.e. the support of the wave function is a small piece of the momentum space.  I say that the state $T_{\tau} \Lambda (f_1,\cdots,f_n|-\infty)$ describes a collision of particles with wave functions $(f_1,\cdots,f_n)$ if these functions do not overlap. In this case I assume that corresponding operators $L(f_i,\tau)$ almost commute for  $\tau \to -\infty$, i.e. their commutator in this limit vanishes: 
\begin {equation} \label{eqLL}
\lim_{\tau \to -\infty} ||[ L(f_i,\tau), L(f_j,\tau)]||=0.
\end {equation}

Why do I assume that? 
  When $f_i$ and $f_j$ do not overlap then in the limit $\tau \to -\infty$, then the essential supports of the functions $T_{\tau}f_i$ and $T_{\tau}f_j$  are far away. In this case, from the point of view of physics, it is natural to think that the corresponding operators almost commute. 

The condition (\ref{eqLL}) can be derived from the requirement that the commutators of the two operators $L$ that depend on the functions $\phi_{\it a}(\bx)$ and $\psi_{\it a}(\bx)$ satisfy the inequality
\begin {equation}
\label{eqLLL}
 ||[L(\phi),L(\psi)]||\leq \int d{\bx}d{\bx}' D^{ab}(\bx-\bx')|\phi_{\it a}(\bx)|\cdot |\psi_{\it b}(\bx')|
\end{equation}
where $D^{ab}(\bx)$ tends to zero faster than any power when $\bx\to \infty.$
Under these conditions, if the sets $U_{f_i}$ for each pair of functions do not overlap, then the commutators  in the formula (\ref{eqLL}) are close to zero. This means that I can permute the operators $L$ in the formula (\ref{eqInLam}) for $in$-state. It follows that $in$-states are symmetric (they do not change when the arguments $f_i$ are rearranged).

Let us now prove that the limit in question exists. To do this, we additionally impose the condition of the smallness of the commutator $[\dot L(f_i,\tau), L(f_j,\tau)]$ at $\tau \to -\infty$, where the functions $f_i, f_j, i\neq j$ do not overlap. This is again an axiom.  More precisely, we impose  the condition 
\begin{equation}\label{LDL}||[\dot L(f_i,\tau), L(f_j,\tau)]||\leq c(\tau),\end {equation} 
where $ c(\tau)$ is summable:
$$ \int |c(\tau)|d\tau<\infty.$$  
We can assume, for example, that $c(\tau)\sim 1/\tau^a$, where $a>1$. 

Now I will give a very simple proof that the $in$-state does exist (the expression $\Lambda (\tau)=\Lambda (f_1, \cdots, f_n|\tau)$ has a limit at $\tau\to -\infty$).
I will prove that $\dot\Lambda (\tau)$ is summable
hence the expression
\begin {equation} \label{eqLt}
\Lambda (\tau_2)-\Lambda (\tau_1)=\int_{\tau_1}^{\tau_2}\dot\Lambda (\tau)d\tau.
\end{equation}
tends to zero as
 $\tau_1,\tau_2\to-\infty$. If  (\ref{eqLt}) tends to zero, then $\Lambda (\tau)$ has a limit. This follows from the completeness of the space in which this vector lies.

Now we need to prove that $\dot \Lambda (\tau)$ is small. Recall that in the definition of $\Lambda (\tau)$ (formula (\ref{eqInLam})) I repeatedly applied the operator $L$ to the state $\omega$. Let us differentiate this expression by $\tau$  applying
 the Leibniz rule.
We obtain several summands, each of which contains a derivative of one of the factors  $L$.  Now we will move $\dot L$ to the right, eventually moving it to the very last place. When I get to the very end, I will use the equality $\dot L \omega=0$. As a result, $\dot \Lambda (\tau)$ will be a summable function of $\tau$, since the commutators $[\dot L(f_i,\tau), L(f_j,\tau)]$ are summable.

Thus we derived the existence of limits from (\ref{LDL}). This is a very important thing. It proves that I can consider the scattering of particles in my picture.  I required very little, but my axioms are sufficient to prove the existence of a limit, to prove that there is a notion of scattering.

The conditions I imposed on $L$ in the case of the geometric approach are 
axioms, while in the case of the algebraic approach similar conditions can be obtained as a consequence of more physical requirements (for example, from the asymptotic commutativity of the algebra $\cal A$). All the above reasoning is valid in the algebraic approach as well. In the algebraic approach, I can impose conditions 
\begin {equation} \label {BDB}||[ \dot B(f_i,\tau), B(f_j,\tau)]||\leq c(\tau),\end {equation}
where $c(\tau)$ is a summable function. The vector $$\Psi (\tau)=B(f_1,\tau)...B(f_n,\tau)\theta$$ will have a limit in the Hilbert space $\bar {\cal H}$ at $\tau \to -\infty$. We can prove this directly by the same method or derive from the existence of the limit (\ref{eqInLam}).
( We should work in the completion of the space $\cal H$  to apply the convergence condition.)

I want to generalize that statement a little bit.  One can a
rgue that  the vector
\begin {equation} \label{eqPsiBB}
\Psi (f_1,\tau_1, ..., f_n,\tau_n)= B(f_1, \tau_1) ...B(f_n,\tau_n)\theta
\end{equation}
has a limit in $\bar {\cal H}$, denoted 
$$\Psi (f_1, ..., f_n| -\infty)$$
as  $\tau_j\to-\infty.$  ( Previously I proved this statement in the case when all times $\tau_j$ are equal.)

The proof can be based on the assumption
$$ ||[ \dot B(\phi),B(\psi)]||\leq \int d{\bx}d{\bx}' D^{ab}(\bx-\bx')|\phi_{\it a}(\bx)|\cdot |\psi_{\it b}(\bx')|,$$
where $D^{ab}(\bx) \to 0$ is faster than any degree when $\bx\to \infty$ (this condition is similar to the condition (\ref{eqLL})).

Analogous statements can be proved if in this assumption or in  (\ref {BDB}) commutators are replaced by anticommutators.

Now I will define the notion of an asymptotic bosonic Fock space $\mathcal {H}_{as}$, assuming that operators $B$ at large distances commute. I will define an asymptotic bosonic Fock space $\mathcal {H}_{as}$ as a Fock representation of the canonical commutative relations:
$$[b(\rho),b(\rho')]=[b^+(\rho),b^+(\rho')]=0, \;\;[b(\rho),b^+(\rho')]=\langle \rho,\rho'\rangle,$$
where $\rho, \rho'\in \textgoth {h}$.

In the case when we consider anticommutators instead of commutators, the bosonic Fock space must be replaced by a fermionic Fock space (by Fock representation of canonical anticommutation relations).

The action of spatial and temporal translations on the elementary space $\textgoth {h}$ can be extended to the Fock space. This is clear because the $n$-particle part of the asymptotic space  is  $n$-th symmetric or antisymmetric power of $\textgoth{h}.$ Corresponding infinitesimal automorphisms  (asymptotic Hamiltonian
and asymptotic momentum operator) are quadratic with respect to creation and annihilation operators; they coincide with Hamiltonian and momentum operator on $\textgoth{h}$ considered as the one-particle subspace of Fock space. The joint spectrum  of asymptotic Hamiltonian and momentum operator coincides with the spectrum of non-interacting bosons or fermions.

Now I will define the M\o ller matrix ( half of the scattering matrix).
The   M\o ller matrix $S_{\pm}$ transforms a vector
$b^+(f_1) ...b^+(f_n)|0\rangle$ from bosonic or fermionic Fock space
 into the state
$\Psi (f_1, ..., f_n| \pm\infty)$.
It is important that the $in$-state is symmetric or antisymmetric. The fact that one can rearrange $f_i$ is essential because otherwise this definition would make no sense since an n-particle subspace in a Fock space is a symmetric or antisymmetric tensor product of the space $\textgoth{h}$. The M\o ller matrices are defined on a dense subset of Fock space.
 In the next lecture, I will prove that it follows from cluster property that $S_-$ and $S_+$ are isometric embeddings of ${\cal H}_{as}$ into $\bar H$, hence they can be extended  to the Fock space considered as a Hilbert space.  If both  M\o ller matrices are not only isometric but also unitary, that is, they are surjective mappings of the Fock space to the entire $\bar H$  then we say that the theory has an interpretation in terms of particles. This means that almost every state is an $in$-state (linear combinations of $in$-states are dense everywhere). In other words, almost every state in the limit $\tau \to -\infty$ evolves to a set of distant particles; similar statement is true for $\tau\to +\infty.$

We obtain a picture similar to the so-called soliton resolution conjecture in classical theory (Section 6.1).

M\o ller matrices commute with translations.  This follows from the formula (\ref{eqTtauLam}), which implies that the action of the time shift of the $in$-state corresponds to the time shift of arguments. The time shift of arguments corresponds to the time shift in Fock space, hence the formula (\ref{eqTtauLam}) says that the time shift in  Fock space corresponds to the time shift in Hilbert space $\bar {\cal H}$.  The fact that M\o ller matrices commute with spatial translations is even easier to prove.

If the theory has particle interpretation the M\o ller matrix $S_-$ specifies unitary equivalence between  the
Hamiltonian and momentum operator in $\bar{\cal H}$ and corresponding operators in the asymptotic Fock space. The same is true for the M\o ller matrix $S_+.$

The scattering matrix (S-matrix) can be defined by the formula
$$S=S_+^{-1}S_-.$$ 
I wrote a similar formula in the soliton picture (Section 6.1). The
 scattering matrix is the main object in quantum field theory. 

Now I will define $in$-operators $a^+_{in} $ using the  limit of operators $B (f,\tau)$ at $\tau\to -\infty $:
\begin{equation}\label{eqININ}
 a^+_{in}(f)= \lim_{\tau\to -\infty} B (f,\tau).
\end{equation}

Why is this a legitimate definition? Let's turn to the formula (\ref{eqPsiBB}), where the operators $B(f_i, \tau_i)$ stand with different times. This means that for one of the arguments I can go to the limit earlier than for the other arguments. It should be emphasized that the $in$-operator is not always defined, but at least if in the formula (\ref {eqININ})  all functions $ f_1, ..., f_n$ do not overlap, the $in$-operator $ a^+_{in}(f_1)$ is defined on the vector $\Psi (f_2, ..., f_n| \pm\infty$ and maps it into the vector  $\Psi (f_1, ..., f_n| \pm\infty.$

In our definition, $in$-operators depend linearly on the functions $ f $. These operators can be regarded as generalized functions; we introduce  the following notation:
$$a^+_{ in}(f)=\int d\bp f^k(\bp)a^+_{ in,k}(\bp),$$ 
where $a^+_{ in,k}(\bp)$ is a generalized function, and the index $k$ specifies the particle type.

 We define $out$-operators in the same way, but  $\tau $ must tend to plus infinity:
$$a^+_{out}(f)= \lim_{\tau\to+\infty} B (f,\tau). $$

Notice, that $in$-operators are related to operators in asymptotic space by formulas:
$$a_{in}^+(\rho) S_-=S_-b^+(\rho),\;\;\; S_-|0\rangle=\theta.$$
These formulas can be considered as an alternative definition of $in$-operators. In the same way one can define  operators  $a_{in}$ and $a_{out}$ associated with annihilation operators in Fock space:
$$a_{in}(\rho) S_-=S_-b(\rho),\;\;\; a_{out}(\rho) S_+=S_+b(\rho).$$

There is an obvious connection between the definitions in the geometric and algebraic approaches.  
If the geometric approach is considered within the algebraic approach, then the operator $L(f,\tau)$ in the space of states corresponds to the operator $B(f,\tau)$ in $\bar {\cal H}$ according to the formula $L(f,\tau)=\tilde {B}(f, \tau)B(f,\tau)$.

The  state $\Lambda (f_1,\dots, f_n|\tau)$  corresponds to the vector $\Psi (f_1,\dots, f_n|\tau)$, and  $\Lambda (f_1,\dots,f_n|-\infty)$ ($in$-state) state corresponds to  $\Psi (f_1,\dots,f_n|-\infty).$

The analog of the M\o ller matrix in the geometric approach is denoted as $\tilde S_-$. While the M\o ller matrix  $S_-$ is a linear operator, $\tilde S_-$ is a non-linear operator. 
For theories that can be formulated algebraically, $S_-$ maps a symmetric power $\textgoth {h}$ treated as a subspace of Fock space into $\bar {\cal H}$.  Taking the composition of this mapping with the natural mapping $\bar {\cal H}$ to the cone of states $\cal C$ we obtain
 ${\tilde S}_-$.  
 
 When $L$ is quadratic or hermitian, it induces a multilinear mapping of the symmetric power of the cone $C (\textgoth {h})$ corresponding to $\textgoth {h}$ into the cone $\cal C.$

 The scattering matrix describes a collision of particles. The connection of the scattering cross-section with the scattering matrix is explained in general courses in quantum mechanics and quantum field theory. I will explain the connection with the inclusive cross-section, and it is simpler. 
 
 First, I must introduce the notion of the inclusive cross-section.
The scattering cross-section is related to the transition probability of, say, a pair of particles to $n$ particles 
$(M,N)\to (Q_1 ...,Q_n)$.  We will consider the process when at the end we obtain particles $(Q_1 ...,Q_n)$ plus something else:
$$(M,N)\to (Q_1,...,Q_m, R_1,..., R_n)$$ 
The inclusive cross-section  is defined as the probability of such a process. One can obtain it from the usual cross-section summing (more precisely, integrating) over $R_1, ..., R_n$.  This is true in a theory having interpretation in terms of particles (this means that everything decays into particles), but I can define the inclusive cross-section even if there is no such interpretation: we can consider a process
$(M,N)\to (Q_1,...,Q_n+ something)$, even if we do not know what is $something.$

 In the geometric approach, only the inclusive cross-section makes sense. In the algebraic approach,  one can work with the usual cross-section but it is possible (and sometimes easier)  to work with the inclusive cross-section.
 
  I consider an arbitrary state $\nu $ and write the following formula for the probability density:
\begin{equation}\label{eqNua}
 \nonumber
 \nu (a^+_{out,k_1}({\bp}_1)a_{out,k_1}({\bp}_1)\dots a^+_{out,k_m}({\bp}_m)a_{out,k_m}({\bp}_m))
\end{equation}
 The expressions $a^+_{out, k_i }({\bp_i })a_{out, k_i }({\bp_i })$ are in fact the numbers of particles with momentum $\bp_i $, so the formula (\ref{eqNua}) represents the probability density in the momentum space of finding $m$ outgoing particles of types $k_1,\dots,k_m$ with momenta $\bp_1,\dots,\bp_m$. I do not look at other particles. 

So far $\nu$ has been any state, but now consider an $in$-state as $\nu$ 
$$\nu=\Lambda (g_1,..., g_n|-\infty)= \lim _{\tau \to -\infty}L(g_1,\tau)...L(g_n,\tau)\omega$$
The $in$-state is determined by the incoming particles. When I defined the inclusive cross-section, I collided two particles, but it is possible to collide several. If in an $in$-state we measure the number of outgoing particles as in the formula (\ref{eqNua}), we get the inclusive cross- section by definition. So if we calculate the expression (\ref{eqNua}) when $\nu$ is an $in$-state, we get an inclusive cross-section.

 We will now represent the answer in a different form. 
 Consider the following expression:

\begin{equation}\label{eqQLom}
\langle 1 | L(g'_1,\tau')...L(g'_{n'},\tau')L(g_1,\tau)...L(g_n,\tau) |\omega\rangle
\end{equation}
assuming that $g'_i$ and $g_j$ do not overlap, and the times tend to infinity, with $\tau '\to +\infty, \tau\to -\infty$.
Acting by operators $L$ on the state $\omega$ we obtain a linear functional on the algebra, so we can calculate its value on the unit element of the algebra. Note that I have used what are called {\it bra-ket notations}, where from the left and from the right are elements from dual spaces. So, consider the expression (\ref{eqQLom}) and denote by $Q$ its limit as $\tau'\to +\infty, \tau\to -\infty$ . Taking the limit $\tau\to -\infty$, we obtain
\begin{equation}\label{eqQLnu}
Q=\lim _{\tau'\to +\infty}\langle 1| L(g'_1,\tau')...L(g'_{n'},\tau')\nu\rangle,
\end{equation}
where $\nu=\Lambda(g_1,...,g_n|-\infty)$. 
Since I assumed that the functions do not overlap, all commutators tend to zero and $Q$ does not change when $g'_1,...,g'_{n'} $ are rearranged.

Now let's look at these formulas in the algebraic approach.  Then $L(g,\tau)=\tilde {B}(g, \tau)B(g,\tau)$. I have the formula $(\tilde M N\nu)(X)=\nu (M^*XN)$. (The operator
  $\tilde M$  multiplies the argument by $M^*$ from the left, the operator $N$ multiplies the argument from the right.) In the formula (\ref{eqQLnu}) these operators are applied to $\nu$. They simply change the argument of $\nu$. In addition, $\langle 1|\sigma\rangle=\sigma(1)$. The result is the expression
$$Q=\lim_{\tau'\to+\infty}\nu (B^*(g'_{n'},\tau')...B^*(g'_1,\tau')B(g'_1,\tau')...B(g'_{n'},\tau')).$$

In the limit $\tau'\to +\infty$ operators $B$ tend  to $in$- and $out$-operators :
$$\lim _{\tau'\to +\infty}B(g,\tau')=a_{out}^+(g), \;\;\;\lim _{\tau'\to +\infty}B^*(g,\tau')=a_{out}(g).$$ 
Using this fact  and the remark that all operators commute in the limit, we obtain the following expression for $Q$:
$$Q=\nu(a_{out}(g'_{n'})...a_{out}(g'_1)a_{out}^+(g'_1)...a_{out}^+(g'_{n'}))).$$
  
I'll call the expression 
$$Q=Q(g'_1,...,g'_{n'},g_1,...,g_n)$$
inclusive scattering matrix.
This expression is quadratic with respect to its arguments. I can switch from quadratic expressions to bilinear expressions - then the number of arguments will double. The resulting expression will also be called the inclusive scattering matrix.   You can get an inclusive cross-section from it. This is not a quite trivial
process. The problem is that in the definition of the inclusive scattering matrix, I considered it as a functional on non-overlapping families of functions.  This functional is linear or antilinear, so it can be regarded as a generalized function, but the arguments of the generalized function (momenta) must be different. In the expression for the inclusive cross-section the momenta can coincide, hence we should take some limits in matrix elements of inclusive scattering matrix to obtain the inclusive cross-section.

In the geometric approach, I can define an inclusive scattering matrix by taking along with $\omega\in \cal L$ some translation-invariant state $\alpha\in \cal L^*$:
\begin{equation}\label{eqAlOm}
 \nonumber
\lim _{\tau'\to+\infty,\tau\to -\infty}\langle \alpha|L(g_1,\tau')...L(g_m, \tau')\times 
L(f_1,\tau)...L(f_n, \tau)|\omega\rangle.
\end{equation}
Such a formula can also be applied in an algebraic situation.  
In it the states $\alpha$ and $\omega$ enter symmetrically. One can formulate it this way: the formula (\ref{eqAlOm}) gives the scalar product of the $out$-state in $\cal L^*$ and the $in$-state in $\cal L$. In other words, the same formula will give
 inclusive scattering matrix of elementary excitations of the  state $\omega$ and inclusive scattering matrix of elementary excitations of the state $\alpha$. 
This is a kind of duality, in my opinion, absolutely mysterious. In the algebraic approach, one can also consider this duality.  

It is important to notice that we can hope to have an interpretation in terms of particles only for elementary excitations of the ground state. Only in this case the conventional scattering matrix makes sense. However, the inclusive scattering matrix and inclusive cross-section make sense also for almost stable quasiparticles (almost stable excitations of any translation-invariant stationary state). 
\newpage
\section{Lecture 8}

\subsection {Link to local quantum field theory. Cluster property.}

What I tell in this course is very different from what is usually told in textbooks on relativistic quantum field theory - they consider local theories. The main idea of the further presentation is to emphasize that locality is not essential in most cases, and the fields themselves are irrelevant. I do not know what should be called fields in the approach I speak about, though all quantum field theory is here. 

I want to begin by establishing a connection between what I am telling you and what is commonly referred to as local relativistic quantum field theory. 

In the axiomatic approach to local theory, there are different systems of axioms starting from Wightman's axioms, where local fields which are generalized operator functions as main objects. This is not very convenient: the fields are local, but they are generalized functions.  If you integrate them, you get ordinary operators. They are no longer local, but in a sense, they are quasilocal (concentrated in some domains). I do not discuss Wightman's axioms.

I will speak about the system of axioms, which belongs to Araki, Haag, and Kastler.  They consider fields concentrated in some open subset of Minkowski space. It is assumed that such fields form an algebra of operators acting in a Hilbert space; the algebra should be closed with respect to weak convergence (this is not essential). These operators should act in the representation space $\cal E$ of a unitary representation of the Poincaré group $\cal P$. 

It is assumed that for each bounded domain (bounded  open subset) $\cal O$ of Minkowski space, we have an algebra of operators $\cal A(\cal O)$ acting in Hilbert space $\cal E$  such that
\begin{itemize}
\item when the domain becomes larger: ${\cal O}_1\subset {\cal O}_2$, then the algebra becomes larger: ${\cal A}{(\cal O}_1)\subset {\cal A}{(\cal O}_2)$;
\item the action of the Poincaré group $\cal P$ on algebras $\cal A(\cal O)$ agrees with the action on domains: $ {\cal A}(g{\cal O})=g{\cal A}({\cal O})g^{-1}$ if $g\in \cal P$;

\item if the space-time interval between points of domains ${\cal O}_1$ and ${\cal O}_2$ is space-like, then the operators belonging to the algebra ${\cal A}{(\cal O}_1)$  commute with operators belonging to the algebra ${\cal A}{(\cal O}_2)$ (roughly speaking, this means that we cannot have a causal relation between observables separated by space-like interval);

\item the ground state $\theta$ of the energy operator is invariant with respect to the Poincaré group
 (having the Poincaré group representation, we can consider the energy operator (Hamiltonian) and momentum operators as infinitesimal generators, respectively, of temporal and spatial translations);
 \item  the vector corresponding to the ground state is cyclic with respect to the union $\cal A$ of all algebras $\cal A(\cal O)$.
\end{itemize}
This is the axiomatics of relativistic local quantum field theory.
\\

In this axiomatics, a particle is defined as an irreducible subrepresentation of the Poincaré group representation in the space $\cal E $.

Let us return now to the definition of scattering in the algebraic approach. The consideration in Lecture 7  is based on axioms, which are not easy to check.
Now I will impose requirements, which are much easier to check. In particular,  they are fulfilled in relativistic local theory. 

My starting point, as before, is an associative algebra $\cal A$ with involution ( $*$-algebra). Space-time translations are automorphisms of this algebra. 

Recall that  non-normalized states correspond to positive linear functionals on the algebra $\cal A$; they 
form a cone $\cal C$. I will work  with non-normalized states. A translation-invariant stationary state will always be denoted as $\omega \in \cal C$.  Excitations of $\omega$ states are elements of the pre-Hilbert space $\cal H$, which is constructed from $\omega$ using the GNS construction. 

In the  algebra $\cal A$ with which I started, there is no norm, but since it is represented in the pre-Hilbert space $\cal H$ and its completion, the Hilbert space $\bar{\cal H}$, one can consider a normed algebra ${\cal A}(\omega) $ consisting of operators $\hat A$.  Moreover, I can work with the completion of the algebra
 ${\cal A}(\omega) $  with respect to this norm, but this is not necessary.

Now let me take an element $A$ of the algebra $\cal A$ which is represented by a bounded operator $\hat A$ in Hilbert space $\bar {\cal H}$. I can consider temporal and spatial translations of this operator. The result will be denoted by $\hat A(\bx,\tau)$. Moreover, I can average such an operator with a smooth and fast decreasing function $ \alpha (\bx,\tau)$:
\begin{equation}\label{SMO} 
B=\int d\tau d\bx \alpha (\bx,\tau)\hat A(\bx,\tau).\end{equation}
It is possible to shift the operator $ B$ in time and space:
$$ B(\bx,\tau)=\int d\tau',d\bx'\alpha (\bx-\bx',\tau-\tau')\hat A(\bx',\tau').$$
One can differentiate under the sign of the integral. Since the function $\alpha (\bx,\tau)$  is assumed to be smooth, one can differentiate as many times as one like. I will always work with operators of the form (\ref{SMO}) and will call them smooth. 

I will consider asymptotically commutative algebras.
In other words, I will require that the commutator of a shifted operator with another operator becomes small at large spatial shifts. This can be formalized in different ways. I will do it in such a way that it is instantly clear that in the axiomatics of Araki, Haag, and Kastler my condition is satisfied. Namely, I will require that the norm $||[B_1(\bx,\tau), B_2]||$ of the commutator of a shifted operator with another operator corresponding to an element of the algebra $\cal A$ decreases faster than any power of $||\bx||$ when $\bx\to \infty$. I will impose the same condition on $||[\dot B_1(\bx,\tau), B_2]||,$ where the dot denotes the time derivative. All operators, let me remind you, are smooth. 

In the axiomatics of Araki, Haag, and Kastler this is always fulfilled, because there after a large spatial shift the space-time interval between the corresponding domains, becomes space-like and therefore we can say that starting from some point the commutator I consider is equal to zero (and thus decreases faster than any power).

Another definition of asymptotic commutativity is the condition
$$||[B_1(\bx,\tau), B_2]||\leq \frac {C_n(\tau)}{1+||\bx||^n},$$
where $C_n(\tau)$ is a polynomial and $n$ is arbitrary (strong asymptotic commutativity).  This condition is satisfied in the Araki, Haag and Kastler axiomatics if the mass spectrum is bounded from below by a positive number.

In addition to the asymptotic commutativity, I want to impose the cluster property on the state $\omega.$  In its simplest form, this means that 
  $$ \omega (A(\bx,t)B)=\omega(A)\omega(B)+\rho(\bx,t),$$
where $\rho(\bx,t)$ is small for large $\bx.$

To formulate the cluster property in general form, I need the notion of a correlation function, which is a generalization of the Wightman function from relativistic quantum field theory. 

I take some elements $A_1 \dots A_r \in\cA$, and shift them in both space and time.  By multiplying them, I get an element of the algebra and after that, I apply $\omega$ or, what is the same, I take the average (the expectation value) of this product in the state $\omega$. The result is: 

$$w_n(\bx_1, t_1,\dots \bx_n,t_n)= \omega (A_1(\bx_1, t_1)\cdots A_n(\bx_n,t_n))=\langle A_1(\bx_1, t_1)\cdots A_n(\bx_n,t_n)\rangle.$$
This is a correlation function. 

It is useful to define the notion of a truncated correlation function 
$$w_n^T(\bx_1, t_1,\dots \bx_n,t_n)\equiv \langle A_1(\bx_1, t_1)\cdots A_n(\bx_n,t_n)\rangle ^T.$$
 This is done somewhat formally by using an inductive formula linking truncated correlation functions to regular correlation functions:
 $$w_n({\bx}_1,\tau_1,k_1\dots,{\bx}_n,\tau_n,k_n) = 
 \sum_{s=1}^n \sum_{\rho \in R_s} w_{\alpha_1}^T(\pi_1) \dots w_{\alpha_s}^T(\pi_s).$$
Here $R_s$ denotes the set of all partitions of the set $\{1,...,n\}$ into subsets $s$ denoted by $\pi_1,...,\pi_s$, the number of elements in the subset $\pi_i$ is denoted by $\alpha_i$, and $w_{\alpha_i}^T(\pi_i)$ denotes the truncated correlation function with arguments ${\bx}_a,\tau_a,k_a$, where $a\in\pi_i$.
This formula expresses correlation functions in terms of truncated functions for all possible partitions of the set of indices. 

When there are only two operators, the truncated correlation function has the form
$$w^T_2({\bx}_1,\tau_1,k_1,{\bx}_2,\tau_2,k_2) = \omega (A_1(\bx_1, t_1) A_2(\bx_2,t_2))-\omega (A_1(\bx_1, t_1)) \omega (A_2(\bx_2,t_2)).$$

Since $\omega$ is translation-invariant and stationary, both usual and truncated correlation functions depend only on the differences $\bx_i-\bx_j, t_i-t_j$. 
We say that the cluster property is satisfied if the truncated correlation functions become small at $\bx_i-\bx_j\to \infty$. Smallness can be understood in different ways, but I mean the strongest condition: at fixed $ t_i$ they tend to zero faster than any power of the difference $d=\min \|\bx_i-\bx_j\|$. More precisely, I assume that
$$ |w_n^T(\bx_1, t_1,\dots \bx_n,t_n)|\leq \frac {C_s(t)}{d^s},$$
where $s$ is any natural number, and $C_s(t)$ is a polynomial function of times $ t_i$.

 We can go to momentum representation by applying Fourier transform with respect to spatial variables.
The invariance with respect to spatial translations leads to the appearance of $\delta$-function of the sum of momenta $\bp_i$. It follows from the cluster property that the truncated correlation function in momentum representation is a smooth function of  momenta multiplied by the $\delta$-function:
$$\nu_n (\bp_2,\dots,\bp_n, t_1,\dots,t_n)\delta (\bp_1+\dots +\bp_n).$$
 ( The Fourier transform of a fast-decreasing function is smooth.)

In relativistic quantum theory, the cluster property is satisfied if the particle masses are bounded from below by a positive number (mass gap).

  \subsection {Green's functions. Connection to the scattering matrix}
  A correlation function is defined as $\omega(M)$ where $M=A_1(\bx_1, t_1) \dots A_r(\bx_r,t_r).$
  In the definition of  Green's function, we replace $M$ by chronological product where the same factors are ordered by time in descending order. This is what is called chronological product. (It is not defined when some times coincide, but this will be irrelevant  in our considerations.) We can say that Green's function is the average (=expectation value) of chronological product with respect to  $\omega.$).  Equivalently, we can say that we are taking the expectation value of this product with respect to he vector $\theta$ corresponding to $\omega$ in the GNS construction.
  
   We obtain the function 
$$G_n=\omega (T(A_1(\bx_1, t_1) \dots A_r(\bx_r,t_r)))=
 \bra \theta| T(\hat A_1(\bx_1, t_1) \dots \hat A_r(\bx_r,t_r))|\theta \ket, $$
which is called the Green's function  
 in $(\bx,t)$-representation (in coordinate representation). 

As always, we can go to the momentum representation by taking the Fourier transform over $\bx$. This will be what is called the $(\bp,t)$-representation (momentum and time). You can also take the (inverse) Fourier transform with respect to the time variable and then the Green's functions will be in the $(\bp,\epsilon)$-representation, where the main variables are momenta and energies. I will need all these representations.

Due to translational invariance, Green's function in the $(\bx,t)$-representation depends on the differences $\bx_i-\bx_j, t_i-t_j$ and, therefore, we have the factor $\delta(\bp_1 + \dots + \bp_r)$  in the $(\bp,t)-
$representation, which corresponds to the momentum conservation law.
In the $(\bp,\epsilon)$-representation we have also the factor $\delta(\epsilon_1+ \dots +\epsilon_r) $, corresponding to the energy conservation law. 

Let us consider poles of Green's function in   $(\bp,\epsilon)$-representation. It should be noted that I always ignore  $\delta$-functions when talking about the poles. 
In particular, when Green's function includes only two operators, in the $(\bp,\epsilon)$-representation we have two momenta, two energies and  $\delta$-functions  depending on momenta and energies:
$$G(\bp_1,\epsilon_1|A,A')\delta(\bp_1+\bp_2)\delta (\epsilon_1+\epsilon_2).$$
 The function $G(\bp_1,\epsilon_1|A,A')$ depends on the variable $\bp_1$ and the variable $\epsilon_1$. It is important to note that the poles of such two-point Green's function with respect to energy at fixed momentum correspond to particles. These poles depend on momentum, and the corresponding function $\varepsilon (\bp)$ gives the dispersion law for particles (dependence of energy on momentum).  These well-known facts can be easily deduced from
of the reasoning that will be used below.

I will prove that in order to find the scattering amplitudes one should consider the asymptotic behavior of Green's function in $(\bp,t)$-representation when $t\to \pm \infty$. This is the first and basic observation. And the other observation is that this asymptotic behavior in the $(\bp,t)$-representation is governed by the poles in the $(\bp,\epsilon)$-representation. More precisely, the asymptotics is described by the residues in these poles. This is called the ``on-shell value of Green function''. 

There is a well-known mathematical fact: If the asymptotic behavior of a function $\rho(t)$ at $t\to \pm \infty$ has the form $e^{-itE_{\pm}}A_{\pm}$ or, put another way, there is a limit 
$\lim _{t\to \pm \infty}e^{itE_{\pm}}\rho (t)=A_{\pm}$,
then the (inverse) Fourier transform $\rho (\epsilon)$ has poles at the points $E_{\pm}\pm i0$ with residues $\mp 2\pi iA_{\pm}.$ In other words, the limit corresponds to the residues in the poles and the exponents correspond to poles; the poles are slightly shifted in the complex plane either up or down from the real axis.  This is an extremely important observation.

One can either look at the poles in the $(\bp,\epsilon)$- representation or look at the asymptotics in the $(\bp,t)$-representation. We show that the calculation of the scattering amplitudes is reduced to finding out the asymptotic behavior of the Green's functions  $(\bp,t)$-representation. Turning to the  $(\bp,\epsilon)$-representation, we can say that the scattering amplitudes are expressed in terms of the on-shell values of Green's functions. This is the Lehmann, Simanzyk and Zimmermann (LSZ) formula.

Below I will prove the LSZ formula under certain conditions. First of all, I assume that the theory has an interpretation in terms of particles. This means that the M\o ller matrices $S_{\pm}$  are unitary. Both $S_-$ and $S_+$ give unitary equivalence between the free Hamiltonian in the asymptotic space ${\cal H}_{as}$ and the Hamiltonian in the space $\cal H$ obtained with the GNS procedure. Second, I assume that the conservation laws for energy and momentum guarantee the stability of particles. The second condition will be relaxed in the next lecture.

I want to simplify the notation, so I will discuss the case when there is only one type of particle. Recall that I considered a generalized function $\Phi(\bp)$ corresponding to the state of a particle with a given momentum $\bp$, and this state is an eigenvector for both momentum and energy operators. The  Hamiltonian acts on $\Phi (\bp)$ as multiplication by the function $\varepsilon (\bp)$ (dispersion law): 
$$\hat H\Phi(\bp)=\varepsilon(\bp)\Phi(\bp),\;\;\;\hat \bP\Phi(\bp)=
 \bp\Phi(\bp).$$ 
We must remember that $\Phi (\bp)$ does not really exist - it is a generalized function. In order for all of this to make exact mathematical sense, we should integrate it with some test function $\phi(\bp)$ to get a vector
$\Phi(\phi)=\int d\bp\phi(\bp)\Phi(\bp).$

Now I want to make an assumption that the one-particle spectrum does not overlap with the multi-particle spectrum.  

Let us formulate this assumption more precisely. Let us denote by ${\cal H}_0$ the one dimensional subspace containing vector $\theta$, by ${\cal H}_1$ the smallest closed subspace of $\cal H$ containing all vectors $\Phi(\phi)$ (one-particle space) and by ${\cal H}_{\cal M}$ the orthogonal complement of the direct sum ${\cal H}_0+ {\cal H}_1$ (multiparticle space).  A corresponding decomposition exists in asymptotic space. {\it I assume that the joint spectra of the Hamiltonian and momentum operator in these three spaces do not overlap. }

The asymptotic Hamiltonian is free. It (and hence $\hat H$) has a spectrum completely determined by the function $\varepsilon (\bp)$. The energies of multiparticle excitations are simply the sums  $\varepsilon (\bp_1)+\dots+\varepsilon (\bp_n)$, corresponding momenta are $\bp_1+...+\bp_n$. If I want to say that the one-particle spectrum does not overlap with the multi-particle spectrum, I must require  the inequality
$$\varepsilon (\bp_1+...+\bp _n)\neq\varepsilon (\bp_1)+...+\varepsilon (\bp_n).$$
 This means that particles with momentum $\bp_1+...+\bp_n$ cannot decay into particles with momenta $\bp_1,...,\bp_n$.  The conservation laws forbid decay.

Now I will formulate the LSZ formula.  To do this I will fix some elements $A_i\in \cal A$ of the algebra $\cal A$. (Recall that I work with smooth elements, but it is not so important here.) Also, I require that by applying the operator $\hat A_i $ to the vector $\theta$ (which in relativistic quantum theory is interpreted as the physical vacuum) and projecting on one-particle space I get a non-zero vector.  More precisely, I require that the projection of the vector $\hat A_i \theta$ be a one-particle state of the form:
$$\Phi (\phi_i)=\int \phi_i(\bp)\Phi(\bp)d\bp,$$
where $\phi_i(\bp)$ is a function which does vanish anywhere.  The projection of this vector onto the vector $\theta$ must vanish.

Let us consider Green functions containing both the elements  $A_i$ and their adjoint elements $A^*_i$. We take  Green's function in $(\bp,t)$-representation:
$$ G_{mn}=
 \omega
 (T(A_1^*(\bx_1, t_1)\dots A_m^*(\bx_m, t_m)
  A_{m+1}(\bx_{m+1}, t_{m+1})\dots A_{m+n}(\bx_{m+n}, t_{m+n})).$$ 
Then we go to $(\bp,\epsilon)$-representation. It is convenient to change the sign of the variables $\bp_i$ and $\epsilon_i$  for $1\leq i\leq m$. 
We multiply the Green's function in the $(\bp,\epsilon)$- representation by the expression: 
 $$\prod_{1\leq i\leq m} \overline {\Lambda_i(\bp_i)}(\epsilon_i+\varepsilon (\bp_i))
  \prod _{m<j\leq m+n} \Lambda_j(\bp_j)(\epsilon_j-\varepsilon(\bp_j)).$$
  where we introduced the notation $\Lambda_i(\bp)=\phi_i(\bp)^{-1}.$
  
Then we take the limit $\epsilon_i\to -\varepsilon(\bp_i)$ for $1\leq i\leq m$ 
and $\epsilon_j\to \varepsilon(\bp_j)$ for $m<j\leq m+n$.

Only the poles will contribute to the limit.  In other words, the calculation boils down to taking residues of the poles.

We can do this procedure in two steps. First, we multiply the Green's function by
$$\prod_{1\leq i\leq m}(\epsilon_i+\varepsilon (\bp_i))
  \prod _{m<j\leq m+n}\epsilon_j-\varepsilon(\bp_j)).$$ and take the limit  $\epsilon_i\to -\varepsilon(\bp_i)$ for $1\leq i\leq m$ 
and $\epsilon_j\to \varepsilon(\bp_j)$ for $m<j\leq m+n$.

At the  end we multiply by $\Lambda_i(\bp)$ for $i>m$  and by  $\overline{\Lambda_i(\bp)}$ if $i\leq m$. In physics, this is called the renormalization of the wave function. In the case when these factors are not included, I will talk about on-shell Green's functions, and if included, I will say that I consider normalized on-shell Green's functions.

The basic statement in the approach of Lehmann, Simanzyk and Zimmermann is that the normalized on-shell Green's function gives the scattering amplitude. To prove it, I will first consider the case where the operators $\hat A_i$ simply give one-particle states $\hat A_i\theta=\Phi (\phi_i)$ (no need to project). We will call them good operators. At the end of the lecture, I will explain that the general case can be reduced to this particular case.

So far we considered the case when there is only one type of particles. Let us consider the case when there are many types of particles, in other words, there are many  functions $\Phi_k(\bp)$ which are eigenvectorss for both momentum and energy: 
$$\bP\Phi_k=\bp\Phi_k(\bp),\;\;\; H\Phi_k(\bp)=\varepsilon_k(\bp) \Phi_k(\bp),$$
but with different dispersion laws $\varepsilon_k(\bp)$ given by smooth functions.  As always, $\Phi_k(\bp)$ are generalized functions, i.e., we should integrate them  with test functions to get vectors from $\cal H$. I consider test functions from the space $\cal S$ of smooth fast decreasing functions.  
To guarantee that time shifts are well-defined in the space $\cal S$, we should assume that the functions $\varepsilon_k(\bp)$ grow at most polynomially.

As already mentioned, I will work with good operators $B_k\in \cA$ (operators which are smooth and transform vector $\theta$ into one-particle states $\hat B_k\theta=\Phi _k(\phi_k)$). Now I define the operator $\hat B_k(f,t)$ depending on the function $f=f(\bp)$ as follows:
$$\hat B_k(f,t)=\int \tilde f(\bx,t)\hat B_k(\bx,t)d\bx,$$ 
where the function $\tilde f(\bx,t)$ is obtained as the Fourier transform of the function $f(\bp)e^{-i\varepsilon_k(\bp)t}$ with respect to the momentum variable.

Similar operators were considered in Section 7.2. They have the property that  applying  them to $\theta$ we obtain a $t$-independent one-particle state
\begin {equation}\nonumber
 \hat B_k(f,t)\theta=\Phi _k(f\phi_k)
 \end {equation}
 hence
 \begin {equation} \label{dotB_kf} 
 \dot{\hat B}_k(f,t)\theta=0,
 \end {equation}
 where the dot stands for the time derivative.
(In Section 7.2, the function $\phi_k$ was equal to $1$.) In general, what was said in Section 7.2 can be repeated here as well.
 The fact that the resulting state is independent of time is the result of a formal calculation.  The calculations become quite simple if we introduce operators
 $$ \hat B_k(\bp,t)=\int d\bx e^{-i\bp\bx}\hat B_k(\bx,t).$$

If this operator is applied to $\theta$, we obtain a one-particle state that does not depend on $t$.

Now I'm repeating the considerations of Section 7.2, but the notations have changed because I do not want to work with elementary spaces. I write the indices explicitly. 

I introduce a vector
\begin{equation}\label{eqPsiDef}
\Psi(k_1,f_1,\dots,k_n,f_n|t_1,\dots,t_n)=\\
\hat B_{k_1}(f_1,t_1) \cdots \hat B_{k_n}(f_n,t_n)\theta,
\end{equation}
where it is assumed that the functions $f_1,\dots,f_n$ have compact supports.

Now,as in Section 7.2, I consider vectors $\bv_i(\bp)=\nabla \varepsilon _{k_i}(\bp)$, which can be interpreted as velocities.  I denote by $U_i$ an open  set containing all possible velocities $\bv_i(\bp)$  where $\bp$ belongs to the support of the function $f_i$. I require that all these sets do not overlap, then I will call the functions $f_1,...,f_n$ non-overlapping. This means that all classical velocities are different and therefore the wave packets are moving in different directions. Then, as I explained in Section 6.3, in the coordinate representation the corresponding wave functions almost do not overlap (essential supports do not overlap). 

Now I will take the limit $t_i\to \infty$.  I will prove that the vector $\Psi(k_1,f_1,\dots,k_n,f_n|t_1,\dots,t_n)$ has a limit, which we denote by
$$\Psi(k_1,f_1,\dots,k_n,f_n|\pm\infty).$$
 
 The proof uses the same reasoning as in Section 7.2.  Again I assume that $t_i=t$ ( all times coincide). In order to prove that there is a limit, I should prove that the derivative with respect to $t$  is a summable function. This condition is satisfied. By definition, the vector $\Psi$ is the result of repeatedly applying  the operators $\hat B_{k_i}(f_i,t_i)$ to $\theta$. When I differentiate this expression with respect to $t$, I have a dot (denoting the time derivative)  over one of the operators $\hat B$. I can move the operator with a dot  to the right using the asymptotic commutativity and (\ref {ESS}) (but only if I work with non-overlapping functions). I get additional summands that are summable functions of $t$. This operator with a dot applied to $\theta$ gives zero due to (\ref {dotB_kf}), so there is a limit. 

Since the limit exists, I can define M\o ller matrices. To do this, I introduce the asymptotic space $\cH_{as}$ as a Fock representation of the operators $a^+_k(f),a_k(f)$ and define the M\o ller matrices $S_-$ and $S_+$ as operators defined on a subset of $\cH_{as}$ and taking values in $\bar {\cH}$ by the formula:
\begin{equation}\nonumber
  \Psi (k_1,f_1,\dots,k_n,f_n | \pm \infty) = S_{\pm}(a^+_{k_1}(f_1\phi_{k_1}) \dots a^+_{k_n}(f_n \phi_{k_n}) |0\ket),
 \end{equation}
where $ |0\ket$ is the Fock vacuum. This is the same formula as in the last lecture with the difference that now we have factors $\phi_{k_i}. $ ( Recall  that a good operator $\hat B_{k_i}$ acting on   $\theta$ gives $\Phi_{k_i}(\phi_{k_i})$.)The M\o ller matrices are defined on a dense subspace of the asymptotic Hilbert space $\cH_{as}$.

It can be proved, and I will do it now, that  M\o ller matrices give isometric embeddings of the asymptotic space  $\cH_{as}$ into the space $\bar{\cH}$. The physical meaning of  M\o ller matrices  can be understood from the following formula (which in other notations was written in the last lecture):
  $$
  e^{-iHt} \Psi (k_1,f_1,\dots,k_n,f_n|\pm\infty)=
  S_{\pm}(a^+_{k_1}(f_1\phi_{k_1}e^{-i\varepsilon_{k_1}t})\dots a^+_{k_n}(f_n\phi_{k_n} e^{-i\varepsilon_{k_n}t})|0\ket).
 $$

 This formula means that when we consider evolution in the space $\bar{\cH}$, in the limit $t\to \pm\infty$ the action of the evolution operator on the vector $\Psi$  corresponds in the asymptotic space to the evolution governed by a free Hamiltonian.  In other words, the evolution of the vector $\Psi$  for large $t$ corresponds  to the evolution of a system of $n$ distant particles with non-overlapping wave functions $f_1\phi_{k_1}e^{-i\varepsilon_{k_1}t},\dots,f_n\phi_{k_n} e^{-i\varepsilon_{k_n}t}$.
 
Our definition of $S_{\pm}$ can be ambiguous. For example, we can use different good operators and it is not clear whether we get the same answer.  However, we can prove that the answer does not depend on our choice. I will derive from cluster property that $S_{\pm}$ are isometric operators. They preserve the norm and preserve the scalar product.  Such operators cannot be multivalued. ( If two vectors coincide, the distance between them is 0, hence two coinciding vectors must go to coinciding vectors.) At the same time we can see that the vector $\Psi (k_1, f_1, ..., k_n,f_n|\pm\infty)$ does not change when the arguments $(k_i,f_i)$ and $(k_j,f_j).$ are permuted.

The main line of proof is as follows.  

To  define M\o ller matrices  we used vectors  $\Psi(k_1,f_1,\dots,k_n,f_n|t_1,\dots,t_n)$  specified by the formula (\ref{eqPsiDef}).  Note that according to (\ref{eqPsiDef}) such a vector is obtained by repeatedly applying the operators $B$ to $\theta$. Iit is easy to see that, the scalar product of two such vectors can be expressed in terms of correlation functions defined as the average values (expectation values)  of products of the operators  $B$ and $B^*.$  The correlation functions are expressed in terms of truncated correlation functions and in truncated correlation functions only two-point correlation functions survive in the limit, $t\to \pm \infty$ if I require cluster property.  This remark relates the scalar product of two vectors of the form  $\Psi(k_1,f_1,\dots,k_n,f_n|\pm\infty)$ to the scalar product in the asymptotic space.

This allows me to say that a M\o ller matrix is an isometric mapping.

Once I have introduced the notion of  M\o ller matrix, I can introduce the notions of $out$-operator and $in$-operator:
\[a_{in} (f)S_-=S_-a(f),\;\; a_{in}^+(f)S_-=S_-a^+(f),\]
\[a_{out} (f)S_+=S_+a(f),\;\; a_{out}^+(f)S_+=S_+a^+(f).\]

 (Here again, I do not write an index describing the type of particles.) 

It is easy to check that
\begin{equation}\label{eqLimain}
\nonumber
  a^+_{in}(f\phi)= \lim_{t\to-\infty} \hat B(f,t),\
  a^+_{out}(f\phi)= \lim_{t\to\infty} \hat B(f,t).
  \end{equation}

The limit in (\ref{eqLimain}) exists on the set of all vectors of the form $\Psi(k_1,f_1,\dots,k_n,f_n|\pm)$ provided that $f,f_1,...,f_n$-is a non-overlapping family of functions. Interestingly, when the dimension of the space $d\geq3$, under some conditions this limit exists without the non-overlapping condition. (This is insignificant for us, because the non-overlapping condition gives a limit on a dense subset, and this is sufficient.)

Now, on the basis of what I have said, I can write out explicitly how the operators we have defined act on the $in$-states:

 \[a^+_{in}(f\phi)\Psi (f_1,\dots,f_n|-\infty)=\Psi (f,f_1,\dots,f_n|-\infty),\]
 \[a^+_{out}(f \phi)\Psi (f_1,\dots,f_n|\infty)=\Psi (f,f_1,\dots,f_n|\infty).\]
 \[a_{in}(f)\Psi (\phi ^{-1}f,f_1,\dots,f_{n}|-\infty)= \Psi (f_1,\dots,f_n|-\infty),\]
 \[a_{out}(f)\Psi (\phi ^{-1} f,f_1,\dots,f_{n}|\infty)= \Psi (f_1,\dots,f_n|\infty).\]

These formulas can be seen as definitions of operators $a_{in}$ and $a_{out}$. Roughly speaking, the operators $a^+_{in/out}$ add one function to $\Psi (f_1,\dots,f_n|\mp\infty)$, and their associated operators destroy one of these functions.

If the operators $S_+$ and $S_-$ are unitary, we say that the theory has an interpretation in terms of particles.
In this case (and in the more general case when the image of $S_-$ coincides with the image of $S_+$), we can define the scattering matrix ($S$-matrix):
\[S=S_+^*S_-.\]
as a unitary operator in the asymptotic space $\cH_{as}$. The asymptotic space is a Fock space. It has a generalized basis 
$$|\bp_1,\dots, \bp_n \ket=\frac {1}{n!} a^+(\bp_1)\dots a^+(\bp_n) |0\ket.$$ 

In this basis, the matrix elements of the unitary operator $S$ (scattering amplitudes) can be expressed in 
in terms of $in$- and $out$-operators. These are the same matrix elements whose squares give the effective scattering cross-sections.  We get the following formula:
\begin{equation}\label{eqSmn}
S_{mn}(\bp_1,\dots,\bp_m|\bq_1,\dots,\bq_n)=\\
\bra a^+_{in}(\bq_1)\dots a^+_{in}(\bq_n)\theta,a^+_{out}(\bp_1) \dots a^+_{out}(\bp_m)\theta\ket
 \end{equation}
This follows directly from the definition of $in$- and $out$-operators.  
In the formula (\ref{eqSmn}) and in the following ones, I will omit the numerical coefficients $(m!)^{-1} (n!)^{-1}.$

The above formula is proved only for the case when all the momentum values $\bp_i,\bq_j$ are different. 
(More precisely, we must assume that all vectors $\bv (\bp_i)=\nabla \varepsilon (\bp_i), \bv (\bq_j)=\nabla \varepsilon (\bq_j)$ are different. When the function $\varepsilon (\bp)$ is strictly convex, it is sufficient to assume that $\bp_i,\bq_j$ are different.  This is not an essential constraint, but it is there.) The formula (\ref{eqSmn}) should be understood in the sense of generalized functions. This means that the set of functions $f_i(\bp_i),g_j(\bq_j)$ with non-overlapping subsets $\overline{U(f_i)}, \overline {U(g_j)}$ should be taken as test functions.

Let us now present the formula (\ref{eqSmn}) in different ways:
 $$S_{mn}(f_1,\dots , f_m|g_1,\dots ,g_n)=$$ 
 $$=\int d^m\bp d^n\bq \prod f_i(\bp_i)\prod g_j(\bq_j) S_{mn}(\bp_1,\dots,\bp_m|\bq_1,\dots,\bq_n)=$$
 $$=\bra a^+_{in}(g_1)\dots a^+_{in}(g_n)\theta, a^+_{out}(\bar f_1)\dots a^+_{out}(\bar f_m)\theta.$$

Recalling that we defined  the S-matrix (scattering matrix) taking limits  $t\to \pm \infty$ and using the formulas (\ref{eqLimain}), we arrive at the following representation:
\[S_{mn}(f_1,, f_m|g_1,\dots,g_n)=\] 
\[\lim_ {t\to \infty,\tau\to -\infty} \bra \theta|\hat B(\bar f_m \phi ^{-1},t)^* \dots \hat B(\bar f_1 \phi^{-1},t)^* 
 \hat B(g_1 \phi^{-1},\tau) \dots \hat B(g_n\phi^{-1},\tau))|\theta\ket=\]
\[\lim_ {t \to \infty, \tau \to -\infty} \omega(B(\bar f_m\bar \phi ^{-1},t)^* \dots B(\bar f_1 \bar \phi^{-1},t)^* B(g_1 \phi^{-1},\tau) \dots B(g_n\phi^{-1},\tau)),\]
  where $B(f,t)^*=\int d\bx B^*(\bx,t)\overline {\tilde f(\bx,t)}.$

We can also write a more general formula
 \[S_{mn}(f_1,\dots , f_m|g_1,...,g_n)= \] \[ \lim_ {t_i\to \infty,\atop{\tau_j\to -\infty}} \omega(B_m(\bar f_m \phi_m ^{-1},t_m)^* \dots B_1(\bar f_1 \phi_1^{-1},t_1)^* 
  B_{m+1}(g_1 \phi_{m+1}^{-1},\tau_1)\dots B_{m+n}(g_n \phi_{m+n}^{-1},\tau_n)) ,\]
where all $B_i$ are different good operators and $B_i\theta=\Phi(\phi_i)$. 

A very important observation: it follows from the non-overlapping condition that in the limit $ t_i\to \infty,\tau_j\to -\infty $ the order of factors is irrelevant both in the group with times tending to +infinity and in the group with times tending to - infinity. This means that for large times    I can rearrange these operators. In particular, I can consider them ordered by time. This means that I can regard the expression under the sign of limit as Green's function. This is the end of the proof of the statement that the matrix element of the scattering matrix can be expressed in terms of the asymptotic behavior of Green's function.

We can express 
  the operators $\hat B_k(f,t)$ in terms of $\hat B_k(\bp,t)$ and get the following result:
$$ S_{mn}(f_1,\dots , f_m|g_1,\dots ,g_n)=$$
$$\int d^{m+n}\bp 
 \lim _{t_i\to\infty,\atop{\tau_j \to -\infty}}\bra \theta| f_m \bar \phi_m^{-1} e^{i\varepsilon_m(\bp_m)t_m}\hat B_m (\bp_m,t_m)^* ...f_1 \bar\phi_1^{-1} e^{i\varepsilon_1(\bp_1)t_1} \hat B_1 (\bp_1,t_1)^*\times$$
   $$ g_1 \phi_{m+1}^{-1}e^{-i\varepsilon _{m+1}(\bp_{m+1})\tau_1} \hat B_{m+1}(\bp_{m+1},\tau_n) ...  g_n\phi_{m+n}^{-1}e^{-i\varepsilon _{m+n}(\bp_{m+n})\tau_n} \hat B_{m+n}(\bp_{m+n},\tau_n)|\theta\ket.$$

We obtain the following formula for the matrix elements of the scattering matrix :

$$ S_{mn}(\bp_1,\dots,\bp_m| \bp_{m+1}\dots,\bp_{m+n})=$$
$$ \lim _{t_1,...,t_m\to\infty,\atop{t_{m+1},...,t_{m+n} \to -\infty}}\bra \theta| \bar \phi_m^{-1} e^{i\varepsilon_m(\bp_m)t_m}\hat B_m (\bp_m,t_m)^* ... \bar \phi_1^{-1} e^{i\varepsilon_1(\bp_1)t_1} \hat B_1 (\bp_1,t_1)^*\times$$
$$\phi_{m+1}^{-1}e^{-i\varepsilon _{m+1}(\bp_{m+1})t_{m+1}} \hat B_{m+1}(\bp_{m+1},t_{m+1}) ... \times$$
$$\phi_{m+n}^{-1}e^{-i\varepsilon _{m+n}(\bp_{m+n}), t_{m+n}} \hat B_{m+n}(\bp_{m+n},t_{m+n})|\theta \ket,$$

This formula tells me that starting with good operators, I can express the scattering matrix in terms of Green's functions in $(\bp,t)$-representation, or, more precisely, in terms of their asymptotics at $ t_i\to \infty $ for $i\leq m$ and $t_j\to -\infty $ for $i>m$. I got factors $\phi_i^{-1}$.  These are exactly the same $\Lambda$ that were introduced in order to get normalized Green's functions. The fact that I am considering asymptotics means that I am taking on-shell Green's functions in the energy representation. The fact that I got factors $\phi_i^{-1}$ means that I get normalized Green's functions, and that's the end of the story.

 I gave proof of  LSZ  formula for good operators. From it, as I said, one can draw a conclusion that it is true for a much broader class of operators.  
 I will explain this in a situation where there is only one type of particle.

 In the approach of Lehmann, Simanzyk and Zimmermann the operators $A_i$ are almost arbitrary. It is only necessary that the projection of the vector $\hat A_i\theta$ on the one-particle states is nonzero, and the projection of this vector on the vector $\theta$ vanishes.
  
What is important to me is that in the definition of the on-shell Green's 
function, these operators can be replaced by smooth operators
$A'_i= \int \alpha (\bx,t)A_i(\bx,t)d\bx dt.$
It is easy to check that this does not change normalized on-shell Green's functions. The proof is based on the remark that $A'_(\bx,t)$ can be obtained as a convolution of $\alpha( \bx,t)$ with $A_i(\bx,t)$. This is the first observation. And the second observation is that for an appropriate choice of $\alpha$ one can consider $A'_i$  as good operators.  
Namely, I can take $\alpha (\bx,t)$ in such a way that the support of its Fourier transform $\hat\alpha(\bp,\omega)$ does not intersect with the multiparticle spectrum and does not contain zero. (I assume that the one-particle spectrum does not intersect the multiparticle spectrum.)  In this case, I automatically obtain a good operator.

 Let me sketch the proof of this fact. I already said that the operator $A'_i\bx,t)$  is obtained  from $A_i(\bx,t)$ by convolution with $\alpha(\bx,t)$. In the $(\bp,\epsilon)$-representation the convolution turns into multiplication by the Fourier transform  $\hat\alpha(\bp,\omega)$ of $\alpha (\bx,t)$. If we consider the spectrum of the energy and momentum operators, multiplication by the function  $\hat\alpha(\bp,\omega)$ in the $(\bp,\epsilon)$-representation kills all points of the spectrum where this function is equal to zero. The function we consider kills the multi-particle spectrum, and we get a good operator.

\newpage
\section{Lecture 9}

\subsection{Introduction and reminder}
In this lecture, as in Lecture  8, I will consider scattering theory in the algebraic approach assuming asymptotic commutativity and cluster property. However, instead of the conjecture that the one-particle spectrum does not overlap with the multiparticle spectrum, I will make a weaker assumption.  In addition to the standard LSZ formula, I will prove its analog for the inclusive scattering matrix.

My reasoning is much the same as in the previous lectures, but I will try to make this lecture independent of the previous two lectures.

 I'll use the notion of generalized Green's functions (Section 5.2). 
These functions appear naturally in the Keldysh formalism and in the formalism of L-functionals.
 I want to show how the inclusive scattering matrix is expressed in terms of generalized Green's functions. For this purpose, I will re-prove the LSZ formula. 
I  show how an ordinary scattering matrix is expressed in terms of ordinary Green's functions, but the proofs are constructed in such a way that it is clear that they can be repeated for inclusive scattering matrix and generalized Green's functions.

I will change the notation a little bit.  I will denote the time variable by 
by the letter $\tau$ and write $B(\tau, \bx)$ instead of $B(\bx,\tau)$. The state $\omega$, as before, will be assumed translation-invariant and stationary, 
the corresponding vector in the pre-Hilbert space $\cal H$ will be denoted by $\theta$.

The ordinary Green's function in the $\omega$ state is the average  (the expectation value) of the product of the operators 
$B_i\in \cal A$ :
$$N= T(B_1(\tau_1,{\bx}_1)\dots B_n(\tau_n,{\bx}_n)),$$
where the times are decreasing (chronological order). In the generalized Green's function 
$$G_{n'n}=\omega (MN)$$
we have both a chronological product $N$, in which times are decreasing, and an anti-chronological product
$$M=T^{opp}(B'^*_1(\tau'_1,{\bx}'_1)\dots B'^*_{n'}(\tau'_{n'},{\bx}'_{n'})),$$ 
in which the times are increasing.  I will prove that the inclusive scattering matrix is expressed in terms of the asymptotic behavior of the generalized Green's functions in the representation where the arguments are momenta and time. By the general properties of the Fourier transform this means that the inclusive scattering matrix is expressed in terms of the poles of the generalized Green's functions, where the arguments are energies and momenta. More precisely, it coincides with the normalized generalized Green's function on-shell.

 \subsection{M\o ller matrix}

The starting point of the theory is a $*$-algebra $\cal A$ where time translations $T_{\tau}$ and spatial translations $T_{\ba}$ act as automorphisms together with translation-invariant stationary state $\omega$. Applying Gelfand- Naimark- Segal (GNS) construction to this state we get pre-Hilbert space $\cal H$ in which the one-particle physical states $\int d\bp f(\bp)\Phi(\bp)$  lie.

 I restrict myself to the case when there is only one type of particle. In other words, we consider the generalized vector function $\Phi(\bp)$, which is an  eigenvector both for the energy operator and the momentum operator:
\begin{equation}\label{eqHP}
\nonumber
H\Phi(\bp)=\epsilon (\bp)\Phi (\bp),\;\;\; \bP\Phi(\bp)=\bp\Phi(\bp).
\end{equation}
(Here you can add an index corresponding to the type of particles, and then it will be a general case.) 

Another definition I will use is that of a smooth operator.
I say that the operator $\hat B$, which is given by the formula
 $$\hat B=\int d\tau d\bx \alpha(\tau,\bx) \hat A(\tau,\bx),\;\;\; \alpha\in{\cal S},A\in \cal A$$
is a smooth operator if $\alpha$ belongs to the space $\cal S$ of fast decreasing smooth functions.
If you make a shift in space and time, the operator
$$\hat B(\tau,\bx)=\int d\tau'd\bx'\alpha (\tau-\tau',\bx-\bx') \hat A(\tau',\bx'),$$
will be a smooth function of $\tau$ and $\bx$. 

For the operator $\hat B(\tau,\bx)$  we take the Fourier transform with respect to spatial variables: 
$$\hat B(\tau,\bx)= \int d\bp e^{i\bp\bx} \hat B(\tau,\bp).$$
We consider the Fourier transform in the sense of generalized functions. (More precisely, the operator  $\hat B(\tau,\bp)$ should be considered as a regular function of $\tau$ and a generalized function of $\bp$.)

It is convenient to introduce  notations: 
$$B(\bp,\tau)=e^{i\epsilon(\bp)\tau}\hat B(\tau,\bp),$$
where $\epsilon(\bp)$ is the dispersion law for the particle in question and
$$B(f,\tau) = T_{\tau}B(T_{-\tau}f)T_{-\tau}=\int d\bp f(\bp) B(\bp,\tau)=\int d\bp d\bx e^{i\epsilon(\bp)\tau-i\bp\bx}f(\bp)\hat B(\tau,\bx),$$
where $f(\bp)$ is a smooth function having compact support.

We assume that the function $\dot B(f,\tau)\theta$ is summable:
\begin{equation}\label{GGG}\int d\tau|| \dot B(f,\tau)\theta||<\infty\end{equation}
 and that
$B(f,\tau)\theta$  tends to one-particle state  as  $\tau\to \pm\infty$:
\begin{equation}\label {GOOD}
\lim_{\tau\to\pm\infty}B(f,\tau)\theta=\Phi(g).
\end{equation}
where $g=f\phi.$  We say that an element $B\in \cal A$ obeying (\ref{GGG}) and (\ref{GOOD}) is admissible. We will verify that  in a  theory having particle interpretation almost all elements are admissible.

The following expression where $B_i$ are admissible elements  can be considered as a vector representing  $n$-particle state: 
\begin{equation}\label{eqPsiB}
\Psi (\tau)=\Psi (f_1,...,f_n|\tau)=B_1(f_1,\tau) ...B_n(f_n,\tau)\theta,
\end{equation}

We define the notion of $in$-state taking in this expression  the limit $\tau \to -\infty$,  and the notion of $out$-state taking the limit  $\tau \to +\infty$:
\begin{equation}\label{eqPsiLim}
\Psi (f_1,...,f_n|\pm \infty)=\lim_{\tau\to\pm \infty}\Psi (f_1,...,f_n|\tau) 
  \end{equation}
 (the limit is taken in the Hilbert space $\bar {\cal H}$).  These limits describe the scattering process. 
They exist under some conditions. The simplest one is to require that the commutators of operators $B_k(f_k,\tau)$ and $\dot B_j(f_j,\tau)$ (where the dot denotes differentiation with respect to $\tau$)  tend to zero fast enough at $\tau \to \pm\infty$ :
\begin{equation}\label{eqBdotB}
||[\dot B_k(f_k,\tau), B_l(f_l,\tau)]||\leq \frac {C}{1+|\tau|^a},
  \end{equation}
where $a>1$ is a fixed number. (It is sufficient to require that the left-hand side is a summable function.) 

If 
this condition is satisfied, then the limit (\ref{eqPsiLim})exists.
The proof is similar to the proof presented in Sections 7.2 and 8.2. Let us differentiate $\Psi (\tau)$ represented by the formula  (\ref{eqPsiB}) with respect to time. If the derivative is a summable function of $\tau$ the limit exists. The Leibniz rule produces $n$ summands, each of which contains a derivative $\dot B_i$. Everything is wonderful when this derivative is at the last place because we assumed that the function $\dot B(f,\tau)\theta$ is summable. If the derivative is not in the last place, then due to conditions on commutators (\ref{eqBdotB}) we can move it to the last place and get  a summable function, but we have to pay with commutators, which also are summable functions. In this case, the difference $\Psi (\tau_1)-\Psi (\tau_2)$, represented as an integral of $\dot\Psi (\tau)$, becomes small at $\tau_1,\tau_2 \to \infty$; hence due to completeness of Hilbert space, we obtain existence of the limit (\ref{eqPsiLim}).

 Our main tool was the smallness of the commutator. 
The statement (\ref{eqBdotB}) saying that the operators entering the expression (\ref{eqPsiB}) almost commute at large $\tau$, can be derived from asymptotic commutativity together with the requirement that the functions $f_k(\bp)$ appearing in formula (\ref{eqBdotB}) do not overlap.  Recall that when we considered the behavior of the wave function $f_k$ as a function of $\tau$, we saw that the set of possible velocities plays an important role in the analysis of evolution in $x$-space; see Section 6.3 (We define the set $U_f$ as an open set containing all vectors of the form $\bv=\nabla \epsilon (\bp)$  where $\bp$ belongs to the support of the function $f(\bp)$. We require that the sets $ U_{f_k}$ do not overlap. Roughly speaking, this means that the particles move in different directions and the essential supports of wave functions in coordinate space are far apart.)  I will impose this condition all the time.  (It is not always satisfied, but I require it to be satisfied for families of functions $f_1, ...,f_n$ belonging  to a dense subset of space ${\cal S}^n$.) If the asymptotic commutativity condition is imposed, it follows that the commutators I need are small.

The asymptotic commutativity condition, first of all, means that the commutator of operators $B_k$ at the same time, but at distant spatial points will be small. The condition of smallness can be varied, but at least I need this commutator to be small in the following sense:
\begin{equation}\label{eqBB}
\| [\hat B_k(\tau,\bx),\hat B_l(\tau, \bx')] \|<\frac {C}{1+|\bx-\bx'|^a},
\end{equation}
where $a>1$ is a fixed number. This estimate must be satisfied not only for the operators themselves but also for their time and space derivatives.

Recall now that all operators $\hat B_k$ are smooth, that is, they can be obtained by smoothing some operators
$\hat B_k=\int g_k(t,\bx)\hat A_k(t,\bx)d\bx dt$,
where $g_k\in\cal S$.  The strong asymptotic commutativity condition can be imposed on the operators $\hat A_k$: 
\begin{equation}\label{eqAA}
\|[\hat A_k(t,\bx),\hat A_l(t',\bx')]\|<\frac{C_a(t-t')}{1+|\bx-\bx'|^a}
   \end{equation}
for any $a$.This means that we require that the commutators decrease faster than any power when the spatial distance tends to infinity. The numerator should contain a polynomial function $C_a(t)$. It is not necessary to impose conditions on derivatives - they follow from (\ref {eqAA}).. 

 In order to derive the condition (\ref{eqBB}) from the strong asymptotic commutativity, it is sufficient to note that it can be reduced to an integral of an expression including the product of functions $f_k$ with different indices, which have essential supports distant from each other and use (\ref {ESS}) ( Section 6.3). These integrals will be small because for distant operators the commutators tend to zero faster than any power. It is very easy to make an estimate in this case. 

One can expect that strong asymptotic commutativity takes place when there is a gap in the spectrum of the Hamiltonian, i.e., the spectrum belongs to the ray $(\epsilon,+\infty)$ starting at some positive  $\epsilon$. In the case of relativistic theory, this corresponds to the case when particles have masses bounded below by a positive number. When the mass is zero, there will be no strong asymptotic commutativity, but weaker conditions ($a>1$) can be fulfilled in conformal theories where all anomalous dimensions $>\frac 1 2$.

 I  define the notion of the M\o ller matrix as follows. 
I start with the notion of asymptotic space.  Asymptotic space 
${\cal H}_{as}$ is defined as the representation space of Fock representation of canonical commutative relations:
$$[a(\bp),a^+(\bp')]=\delta (\bp,\bp'), \;\;\;\;[a(\bp),a(\bp')]=[a^+(\bp),a^+(\bp')]=0. \\
$$
(I am working with momentum variables.) 

Instead of  generalized operator functions $a(\bp),a^+(\bp')$, we can consider operators
$a(f)=\int d\bp f(\bp)a(\bp),a^+(f)=\int d\bp f(\bp)a^+(\bp)$ where $f\in\textgoth {h}=\cal S$. Space and time translations act in the Fock space - this is obvious if we remember that this space can be represented as a completion of the direct sum of symmetric powers of the elementary space $\textgoth {h}=\cal S$. Let us define the M\o ller matrices as mappings
$S_{\pm}:{\cal H}_{as}\to \bar {\cal H}$ from the asymptotic space ${\cal H}_{as}$ to $\bar {\cal H}$ (to the completion of  $\cal H$). 
 This mapping is defined as follows:
\begin{equation}\label{eqMoeller}
\Psi (f_1,...,f_n|\pm \infty)=S_{\pm} 
(a^+(g_1)...a^+(g_n)|0\rangle).
   \end{equation}
   where $f_i$ and $g_i$ are related by the formula (\ref{GOOD}): $\lim_{\tau\to\pm\infty}B(f_i,\tau)\theta=\Phi(g_i).$
   This condition is equivalent to the relation
\begin{equation}\label{ONE}
S_{\pm}a^+(g_i)|0\ket=\Psi (f_i|\pm\infty).
\end{equation}

In the left-hand side of the formula (\ref{eqMoeller}), we have an $in$-state or $out$-state that depends on the functions $f_i$.  We impose the condition that the functions $f_i$ and $g_i$ are related in the following way:  $f_i=g_i\phi_i^{-1}$, where $\phi_i$ does not vanish anywhere. 
Various properties of the M\o ller matrix can be derived from the formula (\ref{eqMoeller}). One of them is that the M\o ller matrix commutes with spatial and temporal translations. This follows almost immediately from the definitions (see Section 7.2). 

Let's show that
formula (\ref{eqMoeller}) defines the M\o ller matrix as an isometric mapping $S_{\pm}:{\cal H}_{as}\to \bar {\cal H}$.

First of all, let us prove that M\o ller matrices do not depend on the choice of - operators $B$. 
Recall that the operators $B$ in the formula (\ref{eqPsiB}) can be different. Let us change one of these operators, the last operator $B_n$ (replace it with another operator), leaving the function $g_n$ unchanged.
Then the limit does change.
This follows from (\ref{ONE})  because the change of the last operator is reduced to the change of the one-particle state. 

Now let us change the operator $B_k$ where $k<n$. Since due to asymptotic commutativity the operators $B_i$ have vanishingly small commutators, I can move the operator $B_k$ to the last place. This must be paid for with commutators, but in the limit it does not change anything. After moving $B_k$ to the last place, we can apply the previous reasoning. This proves that the M\o ller matrix does not depend on the choice of operators $B_i$ used in the definition (\ref{eqPsiB}). 

For the M\o ller matrix to be well- defined, the formula (\ref{eqMoeller}) must have symmetry with respect to the variables $g_i$. This symmetry is present because the commutators are small (if there is strong asymptotic commutativity and we are dealing with non-overlapping functions).

If in addition to the asymptotic commutativity we impose the cluster property, we can prove that the M\o ller matrices are isometric (hence they are single-valued maps that do not depend on the choice of admissible operators $B_i$). Recall that they were defined only for a set of non-overlapping functions, but since they are isometric, they can be extended to the whole asymptotic Hilbert space. We obtain that the operators $S_{\pm}$ are isometric embeddings of the asymptotic space ${\cal H}_{as}$ in the space $\bar {\cal H}$. 

{\it  All the above statements can be derived from cluster property without applying the asymptotic commutativity} but with asymptotic commutativity everything is more transparent.

\subsection {The scattering matrix.  The LSZ formula}

Now I want to define the notion of the scattering matrix. 
This notion is reasonable if the theory has an interpretation in terms of particles, which implies that M\o ller matrices are not only isometric but also unitary operators. In this case, they define an isomorphism of the asymptotic space ${\cal H}_{as}$ and the space $\bar {\cal H}$.  This means, roughly speaking, that any (or almost any) state decays into particles over time. (Something similar was discussed in Section 6.1 under the name  of "soliton resolution conjecture".)

Let us now proceed to the calculation of the matrix elements of the scattering matrix.
Let there be some initial state and a final state, which are denoted, respectively, by the letters $i$ and $f$. The scattering matrix is defined by the formula $S=S^*_+S_-$. Note that $S_-$ specifies the mapping ${\cal H}_{as}\to \bar {\cal H}$, and $S^*_+$ acts in the opposite direction, hence the scattering matrix is an operator in asymptotic space. I must take its matrix elements between states in the asymptotic space: $\langle f|S|i\rangle=\bra f|S_+^*S_-||i\ket$. 

Consider states in the asymptotic space that are obtained from the Fock vacuum by applying some number of $a^+$ operators:
$$|i\ket =a^+(g_1)...a^+(g_n)|0\rangle, \;\;\;\;|f\ket=a^+(g'_1)...a^+(g'_m)|0\rangle.$$
Expressing the matrix element in terms of operators $B$, we arrive at the following formula:
\begin{equation}\label{eqfSi}
 \langle f|S|i\rangle=
\lim_{\tau'_k\to+\infty, \atop {\tau_j\to-\infty}}\\
\bra \theta| B'^*_m(f'_m,\tau'_m) ...B'^*_1(f'_1,\tau'_1) B_1(f_1,\tau_1) ...B_n(f_n,\tau_n)|\theta \ket,
  \end{equation}
where $f_j=g_j\phi_j^{-1}$, $f'_k=g'_k\phi'^{-1}_k$.
The formulas (\ref{eqPsiB}), (\ref{eqPsiLim}) are used here with a slight improvement. Previously, in determining the $in$-vector, it was assumed that all times in the formula (\ref{eqPsiB}) are the same. This is sufficient for the definition of the M\o ller matrix, but here it is convenient (though not necessary) to assume that the times are different, but all tend, respectively, to plus or minus infinity. The proof of this can be easily obtained, but I will not carry it out.

A very important observation is that in the formula (\ref{eqfSi}) all times can be ordered in descending order. Clearly, $\tau'$ is greater than $\tau$ because $\tau' \to +\infty$ and $\tau \to -\infty$.
If we take two times $\tau_{i_1}, \tau_{i_2}$, then the commutator of the corresponding operators is small, since the corresponding functions $f_{i_1}, f_{i_2}$ do not overlap. In such a case I can put the operators in descending order of times and get an expectation value of a chronological product or, in other words, a Green's function. 

In order to keep things simple, I use the standard generalized basis  
$|i\ket=|\bp_1,...,\bp_n\ket$ , $|f\ket=|\bp'_1,...,\bp'_m\ket$.
Actually, in physics, the matrix elements of the scattering matrix must be taken exactly in such a basis, that gives scattering amplitudes of particles with given momenta.
Let us now rewrite the formula (\ref{eqfSi}) in this basis. The operators in this formula can be represented as
$B(f,\tau) = \int d\bp f(\bp) B(\bp,\tau)$. On the other hand, recall that we introduced the notation
$B(\bp,\tau)=e^{i\epsilon(\bp)\tau}\hat B(\tau,\bp).$

As a result, we obtain the following expression for the matrix element in the standard basis:
$$\langle f|S|i\rangle=\lim_{\tau'_k\to+\infty, \tau_j\to-\infty}\omega \big(\prod e^{-i\epsilon(\bp'_k)\tau'_k}(\bar{\phi'_k})^{-1}\hat B'^*_k(\tau'_k,\bp'_k)\prod e^{i\epsilon(\bp_j)\tau_j} (\phi_j)^{-1} \hat B_j(\tau_j,\bp_j)\big).$$
In this formula, we can place the numerical factors in front of the sign of the (linear) functional $\omega$.  In the remaining expression  $\omega \big(\prod \hat B'^*_k(\tau'_k,\bp'_k) \hat B_j(\tau_j,\bp_j)\big)$ we can assume that times are ordered, hence we get a Green's function. This is what I needed. 

We see that the scattering matrix can be expressed in terms of the asymptotic behavior of Green's function in the $(\tau,\bp)$-representation. As already explained, the asymptotics in the $(\epsilon,\bp)$-presentation is determined by the poles with respect to the energy variable. This gives the LSZ formula (for admissible operators).

M\o ller matrices commute with time and space shifts and, hence, with energy and momentum operators. If the theory can be interpreted in terms of particles, then they specify unitary equivalence of  operators   $\hat H$ and $\hat \bP$  in the space $\bar {\cal H}$  with free Hamiltonian $\int d\bp\epsilon (\bp) a^+(\bp)a(p)$ and momentum operator $\int d\bp \bp a^+(\bp)a(\bp)$ in the asymptotic space. It follows that the joint spectrum of $\hat H$ and $\hat \bP$ coincides with the spectrum of a free boson.

Let us introduce $in$- and $out$-operators obeying CCR relations by the formulas:
$$a^+_{out}(f)S_+=S_+a^+(f),\;\;\;\;
a^+_{in}(f)S_=S_-a^+(f),\;\;\;\;
a_{out}(f)S_+=S_+a(f),\;\;\;\;
a_{in}(f)S_-=S_-a(f),\;\;\;\;$$

It is obvious that 
$$\hat H= \int d\bp\epsilon (\bp) a_{in}^+(\bp)a_{in}(p)$$
$$\hat \bP=\int d\bp \bp a^+_{in}+(\bp)a_{in}(\bp).$$

Similar formulas are valid for $out$-operators.

It is important to notice that on a dense subset of $S_-({\cal H}_{as})$ we have
\begin{equation}\label{ASS} a^+_{out}(g)=\lim _{\tau\to+\infty}B(f,\tau),\end{equation}
where $g=f\phi$ is defined  by (\ref{GOOD}) (see Section 8.2).

If there is a representation in terms of particles, any vector  in $\bar{\cal H}$ has the following decomposition: 
\begin{equation}\label{DE}\sum_{r\geq 0}\int d\bp_1...d\bp_r c_r(\bp_1,...,\bp_r)
a_{in}^+(\bp_1)...a_{in}^+(\bp_r)\theta.\end{equation}
(This follows from similar decomposition in asymptotic space.)

Let us represent the vector $|\hat B|\theta \ket$ in the form (\ref{DE}).
For simplicity, assume that  $\bra \theta|\hat B|\theta \ket =0$.  Then we can represent ${\hat B}(f,\tau)\theta$ as

$${\hat B}(f,\tau)\theta=\int d\bp\phi(\bp)f(\bp)a_{in}^+(\bp)\theta+$$ 
$$\sum_{r\geq 2} \int  d\bp_1...d\bp_r e^ {-i\tau(\varepsilon (\bp_1)+...\varepsilon (\bp_r)-\varepsilon (\bp_1+...\bp_r))} c_r(\bp_1,...,\bp_r)f(\bp_1+...+\bp_r)a_{in}^+(\bp_1)...a_{in}^+(\bp_r)\theta.$$

In this formula, the summation should be over all $r$, but for $r=1$ due to cancellation in the exponent the time dependence disappears. The contribution of $r=1$ gives the first summand.  In most cases the sum over $r\geq 2$  tends to zero because 
the exponent  $ {-i\tau(\varepsilon (\bp_1)+...\varepsilon (\bp_r)-\varepsilon (\bp_1+...\bp_r))} $ tends to infinity as $\tau\to\infty$
for almost all values of the arguments. ( If we assume that the energy and momentum conservation laws forbid the decay of the particle, as was assumed in Section 8.2, the exponent always tends to infinity).  We obtain the relation (\ref{GOOD}) with $g(p)=\phi(\bp)f(\bp)$  together with the fact that $\Phi(g)$ is a projection of ${\hat B}(f,\tau)\theta$ on one-particle space is equal to $\Phi(g).$
. 

Differentiating the expression for ${\hat B}(f,\tau)\theta$ with respect to $\tau$  we get (\ref {GGG})  by means of similar reasoning.

This means that in theories having particle interpretation almost all elements $B\in \cal A$ are admissible.
(It seems that all operators encountered in physics are admissible.) 

This concludes the proof of the LSZ formula. 

 \subsection {Inclusive scattering matrix}
Now I want to apply these results to express the inclusive scattering matrix in terms of generalized Green's functions. For this purpose I will consider $in$-states. Earlier I considered them as vectors in Hilbert space, but now I will consider them as  positive functionals on algebra. Every vector $\Psi \in\bar{\cal H}$ specifies a positive functional on the algebra. If I apply some operator, for example $B(f,\tau)$ to a vector $\Psi$, then to get a positive functional corresponding to $B(f,\tau)\Psi$ I should apply an operator $L(f,\tau)=\tilde {B}(f, \tau)B(f,\tau)$ to the positive functional corresponding to $\Psi.$
(Recall that the algebra element defines two operators on functionals: I can either multiply the argument from the right or multiply the argument   by the adjoint element from the left.)

In terms of such operators, the $in$-state corresponding to the vector (\ref {eqPsiLim}) and considered as a positive functional can be written in the following form:
$$\nu=\lim_{\tau\to-\infty}L(f_1,\tau)...L(f_n,\tau) \omega.$$
Now consider the following expression:
$$\langle 1 | L(f'_1,\tau')...L(f'_{n'},\tau')
L(f_1,\tau)...L(f_n,\tau) |\omega\rangle.$$

One can prove the existence of a  limit of this expression as at $\tau '\to +\infty, \tau\to -\infty$ assuming that all functions $f'_i$ and $f_j$ do not overlap.(Recall that asymptotic commutativity is always assumed.) We denote this limit by $Q$; it can be represented in the form
$$Q=\lim _{\tau'\to+\infty}\langle 1| L(f'_1,\tau')...L(f'_{n'},\tau')|\nu\rangle$$

 Expressing $Q$  in terms of functions $g_k,g'_k$  related to functions $f_k,f'k$ by the formula (\ref{GOOD}) we obtain an expression  $Q=Q(g'_1,...,g'_{n'},g_1,...,g_n)$ that I call the inclusive scattering matrix. All inclusive cross-sections can be calculated in terms of the inclusive scattering matrix. This is clear from the
 expression
 $$Q=\nu(a_{out}(g'_{n'})...a_{out}(g'_1)a_{out}^+(g'_1)...a_{out}^+(g'_{n'}))).$$
that  follows the relation  $\langle 1|\sigma\rangle=\sigma(1)$ and the formula $\lim _{\tau'\to +\infty}B(f'_k,\tau')=a_{out}^+(g'_k)$ (see (\ref{ASS})). ( Recall that the notion of inclusive cross-section and inclusive scattering matrix was discussed at the end of Section 7.2.)

$Q$ is a non-linear expression with respect to $g$ and $g'$; it is a quadratic (or rather hermitian) expression. Any quadratic expression can be extended to a bilinear form, and a hermitian expression can be extended to a sesquilinear form - for one variable it will be linear, for another it will be antilinear. 

In order to obtain an expression that will be linear (or somewhere antilinear), I introduce the notation $L(\tilde g, g, \tau)=\tilde B(\tilde f,\tau) B(f,\tau)$. The variables $f$ and $\tilde f$ are separated here, and now what used to be treated as a hermitian expression will be treated as a sesquilinear expression depending on a doubled number of variables: 
$$\rho(\tilde g'_1,g'_1,...,\tilde g'_{n'}, g'_{n'},\tilde g_1,g_1,...,\tilde g_n, g_n)=$$
$$\lim_{\tau'\to +\infty, \atop {\tau'\to -\infty}}
\langle 1| L(\tilde g'_1,g'_1,\tau')...L(\tilde g'_{n'}, g'_{n'},\tau')L(\tilde g_1,g_1,\tau)...L(\tilde g_n,g_n,\tau)|\omega\rangle.$$

This expression will also be called the inclusive scattering matrix and I will express it in terms of generalized Green's functions. First of all, we represent it in the form

$$\rho (\tilde g'_1,g'_1,...,\tilde g'_{n'}, g'_{n'},\tilde g_1,g_1,...,\tilde g_n, g_n)=$$
$$\lim_{\tau'\to +\infty,\atop {\tau\to -\infty}} \bra 1|B(f'_1,\tau')...B(f'_{n'},\tau')\tilde B(\tilde f'_{n'},\tau')\tilde B(\tilde f'_1,\tau')\times $$
$$B(f_1,\tau) ...B(f_n,\tau)\tilde B(\tilde f_n,\tau)...\tilde B(\tilde f_1,\tau)|\omega \ket.$$

It is  convenient to use a  more general formula
$$\rho (\tilde g'_1,g'_1,...,\tilde g'_{n'}, g'_{n'},\tilde g_1,g_1,...,\tilde g_n, g_n)=$$
$$\lim_{\tau'_j,\tilde \tau'_j\to +\infty,\atop {\tau_i,\tilde \tau_i\to -\infty}}
\bra 1|B(f'_1,\tau'_1)...B(f'_{n'},\tau'_{n'})\tilde B(\tilde f'_{n'},\tilde\tau'_{n'})\tilde B(\tilde f'_1,\tilde\tau'_1)\times $$
$$B(f_1,\tau_1) ...B(f_n,\tau_n)\tilde B(\tilde f_n,\tilde\tau_n)...\tilde B(\tilde f_1,\tilde\tau_1)|\omega \ket.$$

We can assume that in this expression the times are ordered. (Part of times tends to $+\infty$, another part to $-\infty$. Within each group, due to asymptotic commutativity, we can rearrange the factors in any order, in particular, in order of decreasing times.) Applying the formula (\ref{MN}) we see that what stands under the limit sign can be expressed in terms of
 generalized Green's functions. The inclusive scattering matrix is expressed in terms of the time asymptotics of this function.
 The same reasoning was used for the usual scattering matrix, only the number of arguments doubled.

\newpage
\section{Lecture 10} 
 \subsection {Elimination of redundant states}
In this lecture, I am planning to explain how one can represent quantum mechanics as classical mechanics, in which we have the possibility to measure only a part of observables. From the point of view of physics, this is quite natural.  Our instruments allow us to measure some observables, but it is possible that there exist better instruments that allow us to measure other things.

I will use the geometric approach to quantum theory. In the geometric approach, we start from a set of states, which is a bounded convex closed subset ${\cal C}_0$ of complete topological vector space $\cal L$. 
The evolution operators must belong to some group $\cal V$, which consists of automorphisms of the set of states ${\cal C}_0$ (either all automorphisms or part of them).  The evolution operator $\sigma_A(t)$ satisfies the equation
\begin {equation}
\frac{d\sigma_A(t)}{dt}=A \sigma_A(t),
\end {equation}
where $A\in Lie (\cal V)$ is an element of the Lie algebra  of the group $\cal V$ (``Hamiltonian''). 
This equation must have a solution (that is, the ``Hamiltonian'' must generate a one-parameter subgroup $\sigma_A(t)$ of the group $\cal V$).   I would like to say that ``Hamiltonians'' are observables in the geometric approach, but the observables should give some numbers, and so I will say that the observable is a pair $ (A,a)$ , where $A\in Lie (\cal V)$ is a ``Hamiltonian'' and $a$ is
 is a linear functional invariant with respect to the group $\sigma_A(t)$ generated by the operator $A$ (this is equivalent to the condition $a(Az)=0$). 
 
 In ordinary quantum mechanics, an observable is specified by ta self-adjoint operator $\hat A$, the corresponding ``Hamiltonian'' acts on the density matrices as a commutator (up to a numerical  factor ), the functional $a$ is defined by the formula $a(K)=\mathrm{Tr} \hat AK.$
 
The group $\cal V$ acts naturally on the observables: $A$ is transformed according to adjoint representation, $a$ as a function on $\cal L$. 

Let us investigate whether there are redundant states in our theory.  When there are two states $x,y\in {\cal C}_0$ such that $a(x)=a(y)$ for any observable $(A,a)$, then we can say that one of these two states can be eliminated. 
If we identify those states which give the same answer for all observables, the result will be a new theory without redundant states, which is essentially equivalent to the original theory. 

 \subsection {Quantum mechanics from classical mechanics}
Let us apply these considerations to the case where the classical theory is taken as the starting point.
In this theory, pure states are points of the phase space (symplectic manifold) $M$. Mixed states are probability distributions on $M$; each mixed state can be uniquely represented as a mixture of pure states. Physical observables are real functions on $M$.
A function $a$ specifies a vector field $A$ on $M$ as a Hamiltonian vector field with Hamiltonian $a$. (Identifying the vector field with a first-order differential operator, we can express $A$ in terms of Poisson brackets: $Af = \{a,f\}$.) We assume that by integrating this vector field we obtain a one-parameter group $\sigma_A(t)$ of canonical transformations (symplectomorphisms) of manifold $M$. This group acts also on mixed states and on observables describing the evolution of these objects. The equation of motion for the probability density  (the Liouville equation) has the form $\frac{d\rho}{dt}=-\{a,\rho\}$, and the equation of motion for the observables has the form $\frac{df}{dt}=\{a,f\}.$

Suppose that our devices are able to see only a part of observables and that the set $\Lambda$ of ``observable observables'' is a linear space closed with respect to the Poisson bracket. Let us index this set by elements of  a Lie algebra denoted by $\textgoth{g}.$ (The mapping $\gamma \to a_{\gamma}$ transforming $\gamma\in \textgoth{g}$ to $a_{\gamma}\in \Lambda$ is an isomorphism of Lie algebras $\textgoth{g}$ and $\Lambda$.) 

Hamiltonian vector fields $A_{\gamma}$ with Hamiltonians $a_{\gamma}$ specify an action of Lie algebra $\textgoth{g}$ on $M$.
The assumption that vector fields $A_{\gamma}$ generate one-dimensional subgroups means that this action is induced by the action of simply connected Lie group $G$ having $\textgoth{g}$ as a Lie algebra.

The above considerations can also be applied to infinite-dimensional symplectic manifolds and to infinite-dimensional Lie algebras and Lie groups. However, in the infinite-dimensional case, these considerations are not rigorous.

Define the moment  map $\mu$ from $M$ to $\textgoth{g}^{\vee}$ as a mapping $x\to \mu_x$, where $\mu_x(\gamma)=a_{\gamma}(x)$. (Here $ x\in M, \gamma \in \textgoth{g} $, and $\textgoth{g}^{\vee}$ denotes the space of linear functionals on $\textgoth{g}.$ ) This mapping is $G$-equivariant with respect to the coadjoint action of $G$ on $\textgoth{g}^{\vee}.$ (In other words, it commutes with transformations from the group $G$.)
For each state of the classical system (for each probability distribution $\rho$ on $M$), we can define a point $\nu (\rho) \in \textgoth{g}^{\vee}$ as an integral of $\mu_x$ over $x\in M$ with the measure $\rho$:
$$\nu(\rho)=\int_{x\in M } \mu_xd\rho.$$
The point $\nu (\rho)$ belongs to the convex envelope $N$ of $\mu (M).$ (The convex envelope of a subset $E$ of a topological vector space $\cal L$ is defined as the smallest convex closed subset  of $\cal L$ containing $E.$)

The group $G$ acts naturally in the space of classical states. It follows from the $G$-equivariance of moment map that the mapping $\nu$ is a $G$-equivariant mapping of classical states
into $\textgoth{g}^{\vee}$ with a coadjoint action $G.$

We  say that two classical states (two probability distributions $\rho$ and $\rho'$) are equivalent if
\begin {equation}
\label{eqE}
\int_{x\in M } a_{\gamma}(x)d\rho= \int _{x\in M } a_{\gamma}(x)d\rho'
\end {equation}
for each
$\gamma\in \textgoth {g}.$
In other words, we say that two states are equivalent if the calculations with these states give the same results for each Hamiltonian $a_{\gamma}.$ (Our devices cannot distinguish these two states.)

Let us now prove the following statement:
{\it two states $\rho$ and $\rho'$ are equivalent if and only if $\nu(\rho)=\nu(\rho').$}

First, note that for each $\gamma\in \textgoth {g}$
$$\nu (\rho)(\gamma)=\int_{x\in M }  \mu_x(\gamma)d\rho=\int_{x\in M } a_{\gamma}(x)
d\rho$$
Similarly,
$$\nu (\rho')(\gamma)=\int_{x\in M }  \mu_x(\gamma)d\rho'=\int_{x\in M } a_{\gamma}(x)
d\rho'.$$
These formulas imply our statement.

In the classical theory where only Hamiltonians from the set $\Lambda = \{ a_{\gamma} \}$, where $\gamma\in \textgoth {g}$, are allowed, equivalent states must be identified (we eliminate redundant states). The mapping $\nu$ induces a bijective mapping of the space of equivalence classes to the set $N$, obtained as a convex envelope of the set $\mu(M)$ (``quantum states''). $G$-equivariance of the mapping $\nu$
means that the evolution of classical states is consistent with the evolution of quantum states.

Let us apply our constructions to the complex projective space $\mathbb {C}\mathbb {P}.$ We define this space as
sphere $||x||=1$ in the complex Hilbert space $\cal H$ with identifications $x\sim\lambda x$, where
$\lambda \in \mathbb {C}, |\lambda|=1.$ Group $U$ of unitary operators acts transitively on
$\mathbb {C}\mathbb {P}.$ There exists a single (up to a constant factor) $U$-invariant symplectic structure on this space; this allows us to consider the complex projective space as a homogeneous symplectic manifold. 

Suppose that we can observe only Hamiltonians of the form $a_C(x)=\bra x, Cx \ket$
where $C$ is a self-adjoint operator. The set $\Lambda$ of such Hamiltonians is a Lie algebra with respect to the Poisson bracket. This Lie algebra is isomorphic to the Lie algebra $\textgoth {g}$ of self-adjoint operators where the operation is defined as the commutator multiplied by $i.$ The one-parameter group of unitary operators corresponding to the Hamiltonian $a_C$ is given by the formula
$\sigma_ (t) = e^ {-iCt}.$ 

Moment map transforms the point $x$ into a linear functional on the space of self-adjoint operators, mapping the operator $ C$ into $\mathrm{Tr} K_x C$, where $K_x$ is the projection
$\cal H$ on vector $x$, i.e. $K_x (z) = \bra z, x\ket x. $ (Recall that in our notation the points of the complex projective space are represented by normalized vectors.) The convex envelope of the image of the moment map consists of positively defined selfadjoint operators with unit trace (i.e. it consists of density matrices).

We see that by applying our general construction to a complex projective space we obtain ordinary quantum mechanics. In this case, our considerations are close to ``non-linear quantum mechanics'' of Weinberg who proposed to consider the classical theory on $\mathbb {C} \mathbb {P}$ as a deformation of quantum mechanics.

The orbits of the   group $G$ in a coadjoint representation (in the space $\textgoth {g}^{\vee}$ dual to the Lie algebra $\textgoth {g}$
of the $G$ group) provide a rich source of examples of the above construction. These orbits play an important role in the representation theory. They are homogeneous symplectic manifolds. Quantizing them, one can obtain
unitary representations of the group $G$.

In order to introduce the structure of a symplectic manifold on an orbit, we should note that for functions on $\textgoth {g}^{\vee}$  one can define a Poisson bracket. (Elements of Lie algebra $\textgoth {g}$ can be thought of as linear functions on $\textgoth {g}^{\vee}$ ; for these functions the Poisson bracket is defined as an operation in Lie algebra. Using the properties of the Poisson bracket, we can define the bracket for arbitrary smooth functions).  The space $\textgoth {g}^{\vee}$ gets the structure of a Poisson manifold, but this manifold is not symplectic  (the Poisson bracket is degenerate).  By restricting the Poisson bracket to an orbit, we obtain a nondegenerate Poisson bracket specifying a symplectic structure. 

Elements of the Lie algebra $\textgoth {g}$ define a family  $\Lambda$ of functions on the orbit; we apply the above construction to this family. In the case under consideration, the moment map 
is simply the embedding of the orbit into $\textgoth {g}^{\vee}$.  We see that considering only Hamiltonians from the family $\Lambda$ as '' observable observables" in the classical theory, we obtain a theory in which the set of states is a convex envelope of the orbit. The group $\cal V$ can be identified with the group $G$.

Let us illustrate the above constructions in the case where $G$ is the group $U$ of unitary transformations of  a Hilbert space $\cal H.$ In this case we can identify the elements of the Lie algebra 
$\textgoth {g}$ with bounded self-adjoint operators. With a suitable choice of topology in $\textgoth {g}$ one can identify the dual space $\textgoth {g}^{\vee}$ with a linear space of self-adjoint operators having a trace. To simplify the notations, we assume that the Hilbert space is finite-dimensional, but our considerations can also be applied to the infinite-dimensional case. 

Consider the orbits of $U$ in the space $\textgoth {g}^{\vee}$ (in the coadjoint representation).

If $\dim {\cal H}=n$, the orbit is indexed by  real numbers $\lambda_1,..., \lambda_r$ (eigenvalues of operators belonging to the orbit) and positive integers $k_1,...,k_r$ (multiplicities of eigenvalues). (The multiplicities must satisfy the condition $k_1+...k_r=n$.) The stationary group $U(n)$ of the point belonging to the orbit is isomorphic to the direct product of groups $U(k_i)$, so the orbit is homeomorphic to $U(n)/U(k_1)\times...\times U(k_r)$ (to the flag manifold). 

If $r = 2$, then the orbit is homeomorphic to the Grassmanian. If $r = 2, k_1 = n-1, k_2=1$, we obtain a complex projective space.
(The Grassmanian $G_k (\cal H)$ is defined as the space of all $k$-dimensional subspaces of the space $\cal H$. It can be viewed as a symplectic $U$-manifold.) 
\newpage
\section{ Notations, conventions, definitions}  

Talking about vector spaces we always have in mind vector spaces over  $\mathbb {R}$ or over $\mathbb{C}.$ (If the field is not specified  we have in mind complex vector space.)

An algebra is a vector space over $\mathbb{R}$ or over $\mathbb{C}$ equipped with an operation of multiplication satisfying the distributivity axiom
$$ (a+b)c=ac +bc,  c(a+b)=ca+cb$$
where $a,b$ are elements of the algebra, $c$ is an element of the algebra or a number.

An algebra is unital if there exists an element $1$ obeying $1\cdot a=a\cdot 1=a.$

An algebra is associative if $(ab)c=a(bc).$

A $*$-algebra is an associative algebra with antilinear involution $^*$ obeying $a^{**}=a,
$ and $(ab)^*=b^*a^*.$  A homomorphism or automorphism of $*$-algebra should agree 
with involution. We use notations $\cal A$ for $*$-algebra and $Aut (\cal A)$ for its group of automorphisms.

The algebra of bounded linear operators in Hilbert space is a $*$-algebra with respect to the involution $A\to A^*$
where $A^*$ denotes the operator adjoint ( Hermitian conjugate) to $A.$ We say that a self-adjoint operator $A$
is positive definite if $\bra x, Ax\ket\geq 0.$  (Notice, that in more standard terminology such operators are called positive semi-definite.)

A Lie algebra is an algebra with an operation $[a,b]$ obeying $[a,b]=-[b,a]$ and
$[[ab],c]+[[b,c],a]+[[c,a],b]=0$ (Jacobi identity).

A topological group, vector space, algebra,... is a group, vector space algebra,...equipped with 
topology in such a way that all operations are continuous.

We always assume that a map(=mapping) of topological spaces  (or homomorphism of topological groups, etc) is continuous.  

Notice that sometimes it is convenient to consider topological algebras where the operation of multiplication is defined only on a dense subset. For example, we can consider the Lie algebra of  (not necessarily bounded )  self-adjoint operators in Hilbert space. ( The commutator of two unbounded self-adjoint operators is not necessarily well-defined.)

A derivation of an algebra is a linear operator $D$ obeying Leibniz rule $D(ab)=Da\cdot b+a\cdot Db.$  A commutator of derivations is a derivation, hence we can talk about the Lie algebra of derivations.

 If we are dealing with a topological algebra we can consider derivations defined on a dense subset.  Their commutator is not necessarily well-defined, but still, we can regard the set of such derivations as topological Lie algebra that can be considered as Lie algebra of the automorphism group of the algebra.
 
 We say that the derivation $D$ is an infinitesimal automorphism it can be considered as 
 a tangent vector of a one-parameter subgroup of the automorphism group, i.e. there exists a solution of the equation $dU/dt=DU(t)$ with initial condition $U(0)=1$. (This solution can be written in the form $\exp(Dt).$
 
 The set of infinitesimal automorphisms also can be regarded as topological Lie algebra, which can be interpreted as Lie algebra of the group of automorphisms.
 
 More generally if we have a topological group we can define a Lie algebra of this group either starting with the set of tangent vectors to the curves in the group at the unit element or starting with the set of tangent vectors to one-parameter subgroups.
 
 Notice that the above definitions and statements are not rigorous. For example, I did not specify the topology in the group of automorphisms, I did not give a definition of tangent vector, etc.  One should remember that one can give several reasonable definitions of these and other notions.  I always disregard these subtleties.
 
  We denote by ${\cal S}={\cal S}(\mathbb{R}^n)$ the space of smooth fast decreasing functions on $\mathbb{R}^n$ (Schwartz space).
  More precisely, this space consists of  smooth functions such that 
  \begin {equation} \label {SEMI} \sup |x_1^{\alpha_1}...x_n^{\alpha_n} \partial_1^{\beta_1}...\partial_n^{\beta_n}|
  \end{equation}
  is finite for any choice of non-negative integers $\alpha_i, \beta_j.$
  
  The expressions (\ref{SEMI}) can be regarded as seminorms specifying the topology in $\cal S.$
  
  We consider generalized functions on $\mathbb{R}^n$ (distributions) as continuous linear functionals on $\cal S.$
  
  More generally, generalized functions are defined as linear functionals on some topological vector space of functions on $\mathbb {R}^n$ ( on the space of test functions); one represents such a functional as formal integral: $f(\phi)=\int dx f(x)\phi(x).$
  
  In the above definitions test functions (hence generalized functions) can be regarded as functions taking values in a space $\mathbb{C}^r$ (as vector-valued functions).

  The elementary space $\textgoth{h}$ can be considered as a space of smooth fast decreasing 
  functions on $\mathbb{R}^d$ taking values in $\mathbb{C}^r.$
  
  Spatial translations (=spatial shifts) are denoted by $T_{\ba}$. In coordinate representation, they act on an element of $\textgoth{h}$ as a shift of the argument $\bx$ by $\ba$, in momentum representation they act as multiplication of the function $f(\bk)$ by $e^{i\ba\bk}.$ (Coordinate and momentum representation are related by Fourier transform.)
  
  Temporal translations (=time translations=time shifts) are denoted by $T_{\tau}.$ They should commute with spatial translations; it follows that in momentum representation they act on the elements of $\textgoth{h}$ by the formula $(T_{\tau}f)(\bk)=e^{-i\tau E(\bk)}f(\bk).$ 
  
  The notations $T_{\ba}$ and $T_{\tau}$ are used for spatial and temporal translations not only in the elementary space but also in other situations. Together the space and time translations generate a commutative group denoted by $\cal T.$
  
  We assume that all convex sets we consider are  closed subsets of some complete topological vector space $\cal L.$ Convex envelope of set $E\subset \cal L$ is the smallest convex subset of $\cal L$ containing $E.$
  
  Convex cone $C$ is by definition a closed convex set that is invariant with respect to dilations $x\to \lambda x$ (here $\lambda$ denotes a non-negative real number). Notice, that this definition of a convex cone is not quite standard, usually one imposes some additional conditions.

 \newpage

\vskip .5in
\section {Some problems} (Mostly mathematical problems related to the material of  my lectures)
\vskip .1in

0. A complex vector space $E$ is equipped with non-negative scalar product.  Prove that we can obtain pre Hilbert space factorizing $E$  with respect to vectors with$  \bra x,x\ket =0.$

Hint. Check that these vectors constitute a linear subspace of $E.$

\vskip .1in

Density matrices are defined as positive-definite  self-adjoint operators having unit trace and acting in complex Hilbert space.

1. Prove that  the set of density matrices is convex. Check that extreme points of this set are one-dimensional projectors $K_{\Psi}(x)= \bra x,\Psi\ket \Psi$ where $||\Psi||=1$ (they are in one-to-one correspondence with non-zero vectors of Hilbert space with identification $\Psi\sim \lambda \Psi).$

2. Prove that the set of density matrices in two-dimensional Hilbert space is a three-dimensional ball and set of of its extreme points is a two-dimensional sphere.

The linear envelope $\cal T$ of the set of density  matrices  is the space of all self-adjoint operators belonging to trace class.  (A self-adjoint  operator belongs to trace class if it has discrete spectrum  and the the series of its eigenvalues is absolutely convergent.)  We  consider $\cal T$ as a normed space with the norm $||T||=\sum |\lambda _k| $ where $\lambda_k$ are eigenvalues of $T.$  By definition an automorphism of the set of density matrices is a  bicontinuous linear operator in $\cal T$ generating a bicontinuous map  of the set density matrices.
(One says that a map is bicontinuous if it is continuous and has a continuous inverse.) It is obvious that a unitary operator specifies an automorphism of the set of density matrices by the formula $T\to UTU^{-1}.$

3.Prove that  automorphisms of the set of density matrices are in one-to-one correspondence with unitary operators. 

 I do not know how to  solve  this problem (this does not mean that it is difficult, I did not try). Maybe this fact is proved somewhere, but I do not know any reference.

       In the next problems  the term ''operator''  means ''linear operator".  It is convenient to define $e^{tA}$ where $A$ is an operator as a solution of the equation 
$$ \frac {d U(t)}{dt}= A \cdot U(t)$$
with initial condition $U(0)=1.$ If $(A_1, ..., A_n)$ is a family of commuting operators we define
$e^{\sum t_iA_i}$ as a product $e^{t_1A_1}...e^{t_nA_n}.$

4. Prove that $e^{i\ba \bP}=T_{\hbar \ba},$
 where $\bP$ denotes the momentum operator $\bP=\frac {\hbar}{i}\nabla$ and $T_{\ba}$ stands for translation operator transforming the function $f(\bx)$ into a function $f(\bx+\ba).$
 
 5.  Let us define an operator  ${\cal C}_A$ acting in the space of operators by the formula ${\cal C}_A(X)=[A,X].$ (For definiteness one can assume that $A$ and $X$ are bounded operators in Hilbert space, but this assumption is not important.) Prove that 
 $$e^{t{\cal C}_A}(X)=e^{tA}Xe^{-tA},$$
 or equivalently
 $$e^{tA}Xe^{-tA}=X+t[A,X]+\frac {t^2}{2!} [A,[A,X]]+\frac {t^3}{3!} [A,[A,[A,X]]]+ ...$$
 
 Hint. Differentiate these equalities.
 
6. Let us assume that the commutator  of operators  $X$ and $Y$  is a number $C$ (or, more generally, an operator $C$, commuting with $X$ and $Y$). Prove that 

$$e^Xe^Y=e^{X+Y}e^{\frac {1}{2}C},$$

$$e^Xe^Y=e^Ce^Ye^X.$$

7.  Let us assume that for an operator $A$ acting in Banach space the norms of operators $e^{tA}$ where $t\in \mathbb {R}$ are uniformly bounded (i.e.
$$\sup_{-\infty <t<+\infty}||e^{tA}||$$
is finite). Prove that all eigenvalues of$A$ are purely imaginary.

8. Let $A$ denote an operator acting in finite-dimensional  complex vector space. Assume that  the norms of operators $e^{tA}$ where $t\in \mathbb {R}$ are uniformly  bounded. Prove that  the operator $A$ is diagonalizable (i.e. there exists a basis consisting of eigenvectors of $A$).

Hint. Use Jordan normal form. Prove that all Jordan cells are one-dimensional.


9.   Let us consider Grassmann algebra with generators $\epsilon_1,\epsilon_2,\epsilon_3, \epsilon_4.$ Calculate 
$$(\epsilon_1+\epsilon_2)(\epsilon_2+\epsilon_3) (\epsilon_3+\epsilon_4),$$
 $$\int(\epsilon_1+\epsilon_2)(\epsilon_2+\epsilon_3) (\epsilon_3+\epsilon_4) (\epsilon_4+\epsilon_1)d^4\epsilon,$$
 $$\frac {\partial}{\partial \epsilon_2}(\epsilon_1\epsilon_2\epsilon_3),$$
 $$\cos (\epsilon_1\epsilon_2+\epsilon_3\epsilon_4).$$
 
 10.Let us consider  a unital associative algebra with commuting generators $x_1,...,x_n$ and anticommuting generators $\xi_1,...,\xi_n$ (tensor product of polynomial algebra and Grassmann algebra).  Prove that $d^2=0$ and $\Delta^2=0$
 where
 $$d=\sum\xi_i\frac {\partial}{\partial x_i},$$
 $$\Delta=\sum\frac {\partial}{\partial x_i}\frac {\partial}{\partial \xi_i}.$$
 \vskip .1in
 
  Weyl  algebra is defined as unital associative algebra with generators $a^*_k, a_k$  obeying CCR  ($[a_k,a_l^*]=\delta_{k,l}, [a_k, a_l]=[a^*_k, a^*_l]=0).$ An involution $^*$  in Weyl algebra  transforms $a_k$ into $a^*_k.$
   Fock representation of Weyl algebra=representation with cyclic vector $ |0\ket$ obeying $\hat a_k |0\ket=0.$ Scalar product is defined by the condition that $\hat a_k^*$ is adjoint to $\hat a^k$
  
  Fock space= completion of the space of Fock representation.
 
 11.  Poisson vector  is defined  by the formula $\Psi_{\lambda}=e^{\lambda \hat a^*} |0\ket$
 where ${\lambda \hat a^*}=\sum \lambda_k\hat a_k^*.$
 
 a) Check that Poisson vector is an eigenvector of  all operators $\hat a_k$
 
 b) Find scalar product of two Poisson vectors.

 12. Let us consider the Hamiltonian $\hat H(t)=\omega(t)\hat a^*\hat a$ (harmonic oscillator with time-dependent frequency).
 
 a) Let us suppose that $\omega(t)= \omega_0+\omega \exp (-\alpha t)$ for $t\geq 0$. (Here $\alpha$  is a small positive number.)
 Calculate the evolution of matrix entries of the density matrix in $\hat H$-representation (i.e. express matrix entries of $K(t)$ in terms of matrix entries of $K(0)$). Here $\hat H=\hat H(0).$  
 
 b) Assuming that $\omega$ is a  random variable with a given expectation value and dispersion uniformly distributed on some interval calculate the average of matrix entries of the density matrix $K(t)$ for $t=\frac{const}{\alpha}.$
 
 13.  A molecule  is placed near a microwave oven.  Give a rough estimate of decoherence time for this molecule.
 
 Hint.  Decoherence appears if the phase factors entering the expressions for non-diagonal matrix entries of density matrix are changed significantly by the electric field of the oven. You can calculate
 this change in perturbation theory; the first order contribution comes from dipole momentum. You can find the information about dipole momenta
 of ground state and excited states on the web; you need only the order of magnitude of these momenta and of electric field.
 
 14. Let us define the $L$-functional corresponding to the density matrix $K$  by the the formula 
 $$ L_K (\alpha)=\mathrm{Tr} e^{-\alpha\hat a^*}e^{\alpha^*\hat a}K.$$ 
 (We consider the  case when there is only one degree of freedom,
 $[\hat a, \hat a^*]=1.$)
 
 Calculate the $L$-functional corresponding to the coherent state (to the normalized Poisson vector).
 
 Reminder. Every normalized vector $\Phi$ defines a density matrix $K_{\Phi}$ and $\mathrm{Tr} AK_{\Phi}=\langle A\Phi,\Phi\rangle.$ Coherent state is a normalized eigenvector of $\hat a.$

15. 
 Let us assume that   $\rm {supp}(\phi)$ is a compact set. Then  for large $|\tau|$ we have
$$ |(T_{\tau}\phi)({\bx})|<  C_n (1+|{\bx}|^{2}+\tau^2)^{-n}$$
where $ \frac {{\bx}}{\tau}\notin U_{\phi}$, the initial data $\phi=\phi ({\bx})$ is the Fourier transform of $\phi ({\bk})$, 
and $n$ is an arbitrary integer.

Here  $\rm {supp}(\phi)$ is the closure of the set of points where $\phi (\bk)\neq 0$, 

$U_{\phi}$ is a set of all points of the form $\nabla \epsilon (\bk)$
where $\bk$ belongs to a neighborhood of  $\rm {supp}(\phi)$,

 $T_{\tau}\phi)({\bx})= \int d\bk e^{i\bk\bx-i\epsilon (\bk)\tau}\phi(\bk)$,
 
 the function $\epsilon(\bk)$ is smooth.

 16.  A generalized function $\rho(\epsilon)$ is a sum of of the function $\frac {A}{\epsilon-E-i0}$ and a square integrable function.  Find the asymptotic behavior of its Fourier transform $\rho(t).$
 
 17.  Let us consider a system with  the Hamiltonian $\hat H=\int\epsilon (\bp)\hat a^*(\bp)\hat a(\bp) d\bp$ and momentum operator $\hat \bP=\int  \bp\hat a^*(\bp)\hat a(\bp) d\bp$ where $\hat a, \hat a^*$ obey CCR.
 
 a) Calculate time and space translations in the formalism of $L$-functionals
 
 b) Find the operators $\hat a(\tau,\bp), \hat a^*(\tau, \bp)$ and corresponding operators acing on $L$-functionals.
 
 c)  Prove that the $L$-functional $\omega=\exp (-\int d\bp \alpha^*(\bp) n(\bp)\alpha(\bp))$  specifies a  stationary translation-invariant state. (Here $n(\bp)$ is an arbitrary function.)
 
 d) Calculate two-point generalized Green functions for the state $\omega$ in $(\tau,\bp)$- and $(\varepsilon,\bp$)-representations.

\end{document}